\newif\ifreport\reporttrue
\newif\ifjournal\journalfalse
\documentclass[journal]{IEEEtran}

\newcommand{\paperlink}[2]{}
\newcommand{\confpaperlink}[2]{}
\usepackage{setspace}
\usepackage{cite}
\usepackage{graphicx}
\usepackage[cmex10]{amsmath}
\usepackage{mathtools}
\usepackage{amsfonts}
\usepackage{amssymb}
\usepackage{texdef2015}
\usepackage{url}
\usepackage{epstopdf}
\usepackage{algorithm}
\usepackage{algorithmic}
\usepackage{subfigure}
\usepackage{fixltx2e}
\usepackage{tikz,bm,hyperref}
\usetikzlibrary{arrows,snakes,shapes}
\usetikzlibrary{positioning,shapes}
\usetikzlibrary{decorations.pathreplacing}
\usepackage[framed,unumbered,autolinebreaks,useliterate]{mcode}

\newcommand{\age}{\Delta}

\usepackage{color}
\def\orange{\color{black}}
\def\blue{\color{black}}

\newcommand{\ignore}[1]{}
\newtheorem{coro}{Corollary}

\begin{document}
\IEEEoverridecommandlockouts
\title{Update or Wait: How to Keep Your Data Fresh}
\author{Yin Sun$^\dag$, Elif Uysal-Biyikoglu$^\ddag$, Roy D. Yates$^\star$, C. Emre Koksal$^\dag$, and Ness B. Shroff$^\dag{}^\S$ \\
$^\dag$Dept. of ECE, $^\S$Dept. of CSE, The Ohio State University, Columbus, OH\\
$^\ddag$Dept. of EEE, Middle East Technical University, Ankara, Turkey\\
$^\star$Dept. of ECE, Rutgers University, North Brunswick, NJ\\
~\\
{April 20, 2017}
\thanks{This paper was presented in part at IEEE INFOCOM 2016.}
\thanks{This work was supported in part by DTRA grant HDTRA1-14-1-0058, NSF grants CNS-1446582, CNS-1409336, CNS-1518829,  CNS-1514260, CNS-1422988, and CNS-1054738, ARO grant W911NF-14-1-0368, and ONR grant N00014-15-1-2166. E. Uysal-Biyikoglu was supported in part by TUBITAK and in part by a Science Academy BAGEP award.
}
}
\maketitle
\begin{abstract}

In this work, we study how to optimally manage the freshness of information updates sent from a source node to a destination via a channel. A proper metric for data freshness at the destination is the \emph{age-of-information}, or simply \emph{age}, which is defined as how old the freshest received update is since the moment that this update was generated at the source node (e.g., a sensor). A reasonable update policy is the \emph{zero-wait} policy, i.e., the source node submits a fresh update  once the previous update  is delivered and the channel becomes free, which achieves the maximum throughput and the minimum delay. Surprisingly, this zero-wait policy does \emph{not} always minimize the age. This counter-intuitive phenomenon motivates us to study how to optimally control information updates to keep the data fresh and to understand when the zero-wait policy is optimal. We introduce a general age penalty function to characterize the level of dissatisfaction on data staleness and formulate the average age penalty minimization problem as a constrained semi-Markov decision problem (SMDP) with an uncountable state space. We develop efficient algorithms to find the optimal update policy among all causal policies, and establish sufficient and necessary conditions  for the optimality of the zero-wait policy. Our investigation shows that the zero-wait policy is far from the optimum if (i) the age penalty function grows quickly with respect to the age, (ii) the packet transmission times over the channel are positively correlated over time, or (iii) the packet transmission times are highly random (e.g., following a heavy-tail distribution).
\end{abstract}


\section{Introduction}\label{intro}
In recent years, the proliferation of mobile devices and applications has significantly boosted the need for real-time information updates, such as news, weather reports, traffic alerts, email notifications, stock quotes, social updates, mobile ads, etc. In addition, timely information updates are also critical in real-time monitoring and control systems, including sensor networks used in temperature and air pollution monitoring \cite{Ali2010}, surround monitoring in autonomous vehicles \cite{Gandhi07}, phasor data updates in power grid stabilization systems \cite{Terzija2011}, and so on.

A common need in these real-time applications is to optimize the freshness of the data.
In light of this, a metric of data freshness  called the \emph{age-of-information}, or simply \emph{age}, was defined in, e.g., \cite{Song1990,KaulYatesGruteser-Infocom2012}.
At any time $t$, if the freshest  update delivered to the  destination of information updates (e.g., a monitor or controller)
was generated at time $U(t)$, then the age at the destination is 
\begin{align}
\age(t)=t- U(t). 
\end{align}
Hence, the age is the amount of time elapsed since the moment that the freshest delivered update was generated.

Most existing research on the age-of-information focuses on an ``enqueue-and-forward'' model \cite{KaulYatesGruteser-Infocom2012,2012CISS-KaulYatesGruteser,2012ISIT-YatesKaul,Yates2016,Kam-status-ISIT2013,
KamKompellaEphremides2014ISIT,Kam2016,CostaCodreanuEphremides2014ISIT,Costa2016,Pappas2015,2015ISITHuangModiano,KamISIT2016,Chen2016,NajmISIT2016,age_optimality_multi_server,age_optimality_multi_hop}, where information update packets arrive stochastically at the source node. The source node enqueues these updates and then forwards them to a destination through a channel. It is worth noting that there is a significant difference between \emph{age} and \emph{delay}: A low update frequency results in a short queueing delay because the queue is almost empty, but the destination may end up having stale data  due to infrequent updates. On the other hand, a high update frequency will increase the queueing delay, and the age is also high because the updates are becoming stale during their long waiting time in the queue. Hence, delay is an increasing function of the update frequency, but the age first decreases and then increases with respect to the update frequency \cite{KaulYatesGruteser-Infocom2012}. In \cite{KaulYatesGruteser-Infocom2012, CostaCodreanuEphremides2014ISIT}, it was found that a good policy is to discard the old updates waiting in the queue when a new sample arrives, which can greatly reduce the negative impact of  queueing delay. {In \cite{age_optimality_multi_server,age_optimality_multi_hop},  a  Last Generated First Served (LGFS) policy was shown to  minimize the age process in a strong sense (in the sense of stochastic ordering) for multi-channel and multi-hop networks with an arbitrary (e.g., out-of-order) update arrival process. }   



In this paper, we study a ``generate-at-will'' model {\orange proposed in \cite{2015ISITYates}: As depicted in Fig. \ref{fig:model},}
the source node (e.g., a sensor) has access to the channel's idle/busy state through acknowledgements (ACKs), and in contrast to \cite{KaulYatesGruteser-Infocom2012,2012CISS-KaulYatesGruteser,2012ISIT-YatesKaul,Yates2016,Kam-status-ISIT2013,
KamKompellaEphremides2014ISIT,Kam2016,CostaCodreanuEphremides2014ISIT,Costa2016,Pappas2015,2015ISITHuangModiano,KamISIT2016,Chen2016,NajmISIT2016,age_optimality_multi_server,age_optimality_multi_hop}, is able to generate information updates at any time by its own will. 
Because of queueing, update packets may need to wait in the queue for their transmission opportunity, and become stale while waiting. Hence, it is better not to generate updates when the channel is busy, which completely eliminates the waiting time in the queue. In this case, a reasonable update policy is the \emph{zero-wait} policy, {\orange also called \emph{just-in-time updating} in \cite{2015ISITYates} and the \emph{work-conserving} policy in queueing theory \cite{Kleinrock1975},} that submits a fresh update once the previous update is delivered and an ACK is received.\footnote{\orange This policy was proposed in \cite[Section VII]{KaulYatesGruteser-Infocom2012} to provide a lower bound to the age in the ``enqueue-and-forward'' model.} The zero-wait policy achieves the maximum throughput and the minimum delay. Surprisingly, this zero-wait policy does \textbf{not} always minimize the age-of-information. {\orange In particular, an optimal policy was obtained in \cite{2015ISITYates} for minimizing the time-average age, from which one can deduce that the zero-wait policy is not age-optimal.}
The following example reveals the reason behind this counter-intuitive phenomenon:

\begin{figure}
\centering
\ifjournal
\begin{tikzpicture}[scale=1.5]
\else
\begin{tikzpicture}[scale=1.5]
\fi
\draw (-4.2,0)  circle [radius=0.65] node (source) {Source};
\draw [->, thick] (-3.55,0) -- (-3.2,0);
\draw (-3.25,0.25) -- ++(0.75,0) -- ++(0,-0.5) -- ++(-0.75,0);
\foreach \i in {1,...,3}
  \draw (-2.5-\i*0.2,0.25) -- +(0,-0.5);
\draw (-2.25,0) circle [radius=0.25] node (channel) {};
\draw [->, thick] (-2.25,0.25) -- (-2.25 ,1) -- (-4.2, 1) -- (-4.2,0.65);
\draw (-3.6,1.4) node [below,text width=1.2] {ACK};
\draw (-3.2,-0.25) node [below,text width=1.2] {Queue};
\draw (-2.4,-0.25) node [below,text width=1.2] {Channel};
\draw (-0.85,0) circle [radius=0.65] node (monitor) {Destination};
\draw [->, thick] (-2,0) -- (-1.5,0);
\end{tikzpicture}
\caption{System model.}
\vspace{-0.5cm}
\label{fig:model}
\end{figure}
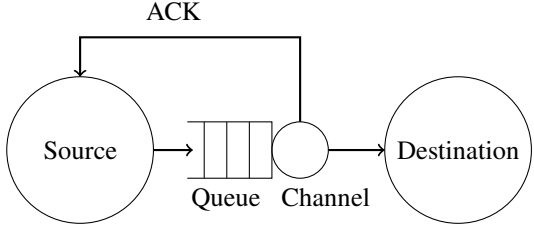

{
\textbf{Example:}
\emph{Suppose that the source node sends a sequence of information updates to a destination. The transmission times of these updates form a periodic sequence
$$0, 0, 2, 2, 0, 0, 2, 2, 0, 0, 2, 2,\dots$$
Suppose that Update $1$ is generated and submitted at time $0$ and delivered at time $0$. Under the zero-wait policy, Update $2$ is also generated at time $0$ and delivered at time $0$. However, despite  its negligible transmission time, Update $2$ has not brought any new information to the destination after Update $1$ was delivered, because both updates were sampled at the same time. Therefore, the potential benefit of the zero transmission time of Update $2$ is wasted! This issue occurs periodically over time: Whenever two consecutive updates have zero transmission time, the second update of the two is wasted. Therefore, a 1/4 of the updates in this sequence are wasted in the zero-wait policy! }

\emph{For comparison, consider a $\epsilon$-wait policy that waits for $\epsilon$ seconds after each update with a zero transmission time, and does not wait after each update with a transmission time of $2$ seconds. Note that the control decisions in the $\epsilon$-wait policy are made causally.
The time-evolution of the age $\age(t)$ in the $\epsilon$-wait policy is shown in Fig. \ref{fig_example}. Update $1$ is generated and delivered at time $0$. Update $2$ is generated and delivered at time $\epsilon$. Update $3$ is generated at time $2\epsilon$ and is delivered at time $2 + 2\epsilon$. Because the transmission time of Update $3$ is $2$ seconds, the latest delivered update  at time $2 + 2\epsilon$ is of the age $2$ seconds. Hence, the age $\age(t)$ drops to $2$ seconds at time $2 + 2\epsilon$. Update $4$ is generated at time $2 + 2\epsilon$ and is delivered at time $4 + 2\epsilon$. At time $4 + 2\epsilon$, the age drops to zero because Update $5$ is generated at this time and is delivered immediately.}

\emph{According to Fig. \ref{fig_example}, one can compute the time-average age of the $\epsilon$-wait policy, which is given by
\begin{align}
&(\epsilon^2/2 + \epsilon^2/2 + 2\epsilon + 4^2/2) / (4 + 2\epsilon) \nonumber\\
=& (\epsilon^2 + 2\epsilon + 8) / (4 + 2\epsilon)~\text{seconds.}\nonumber
\end{align} 
If the waiting time is $\epsilon=0.5$, the time-average age of the $\epsilon$-wait policy is $1.85$ seconds.
If the waiting time is $\epsilon=0$, it reduces to the zero-wait policy, whose time-average age is $2$ seconds. 
Hence, the zero-wait policy is not optimal!} 
\emph{In Section \ref{sec:whenoptimal}, we will see more numerical results showing that the zero-wait policy can be far from the optimum. }

\begin{figure}
\centering
\includegraphics[width=0.4\textwidth]{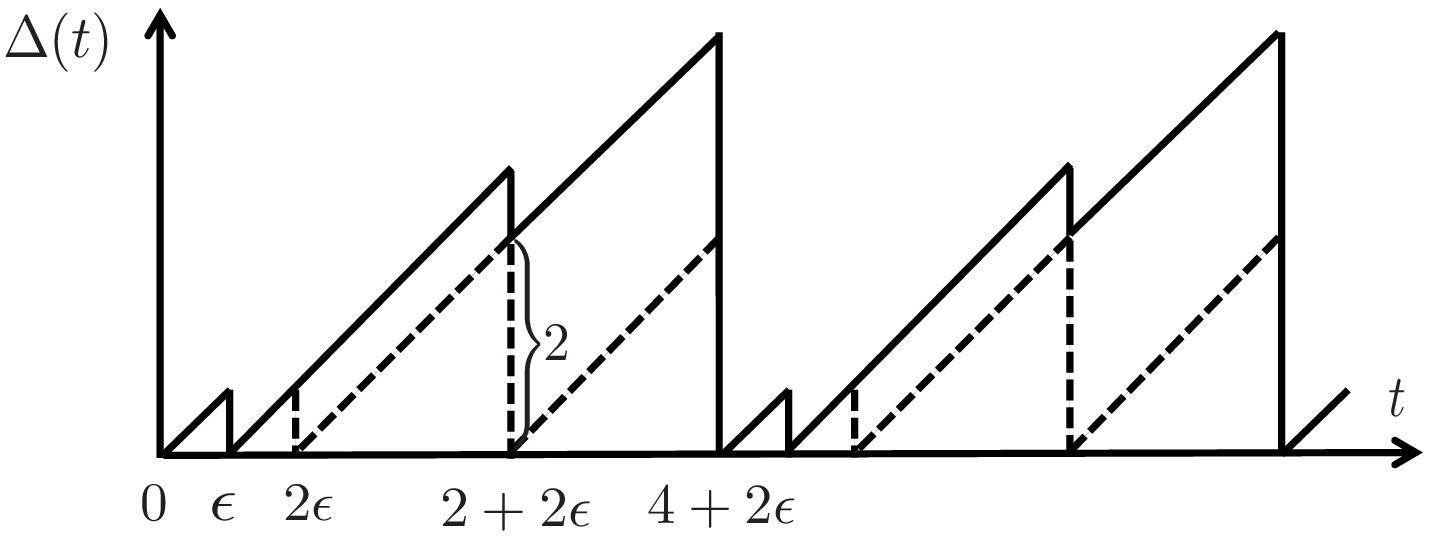}   
\caption{Evolution of the age $\age(t)$ in the $\epsilon$-wait policy in the example.}\vspace{-0.0cm}
\label{fig_example}
\vspace{-0.4cm}
\end{figure}    


}

This example points out a key difference between  data communication systems and information update systems: 
In communication systems, all packets are equally important; however,
in information update systems, an update packet is useful only if it carries some fresh information to the destination. While the theory of data communications is quite mature, the optimal control of information updates remains open.


The aim of this paper is to answer the following questions:
How to optimally submit update packets to minimize the age-of-information at the destination?
When is the zero-wait policy optimal?
To that end, 
the following are the key contributions of this paper:
\begin{itemize}
\item
We {\orange generalize \cite{2015ISITYates}} by introducing two new features: (i) age penalty functions and (ii) correlated (non-\emph{i.i.d.}) packet transmission times.
We define a general age penalty function $g(\age)$ to characterize the level of dissatisfaction for data staleness, where $g(\cdot)$ can be any \emph{non-negative} and \emph{non-decreasing} function. This age penalty model is quite general, as it allows $g(\cdot)$ to be discontinuous and non-convex. In practice, one can choose the age penalty function based on the specific applications; a few examples are provided in Section \ref{sec:formulation}. 
In addition, the packet transmission times are modeled as a stationary ergodic Markov chain with an uncountable state space, which is more general than the \emph{i.i.d.} transmission time processes assumed in related studies.

\item We formulate the average age penalty minimization problem as a constrained semi-Markov decision problem (SMDP) with an uncountable state space. {\orange The set of feasible policies in this problem contains all \emph{causal} policies, such that control decisions are made based on the history and current information of the system, which is more general than the feasible policy space considered in  \cite{2015ISITYates}.}
Despite the difficulty of this problem, we manage to solve it by a  divide-and-conquer approach: We first prove that there exists a {stationary randomized policy} that is optimal for this problem (Theorem \ref{lem_SRoptimal}). Further, we prove that there exists a {stationary deterministic policy} that is optimal for this problem (Theorem \ref{lem_SD}).
Finally, we develop a low-complexity algorithm to find the optimal stationary deterministic policy (Theorem \ref{lem_SD_solution}).


\item We further investigate when the zero-wait policy is optimal.
For the special case of proportional penalty function and \emph{i.i.d.} transmission times,
we devise a simpler solution to minimize the average age (Theorem \ref{beta}), and obtain a sufficient and necessary condition to characterize when the zero-wait policy is optimal (Theorem \ref{iffcondition}). We find that the zero-wait policy is optimal if the transmission times are constant; and is not optimal for many commonly used transmission time distributions in communication and queueing theory, such as exponential distribution, geometric distribution, Erlang distribution, hyperexponential distribution, etc. (Corollary \ref{coro1}).
In addition, 
several sufficient conditions for the optimality of the zero-wait policy are provided for general age penalty functions and correlated transmission time processes (Lemma \ref{just_in_time}).

\item 
Our theoretical and simulation results demonstrate that, in many scenarios, the optimal information update policy is to wait for a certain amount of time before submitting a new update. In particular,  the zero-wait policy is far from the optimum if (i) the age penalty function grows quickly with respect to the age, (ii) the packet transmission times over the channel are positively correlated over time, or (iii) the packet transmission times are highly random (e.g., following a heavy-tail distribution).

\end{itemize}

The rest of this paper is organized as follows. In Section \ref{sec:relate}, we discuss some related work. In Section \ref{sec:model}, we describe
the system model and the formulation of the average age penalty minimization problem. In Section \ref{sec:solution}, we develop the optimal  update policy that minimizes the average age penalty among all causal policies. In Section \ref{sec:whenoptimal}, we provide sufficient and necessary conditions  for the optimality of the zero-wait policy
Finally, in Section \ref{conclusion}, we conclude the 
paper.

{\blue
\section{Related Work}\label{sec:relate}

The age-of-information was defined as a metric of data freshness as early as 1990s in the studies of real-time databases \cite{Song1990,Segev:1991,Adelberg:1995,Cho:2000}. In recent years, queueing theoretic techniques were used to evaluate the age-of-information in various system settings. The average age was analyzed for First-Come First-Served (FCFS) systems \cite{KaulYatesGruteser-Infocom2012},  Last-Come First-Served (LCFS) systems  \cite{2012CISS-KaulYatesGruteser}, multi-source networks \cite{2012ISIT-YatesKaul,Yates2016}, and multi-channel networks \cite{Kam-status-ISIT2013,KamKompellaEphremides2014ISIT,Kam2016}. A peak age metric was introduced in \cite{CostaCodreanuEphremides2014ISIT} and studied for a multi-class M/G/1 queueing system \cite{2015ISITHuangModiano}. 

In \cite{CostaCodreanuEphremides2014ISIT,Pappas2015,Costa2016}, a packet management policy was shown to reduce the age, in which the old updates waiting in the queue are discarded when a new sample arrives. 
In \cite{KamISIT2016,Kam2016Milcom}, it was shown that controlling buffer size and packet deadline can improve the age. The age performance in the presence of errors was analyzed in \cite{Chen2016}. Gamma-distributed transmission times was considered in \cite{NajmISIT2016}. The age-of-information under energy-harvesting constraints was studied in \cite{Bacinoglu2015,Bacinoglu2017}. Source coding techniques for reducing the age were evaluated in \cite{Zhong2016}. In \cite{age_optimality_multi_server,age_optimality_multi_hop}, it was proven that the LGFS policy achieves a smaller age process (in a stochastic ordering sense) than any other causal policy for multi-channel and multi-hop networks with an arbitrary (e.g., out-of-order) update arrival process. 
Scheduling of updates broadcasted to multiple users with unreliable channels was optimized in \cite{Kadota2016}. NP-hardness of wireless scheduling for minimizing the age in general networks was investigated in \cite{He2016,He2016-2}. Reducing the age of channel state information in wireless systems was studied in \cite{Costa2015,Costa2015ICC}. 
Experimental evaluation of the age-of-information was conducted via emulation in \cite{Kam2015}.}

{\blue The work that is most relevant  to this paper is \cite{2015ISITYates}. We have generalized the results in \cite{2015ISITYates} by finding the optimal update policy for more general age penalty model, transmission time model, and in a larger feasible policy space. }

\section{ Model and  Formulation}\label{sec:model}
\subsection{System Model}
We consider an information update system depicted in Fig.~\ref{fig:model}, where a source node generates update packets and sends them to a destination through a channel. 
The source node generates and {submits} update packets at successive times $S_0,S_1,\ldots$ 
The source node has access to the idle/busy state of the channel through ACK and is able to generate updates at any time by its own will. 
As we have mentioned before, the source node should not generate a new update when the channel is busy sending previous updates, because this will incur an unnecessary waiting time in the queue. 

\begin{figure}
\centering
\ifjournal
\begin{tikzpicture}[scale=0.3]
\else
\begin{tikzpicture}[scale=0.21]
\fi
\draw [<-|] (0,10)  -- (0,0) -- (14.5,0);
\draw [|->] (15,0) -- (30.5,0) node [below] {\small$t$};
\draw (-2,11) node [right] {\small$\age(t)$};
\fill
(2,0)  circle[radius=4pt]
(8,0)  circle[radius=4pt]
(17,0)  circle[radius=4pt]
(25,0)  circle[radius=4pt];
\draw
(2,0) node [below] {\small$S_1$}
(8,0) node [below] {\small$S_2$}
(17,0) node [below] {\small$S_{n-1}$}
(25,0) node [below] {\small$S_{n}$};

\fill
(0,0)  circle[radius=4pt]
(4,0)  circle[radius=4pt]
(11,0)  circle[radius=4pt]
(20,0)  circle[radius=4pt]
(27,0)  circle[radius=4pt];
\draw
(-1,0) node [below] {\small $D_0\!=\!0$}
(4,0) node [below] {\small$D_1$}
(11,0) node [below] {\small$D_2$}
(21,0) node [below] {\small$D_{n-1}$}
(27,0) node [below] {\small$D_{n}$};
\draw[ thick, domain=0:4] plot (\x, {\x+2.5})  -- (4, {2});
 \draw [ thick, domain=4:11] plot (\x, {\x-2})  -- (11, {3});
\draw[ thick, domain=11:14] plot (\x, {\x-8});
\draw[ thick, domain=16:20] plot (\x, {\x-14}) -- (20, {3});
\draw[ thick, domain=20:27] plot (\x, {\x-17}) -- (27, {2});
\draw[ thick, domain=27:29] plot (\x, {\x-25});
\draw[ ultra thin,dashed] (4, {2}) -- (4, 0);
\draw[ ultra thin,dashed] (11, {3}) -- (11, 0);
\draw[ ultra thin,dashed] (20, {3}) -- (20, 0);
\draw[ ultra thin,dashed] (27, {2}) -- (27, 0);
\draw[ ultra thin,dashed,  domain=2:4] plot (\x, {\x-2});
\draw[ ultra thin,dashed,  domain=8:11] plot (\x, {\x-8});
\draw[ ultra thin,dashed,  domain=17:20] plot (\x, {\x-17});
\draw[ ultra thin,dashed,  domain=25:27] plot (\x, {\x-25});
\draw [|-|] (0,-3) -- (2,-3);
\draw [-|] (2,-3) -- (4,-3);
\draw [-|] (4,-3) -- (8,-3);
\draw [-|] (8,-3) -- (11,-3);
\draw [|-|] (17,-3) -- (20,-3);
\draw [-|] (20,-3) -- (25,-3);
\draw [-|] (25,-3) -- (27,-3);
\draw  [decorate,decoration={brace,amplitude=4pt}] (2,-3) to node [below] [black,midway,below=2pt,xshift=-0pt]{\small $Z_0$} (0,-3);
\draw  [decorate,decoration={brace,amplitude=4pt}] (2+2,-3) to node [below] [black,midway,below=2pt,xshift=-0pt]{\small $Y_1$} (0+2,-3);
\draw  [decorate,decoration={brace,amplitude=4pt}] (6+2,-3) to node [below] [black,midway,below=2pt,xshift=-0pt]{\small $Z_1$} (2+2,-3);
\draw  [decorate,decoration={brace,amplitude=4pt}] (13-4+2,-3) to node [below] [black,midway,below=2pt,xshift=-0pt]{\small $Y_2$} (10-4+2,-3);
\draw  [decorate,decoration={brace,amplitude=4pt}] (20,-3) to node [below] [black,midway,below=2pt,xshift=-0pt]{\small $Y_{n-1}$}  (17,-3);
\draw  [decorate,decoration={brace,amplitude=4pt}] (25,-3) -- node [below] [black,midway,below=2pt,xshift=-0pt]{\small $Z_{n-1}$}  (20,-3);
\draw  [decorate,decoration={brace,amplitude=4pt}]  (27,-3) -- node [below] [black,midway,below=2pt,xshift=-0pt]{\small $Y_{n}$} (25,-3);
\draw  [decorate,decoration={brace,amplitude=3pt}]  (27,2) --  (27,0);
\draw (27.1,1) node [right] {\small$Y_{n}$};
\draw  [decorate,decoration={brace,amplitude=4pt}]  (0,0) --  (0,2.5);
\draw (-0.3,1.25) node [left] {\small$Y_{0}$};
\draw  [decorate,decoration={brace,amplitude=4pt}]  (11,3) --  (11,0);
\draw (11.3,1.5) node [right] {\small$Y_{2}$};
\end{tikzpicture}
\caption{Evolution of the age-of-information $\age(t)$.}
\label{fig:age1}
\vspace{-0.5cm}
\end{figure}
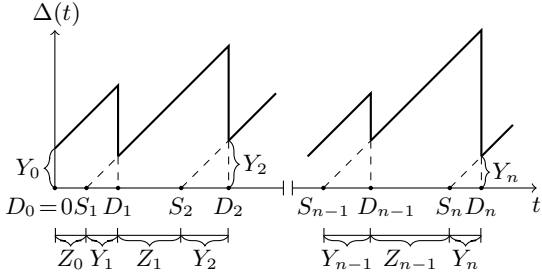

Suppose that Update $i$ is submitted at time $S_i$, and its transmission time is $Y_i\geq0$. Hence, Update $i$
is {delivered} at time $D_i=S_i+Y_i$. We assume that Update $0$ is submitted to an idle channel at time $S_{0}=-Y_0$ and is delivered at $D_0=0$, as shown in Fig. \ref{fig:age1}.
After Update $i$ is delivered at time $D_i$, the source node may insert a {\em waiting time} $Z_i\in[0,M]$ before submitting Update $i+1$ at time $S_{i+1}=D_i+Z_i$, where $M<\infty$ is the maximum waiting time allowed by the system.
The source node can switch to a low-power sleep mode during the waiting period $[D_i,S_{i+1})$.
We assume that the transmission time process $(Y_0,Y_1,\ldots)$ is \emph{a stationary and ergodic Markov chain} with a possibly uncountable state space and a positive mean $0<\mathbb{E}[Y_i]<\infty$.\footnote{The results in this paper can be readily extended to a more general Markovian transmission time process $(Y_0,Y_1,\ldots)$ 
with a longer memory, in which $\bm W_i$ is defined as $\bm W_i = (Y_i, Y_{i+1},\ldots,Y_{i+k})$ for some finite $k$, and the sequence $(\bm W_0,\bm W_1,\ldots)$ forms a Markov chain.} This model generalizes the \emph{i.i.d.} transmission time processes in related studies.
 This Markovian model is introduced to study the impact of temporal correlation on the optimality of the zero-wait policy. In Section \ref{sec:whenoptimal}, we will see that the zero-wait policy is close to the optimum when the transmission times are negatively correlated; and is far from the optimum when the transmission times are positively correlated.


At any time $t$, the most recently received update packet is generated at time 
\begin{align}
U(t)=\max\{S_i: D_i\leq t \}. 
\end{align}
The \emph{age-of-information} $\age(t)$ is defined as \cite{Song1990,KaulYatesGruteser-Infocom2012}
\begin{align}
\age(t)= t - U(t),
\end{align}
which is also referred to as \emph{age}.
The age $\age(t)$ is a stochastic process that increases linearly with $t$ between updates, with downward jumps occurring when updates are delivered. \ifreport
As shown in Fig. \ref{fig:age1}, Update $i$ is sent at time $t = S_i$, and is delivered at time $D_i=S_i+Y_i$ with age $\age(D_i)=D_i-S_i=Y_i$. After that, the age increases linearly and reaches $\age(D_{i+1}^-)=Y_i+Z_i+Y_{i+1}$ just before Update $i+1$ is delivered. Then, at time $D_{i+1}$, the age drops to $\age(D_{i+1})=Y_{i+1}$.
\fi

\subsection{Problem Formulation}\label{sec:formulation}
We introduce an age penalty function $g(\age)$ to represent the level
of dissatisfaction for data staleness or the need for new information update. 
{The function $g: [0,\infty)\rightarrow [0,\infty)$ is assumed to be \emph{
measurable, non-negative, and non-decreasing}. This age penalty model is quite general, as it allows $g(\cdot)$ to be discontinuous and non-convex. In practice, one can specify the age penalty function based on the applications. A few examples are discussed in the following:

\begin{itemize}
\item[1.] Online Learning: Fresh data is critical in online learning which is of great interest to practitioners due to the recent emergence of real-time applications such as advertisement placement and online web ranking \cite{Shwartz2012,He2014}.  One can employ the age penalty functions that grow quickly with respect to the age, such as the power function $g(\age)=\age^a$ and the exponential function $g(\age)=e^{a\age}$ with $a\geq0$, to characterize the desire for data refreshing in these applications.

\item[2.] Periodic Inspection and Monitoring: One can use the stair-step function $g(\age)=\lfloor a \age\rfloor$, where $\lfloor x\rfloor$ is the largest integer no greater than $x$, to characterize the dissatisfaction of data staleness if the information of interest is checked periodically \cite{Chiang2001165,Kottapalli2003}. 
\end{itemize}}

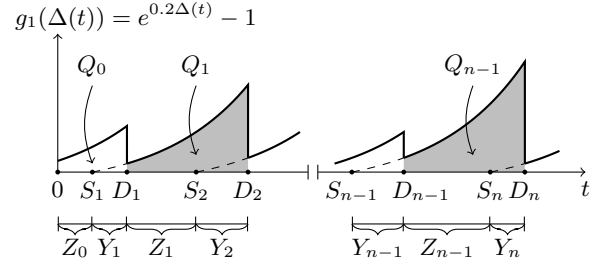
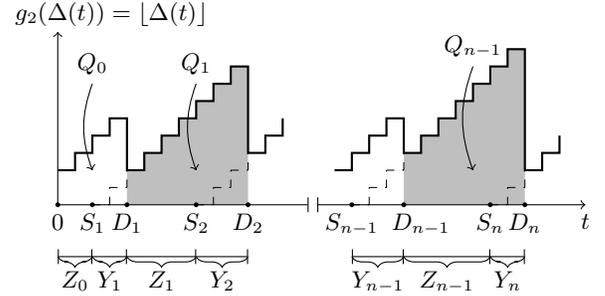
\begin{figure}
\centering
\subfigure[][An exponential age penalty function $g_1(\age)=e^{0.2\age}-1$.]
{
\ifjournal
\begin{tikzpicture}[scale=0.3]
\else
\begin{tikzpicture}[scale=0.23]
\fi
\fill[fill=lightgray, ultra thin] (4,0) -- plot [domain=4:11] (\x, {exp(0.2*(\x-2))-1}) -- (11,0) -- cycle;
\fill[fill=lightgray, ultra thin] (20,0) -- plot [domain=20:27] (\x, {exp(0.2*(\x-17))-1}) -- (27,0) -- cycle;
\draw [<-|] (0,8)  -- (0,0) -- (14.5,0);
\draw [|->] (15,0) -- (30.5,0) node [below] {\small$t$};
\draw (-3,9) node [right] {\small$g_1(\age(t))=e^{0.2\age(t)}-1$};
\fill
(2,0)  circle[radius=4pt]
(8,0)  circle[radius=4pt]
(17,0)  circle[radius=4pt]
(25,0)  circle[radius=4pt];
\draw
(2,0) node [below] {\small$S_1$}
(8,0) node [below] {\small$S_2$}
(17,0) node [below] {\small$S_{n-1}$}
(25,0) node [below] {\small$S_{n}$};

\fill
(0,0)  circle[radius=4pt]
(4,0)  circle[radius=4pt]
(11,0)  circle[radius=4pt]
(20,0)  circle[radius=4pt]
(27,0)  circle[radius=4pt];
\draw
(0,0) node [below] {\small $0$}
(4,0) node [below] {\small$D_1$}
(11,0) node [below] {\small$D_2$}
(21,0) node [below] {\small$D_{n-1}$}
(27,0) node [below] {\small$D_{n}$};

\draw[<-] (2,0.5) to [out=110,in=250] (2,5) node [above] {\small${Q}_0$};
\draw[<-] (8,1) to [out=110,in=250] (8,5) node [above] {\small$Q_1$};
\draw[<-] (24,1) to [out=110,in=250] (24,5) node [above] {\small$Q_{n-1}$};
\draw[ thick, domain=0:4] plot (\x, {exp(0.2*(\x+2.5))-1}) -- (4, {exp(0.2*(4-2))-1});
\draw[ thick, domain=4:11] plot (\x, {exp(0.2*(\x-2))-1}) -- (11, {exp(0.2*(11-8))-1});
\draw[ ultra thin,dashed,  domain=2:4] plot (\x, {exp(0.2*(\x-2))-1});
\draw[ thick, domain=11:14] plot (\x, {exp(0.2*(\x-8))-1});
\draw[ ultra thin,dashed,  domain=8:11] plot (\x, {exp(0.2*(\x-8))-1});
\draw[ thick, domain=16:20] plot (\x, {exp(0.2*(\x-14))-1}) -- (20, {exp(0.2*(20-17))-1});
\draw[ thick, domain=20:27] plot (\x, {exp(0.2*(\x-17))-1}) -- (27, {exp(0.2*(27-25))-1});
\draw[ ultra thin,dashed,  domain=17:20] plot (\x, {exp(0.2*(\x-17))-1});
\draw[ thick, domain=27:29] plot (\x, {exp(0.2*(\x-25))-1});
\draw[ ultra thin,dashed,  domain=25:27] plot (\x, {exp(0.2*(\x-25))-1});
\draw [|-|] (0,-3) -- (2,-3);
\draw [-|] (2,-3) -- (4,-3);
\draw [-|] (4,-3) -- (8,-3);
\draw [-|] (8,-3) -- (11,-3);
\draw [|-|] (17,-3) -- (20,-3);
\draw [-|] (20,-3) -- (25,-3);
\draw [-|] (25,-3) -- (27,-3);
\draw  [decorate,decoration={brace,amplitude=4pt}] (2,-3) to node [below] [black,midway,below=2pt,xshift=-0pt]{\small $Z_0$} (0,-3);
\draw  [decorate,decoration={brace,amplitude=4pt}] (2+2,-3) to node [below] [black,midway,below=2pt,xshift=-0pt]{\small $Y_1$} (0+2,-3);
\draw  [decorate,decoration={brace,amplitude=4pt}] (6+2,-3) to node [below] [black,midway,below=2pt,xshift=-0pt]{\small $Z_1$} (2+2,-3);
\draw  [decorate,decoration={brace,amplitude=4pt}] (13-4+2,-3) to node [below] [black,midway,below=2pt,xshift=-0pt]{\small $Y_2$} (10-4+2,-3);
\draw  [decorate,decoration={brace,amplitude=4pt}] (20,-3) to node [below] [black,midway,below=2pt,xshift=-0pt]{\small $Y_{n-1}$}  (17,-3);
\draw  [decorate,decoration={brace,amplitude=4pt}] (25,-3) -- node [below] [black,midway,below=2pt,xshift=-0pt]{\small $Z_{n-1}$}  (20,-3);
\draw  [decorate,decoration={brace,amplitude=4pt}]  (27,-3) -- node [below] [black,midway,below=2pt,xshift=-0pt]{\small $Y_{n}$} (25,-3);
\end{tikzpicture}
}

\subfigure[][A stair-step age penalty function $g_2(\age)=\lfloor\age\rfloor$.]
{
\ifjournal
\begin{tikzpicture}[scale=0.3]
\else
\begin{tikzpicture}[scale=0.23]
\fi
\fill[fill=lightgray, ultra thin] (4,0) -- (4,2)
 \foreach \x in {4,...,10} { --(\x,\x-2)-- (\x+1,\x-2)}
-- (11,0) -- cycle;
\fill[fill=lightgray, ultra thin] (20,0) -- (20,3)
 \foreach \x in {20,...,26} { --(\x,\x-17)-- (\x+1,\x-17)}
-- (27,0) -- cycle;

\draw [<-|] (0,10)  -- (0,0) -- (14.5,0);
\draw [|->] (15,0) -- (30.5,0) node [below] {\small$t$};
\draw (-3,11) node [right] {\small$g_2(\age(t))=\lfloor\age(t)\rfloor$};
\fill
(2,0)  circle[radius=4pt]
(8,0)  circle[radius=4pt]
(17,0)  circle[radius=4pt]
(25,0)  circle[radius=4pt];
\draw
(2,0) node [below] {\small$S_1$}
(8,0) node [below] {\small$S_2$}
(17,0) node [below] {\small$S_{n-1}$}
(25,0) node [below] {\small$S_{n}$};

\fill
(0,0)  circle[radius=4pt]
(4,0)  circle[radius=4pt]
(11,0)  circle[radius=4pt]
(20,0)  circle[radius=4pt]
(27,0)  circle[radius=4pt];
\draw
(0,0) node [below] {\small $0$}
(4,0) node [below] {\small$D_1$}
(11,0) node [below] {\small$D_2$}
(21,0) node [below] {\small$D_{n-1}$}
(27,0) node [below] {\small$D_{n}$};

\draw[<-] (2,2) to [out=110,in=250] (2,7) node [above] {\small${Q}_0$};
\draw[<-] (8,2) to [out=110,in=250] (8,7) node [above] {\small$Q_1$};
\draw[<-] (24,2) to [out=110,in=250] (24,8) node [above] {\small$Q_{n-1}$};

\foreach \i in {-2} \draw [ thick]   (0,-\i) \foreach \x in {0,...,2} {-- (\x+1,\x-\i)-- (\x+1,\x+1-\i)}
-- (4, {5}) -- (4, {2}) \foreach \i in {2}\foreach \x in {4,...,9} {-- (\x+1,\x-\i)-- (\x+1,\x+1-\i)} --(11, {8}) -- (11, {3}) \foreach \i in {8}   \foreach \x in {11,...,12} {-- (\x+1,\x-\i)-- (\x+1,\x+1-\i)};

\foreach \i in {2} \draw [ ultra thin,dashed]   \foreach \x in {2,...,3} { (\x,\x-\i)-- (\x+1,\x-\i)}
    \foreach \x in {2,...,3} { (\x+1,\x-\i)-- (\x+1,\x+1-\i)};

\foreach \i in {8} \draw [ ultra thin,dashed]   \foreach \x in {8,...,10} { (\x,\x-\i)-- (\x+1,\x-\i)}
    \foreach \x in {8,...,10} { (\x+1,\x-\i)-- (\x+1,\x+1-\i)};

\foreach \i in {14} \draw [ thick] (16,16-\i)  \foreach \x in {16,...,18} {-- (\x+1,\x-\i)-- (\x+1,\x+1-\i)} -- (20, {5}) -- (20, {3}) \foreach \i in {17}   \foreach \x in {20,...,25} {-- (\x+1,\x-\i)-- (\x+1,\x+1-\i)}
--(27, {9}) -- (27, {2}) \foreach \i in {25}   \foreach \x in {27,...,28} {-- (\x+1,\x-\i)-- (\x+1,\x+1-\i)};

\foreach \i in {17} \draw [ ultra thin,dashed]   \foreach \x in {17,...,19} { (\x,\x-\i)-- (\x+1,\x-\i)}
    \foreach \x in {17,...,19} { (\x+1,\x-\i)-- (\x+1,\x+1-\i)};

\foreach \i in {25} \draw [ ultra thin,dashed]   \foreach \x in {25,...,26} { (\x,\x-\i)-- (\x+1,\x-\i)}
    \foreach \x in {25,...,26} { (\x+1,\x-\i)-- (\x+1,\x+1-\i)};
\draw [|-|] (0,-3) -- (2,-3);
\draw [-|] (2,-3) -- (4,-3);
\draw [-|] (4,-3) -- (8,-3);
\draw [-|] (8,-3) -- (11,-3);
\draw [|-|] (17,-3) -- (20,-3);
\draw [-|] (20,-3) -- (25,-3);
\draw [-|] (25,-3) -- (27,-3);
\draw  [decorate,decoration={brace,amplitude=4pt}] (2,-3) to node [below] [black,midway,below=2pt,xshift=-0pt]{\small $Z_0$} (0,-3);
\draw  [decorate,decoration={brace,amplitude=4pt}] (2+2,-3) to node [below] [black,midway,below=2pt,xshift=-0pt]{\small $Y_1$} (0+2,-3);
\draw  [decorate,decoration={brace,amplitude=4pt}] (6+2,-3) to node [below] [black,midway,below=2pt,xshift=-0pt]{\small $Z_1$} (2+2,-3);
\draw  [decorate,decoration={brace,amplitude=4pt}] (13-4+2,-3) to node [below] [black,midway,below=2pt,xshift=-0pt]{\small $Y_2$} (10-4+2,-3);
\draw  [decorate,decoration={brace,amplitude=4pt}] (20,-3) to node [below] [black,midway,below=2pt,xshift=-0pt]{\small $Y_{n-1}$}  (17,-3);
\draw  [decorate,decoration={brace,amplitude=4pt}] (25,-3) -- node [below] [black,midway,below=2pt,xshift=-0pt]{\small $Z_{n-1}$}  (20,-3);
\draw  [decorate,decoration={brace,amplitude=4pt}]  (27,-3) -- node [below] [black,midway,below=2pt,xshift=-0pt]{\small $Y_{n}$} (25,-3);
\end{tikzpicture}
}
\caption{Evolutions of two age penalty functions.}
\vspace{-0.5cm}
\label{fig:age2}
\end{figure}

Two age penalty functions are depicted in Figure~\ref{fig:age2}. 
To analyze the average age penalty, we decompose the area under the curve $g(\age(t))$ into a sum of disjoint components: Consider the time interval $[0,D_n]$, where $D_n = \sum_{i=0}^{n-1} (Z_{i}+Y_{i+1})$.
In this interval, the area under $g(\age(t))$ can be seen as the concatenation of the areas $Q_i$, $0\leq i\leq n-1,$ such that
\begin{align}
\int_0^{D_n} g(\age(t))dt=\sum_{i=0}^{n-1} Q_i,\nonumber
\end{align}
where
\begin{align}\label{eq_Q_i}
Q_i = \int_{D_{i}}^{D_i+Z_{i}+Y_{i+1}} g(\age(t)) dt = \int_{Y_{i}}^{Y_i+Z_{i}+Y_{i+1}} g(\tau) d\tau.
\end{align}
In the second equation of \eqref{eq_Q_i}, we have used the fact that $\age(t) = t - S_i = t - (D_i-Y_i)$ for $t \in [D_i, D_i + Z_i +Y_{i+1}]$. 
Let us define
\begin{align}\label{eq_q_def}
q(y,z,y') = \int_y^{y+z+y'} g(\tau) d\tau.
\end{align}
Then, $Q_i$ can be expressed as $Q_i =q(Y_i,Z_i,Y_{i+1})$. Since $g(\age)$ is non-negative, the function $q(y,z,y')$  is non-decreasing in $z$. We assume 
\begin{align}\label{eq_bound}
\mathbb{E}\left[q(Y_{i},M,Y_{i+1})\right]<\infty,
\end{align}
which implies $\mathbb{E}\left[q(Y_{i},Z_i,Y_{i+1})\right]<\infty$ for all $Z_i\in[0,M]$.


Our goal is to minimize the average age penalty by controlling the sequence of waiting times $(Z_0,Z_1,\ldots)$.
Let $\pi\triangleq(Z_0,Z_1,\ldots)$ denote an information update policy.
We consider the class of \emph{causal} policies, in which control decisions are made based on the history and current information of the system, as well as the distribution of the transmission time process $(Y_0,Y_1,\ldots)$. Specifically, $Z_i$ is determined based on the past realizations of $(Y_0,Y_1,\ldots,Y_i)$, without using the realizations of future transmission times
 $(Y_{i+1}, Y_{i+2},\ldots)$; but the conditional distribution of $(Y_{i+1}, Y_{i+2},\ldots)$  given $(Y_0,Y_1,\ldots,Y_i)$ is available.
Let $\Pi$ denote the set of all causal policies satisfying $Z_i\in[0,M]$ for all $i$.


The average age penalty per unit time is defined by\footnote{There are two widely used definitions of average cost per unit time in infinite-horizon undiscounted semi-Markov decision problems (SMDP) \cite{Ross1970,Mine1970,Hayman1984,Feinberg1994,Bertsekas2005bookDPVol1}: In one definition, the average cost is the limit of expected total cost over a finite deterministic horizon divided by the length of the horizon, i.e., 
$\limsup_{T\rightarrow\infty}\frac{1}{T}\mathbb{E}\big[\int_0^{T} g(\age(t))dt\big]$; in the second definition, the average cost is the limit of the expected total cost over a
finite number of stages divided by the expected cumulative time of these stages, as in \eqref{eq_integral1}. These two definitions are both reasonable \cite{Ross1970,Feinberg1994,Bertsekas2005bookDPVol1}. They are equal under stationary randomized policies in which the generated semi-Markov chain has one ergodic class; see \cite{Ross1970,Mine1970,Hayman1984} for finite and countable state models. In general, however, these criteria are different. In our study, the second definition turns out to be analytically convenient.}
\begin{align}\label{eq_integral1}
\limsup_{n\rightarrow\infty}\frac{\mathbb{E}\left[\int_0^{D_n} g(\age(t))dt\right]}{\mathbb{E}\left[D_n\right]}.
\end{align}
Because the transmission time process $(Y_0,Y_1,\ldots)$ is stationary and ergodic, we can obtain $\mathbb{E}[Y_i] = \mathbb{E}[Y_{i+1}]$ and hence $\mathbb{E}\left[D_n\right]=\mathbb{E}[\sum_{i=0}^{n-1} (Z_{i}+Y_{i+1})] = \mathbb{E}[\sum_{i=0}^{n-1} (Y_{i}+Z_{i})]$.
Using this, the optimal information update problem for minimizing the average age penalty can be formulated as
\begin{align}\label{eq_DPExpected}
\overline{g}_{\text{opt}}=&\min_{\pi\in\Pi}~ \limsup_{n\rightarrow \infty}\frac{\mathbb{E}\left[\sum_{i=0}^{n-1} q(Y_i,Z_i,Y_{i+1})\right]}{\mathbb{E}[\sum_{i=0}^{n-1} (Y_{i}+Z_{i})]} \\
&~\text{s.t.}~~ \liminf_{n\rightarrow \infty} \frac{1}{n} \mathbb{E}\left[\sum_{i=0}^{n-1} (Y_{i}+Z_{i})\right]\geq \frac{1}{f_{\max}},
\label{eq_DPExpected_con}
\end{align}
where $\overline{g}_{\text{opt}}$ is the optimum objective value of Problem \eqref{eq_DPExpected}, the expectation $\mathbb{E}$ is taken over the stochastic  process $(Y_0,Y_1,\ldots)$ for given policy $\pi$, and ${f_{\max}}$ is the maximum allowed average update frequency due to a long-term average resource constraint (i.e., on the power resource or CPU cycles) spent on generating information updates. 
We assume $M >{1}/{f_{\max}}$ such that Problem \eqref{eq_DPExpected} is always feasible and $\overline{g}_{\text{opt}}<\infty$. In this paper, we will study Problem  \eqref{eq_DPExpected} both with and without the constraint \eqref{eq_DPExpected_con}. In Section \ref{sec:whenoptimal}, sufficient and necessary conditions will be provided to characterize when the zero-wait policy is optimal for solving Problem \eqref{eq_DPExpected} without the constraint \eqref{eq_DPExpected_con}.



In Problem \eqref{eq_DPExpected}, $Y_{i}$ is the state of an embedded Markov chain, $Z_{i}$ is the control action taken after observing $Y_i$,  $Y_{i} + Z_{i}$ is the  period of stage $i$, and $q(Y_{i},Z_{i},Y_{i+1})$ is the reward related to both stage $i$ and $i+1$.
Therefore, {Problem \eqref{eq_DPExpected} belongs to the class of constrained semi-Markov decision problems (SMDP) with an uncountable state space, which is generally known to be quite difficult.
The class of SMDPs includes Markov decision problems (MDPs) \cite{Bellman1957,Bertsekas2005bookDPVol1} and optimization problems of renewal processes \cite{Neely2013TAC} as special cases. Most existing studies on SMDPs deal with (i) unconstrained SMDPs, e.g.,\cite{Ross1970,Federgruen1983,Klabjan2006,Bertsekas2005bookDPVol1}, or (ii) constrained SMDPs with a countable state space, e.g.,\cite{Beutler1985,Beutler1986,Feinberg1994,Baykal2007}. However,
the optimality equations (e.g., Bellman's equation) for solving unconstrained SMDPs are not applied to constrained SMDPs \cite{Feinberg2012}, and the studies on constrained SMDPs with a countable state space
cannot be directly applied to Problem \eqref{eq_DPExpected} which has an uncountable state space.} 








\section{Optimal Information Update Policy}\label{sec:solution}
In this section, we develop a chain of new theoretical results to solve Problem \eqref{eq_DPExpected} in a divide-and-conquer fashion:
First, we prove that there exists a \emph{stationary randomized} policy that is optimal for Problem \eqref{eq_DPExpected}. Further, we prove that there exists a \emph{stationary deterministic} policy that is optimal for Problem \eqref{eq_DPExpected}. 
Finally, we develop a low-complexity algorithm to find the \emph{optimal} stationary deterministic policy that solves Problem \eqref{eq_DPExpected}. 
\subsection{Optimality of Stationary Randomized Policies}
A policy $\pi\in\Pi_{}$ is said to be a \emph{stationary randomized} policy, if it observes $Y_i$ and then chooses a waiting time $Z_i\in[0,M]$ based only on the observed value of $Y_i$. In this case, $Z_i$ is determined according to a conditional probability measure $p(y,A)\triangleq\Pr[Z_i\in A| Y_i=y]$ that is invariant for all $i=0,1,\ldots$ We use $\Pi_{\text{SR}}$ ($\Pi_{\text{SR}}\subseteq \Pi$) to denote the set of stationary randomized policies such that
\begin{align}
&\Pi_{\text{SR}}\!=\!\{\pi\in\Pi: \text{Given the observation $Y_i=y_i$, $Z_i$ is chosen}\nonumber\\
&\text{according to the probability measure } p(y_i,A) \text{ for all } i\}.\nonumber
\end{align}
Note that $(Y_i,Z_i,Y_{i+1})$ is stationary and ergodic for all stationary randomized policies.
In the sequel, when we refer to the stationary distribution of a stationary randomized policy $\pi\in \Pi_{\text{SR}}$, we will remove  subscript $i$. In particular, {the random variables $(Y_i,Z_i,Y_{i+1})$ are replaced by $(Y,Z,Y')$}, where $Z$ is chosen based on the conditional probability measure $\Pr[Z\in A| Y=y] =p(y,A)$ after observing $Y=y$, and $(Y,Y')$ have the same joint distribution as $(Y_0,Y_1)$.
The first key result of this paper is stated as follows:
\begin{theorem}\label{lem_SRoptimal}
(Optimality of Stationary Randomized Policies) If $M<\infty$, $g: [0,\infty)\rightarrow [0,\infty)$ is measurable and non-negative, $(Y_0,Y_1,\ldots)$ is a stationary ergodic Markov chain with $Y_i\geq0$ and $0<\mathbb{E}[Y_i]<\infty$, and condition \eqref{eq_bound} is satisfied,
then there exists a stationary randomized policy that is 
optimal for Problem \eqref{eq_DPExpected}.
\end{theorem}
\begin{IEEEproof}[Proof sketch of Theorem \ref{lem_SRoptimal}]
For any policy $\pi\in\Pi$, define the finite time-horizon average occupation measures
\begin{align}\label{eq_time_average_exp1}
&{a}_{n,\pi}\!\triangleq\!{\frac{1}{n}\mathbb{E}\bigg[\sum_{i=0}^{n-1} q(Y_i,Z_i,Y_{i+1})\bigg]}- \frac{\overline{g}_{\text{opt}}}{n}{\mathbb{E}\bigg[\sum_{i=0}^{n-1} (Y_{i}+Z_{i})\bigg]}, \\
&{b}_{n,\pi}\triangleq\frac{1}{n} \mathbb{E}\left[\sum_{i=0}^{n-1} (Y_{i}+Z_{i})\right].\label{eq_time_average_exp2}
\end{align}
Let $\Gamma_{\text{SR}}$ be the set of limit points of sequences $(({a}_{n,\pi},{b}_{n,\pi}),$ $n=1,2,\ldots)$ associated with all stationary randomized policies in $\Pi_{\text{SR}}$.
We first prove that $\Gamma_{\text{SR}}$ is convex and compact. Then, we show that there exists an optimal policy $\pi_{\text{opt}}$ of Problem \eqref{eq_DPExpected}, such that the sequence $(({a}_{n,\pi_{\text{opt}}},{b}_{n,\pi_{\text{opt}}}),n=1,2,\ldots)$ associated with policy $\pi_{\text{opt}}$ has a limit point $(a^*,b^*)$ satisfying $(a^*,b^*)\in\Gamma_{\text{SR}}$, $a^*\leq0$, and $b^*\geq \frac{1}{f_{\max}}$. Since $(a^*,b^*)\in\Gamma_{\text{SR}}$, there exists a stationary randomized policy $\pi^*$ achieving this limit point $(a^*,b^*)$. Finally, we show that policy $\pi^*$ is optimal for Problem \eqref{eq_DPExpected}, which completes the proof. The details are available in Appendix \ref{app3}.
\end{IEEEproof}

{The convexity and compactness properties of the set of average occupation measures are essential in the study of constrained MDPs \cite[Sec. 1.5]{Altman1999CMDP}, which dates back to Derman's monograph in 1970 \cite{Derman:1970:FSM:578852}. Recently, it was used in stochastic optimization for discrete-time queueing systems and renewal processes, e.g., \cite{Neely2010book,Neely2013TAC}. The techniques in these studies cannot directly handle constrained SMDPs with an uncountable state space, like Problem \eqref{eq_DPExpected}. One crucial novel idea in our proof is to introduce $\overline{g}_{\text{opt}}$ in the definition of average occupation measures in \eqref{eq_time_average_exp1}, which turns out to be essential in later steps for showing the optimality of the stationary randomized policy $\pi^*$. In addition, we have also used one property of Problem \eqref{eq_DPExpected} in the proof: the observation $Y_{i+1}$ depends only on the immediately preceding state $Y_i$ and not on earlier system states $(Y_0,\ldots, Y_{i-1})$ and control actions $(Z_0,\ldots, Z_{i-1})$.

}

By Theorem \ref{lem_SRoptimal}, we only need to consider the class of stationary randomized policies $\Pi_{\text{SR}}$.
Using this, Problem \eqref{eq_DPExpected} can be simplified to the following functional optimization problem, as shown in Appendix \ref{app3}:
\begin{align}\label{eq_SR}
\min_{\substack{p(y,A)}}&~ \frac{\mathbb{E}[q(Y,Z,Y')]}{\mathbb{E}[Y+Z]} \\
~\text{s.t.}~&~ \mathbb{E}[Y+Z]\geq \frac{1}{f_{\max}}\nonumber\\
&~0\leq Z\leq M,\nonumber
\end{align}
where $p(y,A)=\Pr[Z\in A| Y=y]$ is the conditional probability measure of some stationary randomized policy, and $(Y,Y')$ have the same distribution as $(Y_0,Y_1)$. 

\subsection{Optimality of Stationary Deterministic Policies}
A policy $\pi\in\Pi_{\text{SR}}$ is said to be a \emph{stationary deterministic} policy if $Z_i=z(Y_i)$ for all $i=0,1,\ldots$, where $z: [0,\infty)\rightarrow [0,M]$ is a deterministic function.
We use $\Pi_{\text{SD}}$ ($\Pi_{\text{SD}}\subseteq \Pi_{\text{SR}}$) to denote the set of stationary deterministic policies such that
\begin{align}
&\!\!\!\Pi_{\text{SD}}\!=\!\{\pi\in \Pi_{\text{SR}}\!:\! Z_i\!=\!z(Y_i)~\text{for all}~i \}.\!\!\nonumber
\end{align}

\begin{theorem}\label{lem_SD}
(Optimality of Stationary Deterministic Policies)
If $g: [0,\infty)\rightarrow [0,\infty)$ is measurable and non-decreasing, 
then there exists a stationary deterministic policy that is optimal for Problem \eqref{eq_SR}.
\end{theorem}

\begin{IEEEproof}[Proof sketch of Theorem \ref{lem_SD}]
Since $g(\age)$ is non-decreasing, $q(y,z,y')$ is convex in $z$ for any fixed $y$ and $y'$. Using Jensen's inequality, we can show that for any feasible stationary randomized policy $\pi_1\in \Pi_{\text{SR}}$, there is a feasible stationary deterministic policy that is no worse than policy $\pi_1$. The details are provided in
\ifreport
Appendix \ref{app4}.
\fi
\end{IEEEproof}

Let $\mu_Y$ be the probability measure of $Y_i$, then any bounded measurable function $z: [0,\infty)\rightarrow [0,M]$ belongs to the Lebesgue space $L^2(\mu_Y)$ \cite[Section 3]{Rudin87}, because
\begin{align}
\int_0^\infty |z(y)|^2 d\mu_Y(y)\leq  \int_0^\infty M^2 d\mu_Y(y) =M^2 <\infty.\nonumber
\end{align}
By Theorems \ref{lem_SRoptimal} and \ref{lem_SD}, we only need to consider the class of stationary deterministic policies $\Pi_{\text{SD}}$ and Problem \eqref{eq_DPExpected} is simplified as the following functional optimization problem:
\begin{align}\label{eq_SD}
\min_{\substack{z(\cdot)\in L^2(\mu_Y)}}&~ \frac{\mathbb{E}\left[q(Y,z(Y),Y')\right]}{\mathbb{E}[Y+z(Y)]} \\
~\text{s.t.}~~~~&~ \mathbb{E}[Y+z(Y)]\geq \frac{1}{f_{\max}}\label{eq_constraint}\\
&~0\leq z(y)\leq M,~\forall~y\geq0,\nonumber
\end{align}
where $z(\cdot)$ is the  function associated with a stationary deterministic policy $\pi\in\Pi_{\text{SD}}$, and $(Y,Y')$ have the same distribution as $(Y_0,Y_1)$. The optimum objective value of Problem \eqref{eq_SD} is equal to $\overline{g}_{\text{opt}}$.

\ifreport
\begin{algorithm}
\caption{Two-layer bisection method for Problem \eqref{eq_SD}} \label{alg1}
\begin{algorithmic}[]
\STATE \textbf{given} $l=0$, sufficiently large $u>\overline{g}_{\text{opt}}$, tolerance $\epsilon_1$.
\REPEAT
\STATE $c:= (l+u)/2$.
\STATE \textbf{given} $\zeta_l=0$, sufficiently large $\zeta_u>0$, tolerance $\epsilon_2$.
\STATE $\zeta:=\zeta_l,~\nu:=\zeta+c$.
\STATE Compute $z_{\nu}(\cdot)$ in \eqref{z1_solution}.
\IF{$\mathbb{E}[Y+z_{\nu}(Y)]< \frac{1}{f_{\max}}$}
\REPEAT
\STATE $\zeta:= (\zeta_l+\zeta_u)/2,~\nu:=\zeta+c$.
\STATE Compute $z_{\nu}(\cdot)$ in \eqref{z1_solution}.
\STATE \textbf{if} $\mathbb{E}[Y+z_{\nu}(Y)]\geq \frac{1}{f_{\max}}$, $\zeta_u\!:=\zeta$; \textbf{else}, $\zeta_l\!:=\zeta$.
\UNTIL $\zeta_u-\zeta_l\leq \epsilon_2$.
\ENDIF
\STATE \textbf{if} $f(c)\leq 0$, $u:=c$; \textbf{else}, $l:=c$.
\UNTIL $u-l\leq \epsilon_1$.
\STATE \textbf{return} $z(\cdot):=z_{\nu}(\cdot)$.
\end{algorithmic}
\end{algorithm}
\fi
\subsection{A Low Complexity Solution to Problem \eqref{eq_SD}}
\begin{lemma}\label{lem_quasiconvex}
If $g: [0,\infty)\rightarrow [0,\infty)$ is measurable, non-negative, and non-decreasing, then
the functional $h: L^2(\mu_Y) \rightarrow [0,\infty)$ defined by $$h(z)=\frac{\mathbb{E}\left[q(Y,z(Y),Y')\right]}{\mathbb{E}[Y+z(Y)]}$$ is quasi-convex.
\end{lemma}
\begin{IEEEproof}
\ifreport
See Appendix \ref{app5_0}.
\fi
\end{IEEEproof}

Therefore, Problem \eqref{eq_SD} is a functional quasi-convex optimization problem.
%
In order to solve Problem \eqref{eq_SD}, we consider the following functional convex optimization problem with a parameter $c$:
\begin{align}\label{eq_SD_equavilent}
f(c)=\min_{\substack{z(\cdot)\in L^2(\mu_Y)}}&~{\mathbb{E}\left[q(Y,z(Y),Y')\right]}- c\mathbb{E}[Y+z(Y)]\\
~\text{s.t.}~~~~&~\mathbb{E}[Y+z(Y)]\geq \frac{1}{f_{\max}}\label{eq_SD_equavilent1}\\
&~0\leq z(y)\leq M,~\forall~y\geq0.\nonumber
\end{align}
It is easy to show that $\overline{g}_{\text{opt}}\leq c$ if and only if $f(c)\leq 0$ \cite{Boyd04}. Therefore, we can solve Problem \eqref{eq_SD} by a two-layer nested algorithm, such as Algorithm \ref{alg1}. In the inner layer, we use \emph{bisection} to solve Problem \eqref{eq_SD_equavilent} for any given parameter $c$; in the outer layer, we employ \emph{bisection} again to search for a $c^*$ such that $f(c^*)=0$ and thus $\overline{g}_{\text{opt}}= c^*$. Algorithm \ref{alg1} has low complexity. It requires at most $\lceil\log_2((u-l)/\epsilon_1)\rceil\times$ $\lceil\log_2((\zeta_u-\zeta_l)/\epsilon_2)\rceil$ iterations to terminate and each iteration involves computing $\mathbb{E}[z_\nu(Y)]$ based on \eqref{z1_solution}. The optimality of Algorithm \ref{alg1} is guaranteed by the following theorem:

\begin{theorem}\label{lem_SD_solution}
If $g: [0,\infty)\rightarrow [0,\infty)$ is measurable, non-negative, and non-decreasing, then an optimal solution $\pi_{\text{opt}}$ to Problem \eqref{eq_SD} is obtained by Algorithm \ref{alg1}, where 
the function $z_\nu(\cdot)$ is determined by
\begin{align}\label{z1_solution}
\!\!\!z_\nu(y)=\sup\{z\in[0,M]:\mathbb{E}\left[ g(y\!+\!z\!+\!Y')|Y=y\right] \leq \nu\},
\end{align}
and $(Y,Y')$ follow the same distribution as $(Y_0,Y_1)$.
\end{theorem}
\begin{IEEEproof}[Proof Sketch of Theorem \ref{lem_SD_solution}]
{We use Lagrangian duality theory to solve Problem \eqref{eq_SD}. Different from traditional finite dimensional optimization problems \cite{Boyd04}, Problem \eqref{eq_SD} is an infinite dimensional functional optimization problem. Therefore, the Karush-Kuhn-Tucker (KKT) theorem for infinite dimensional space \cite{infinite_dimensional,convex_analysis} and the calculus of variations are required in the analysis. In particular, since the Lagrangian may not be strictly convex for some penalty functions, one-sided G\^ateaux derivative (similar to sub-gradient in finite dimensional space) is used to solve the KKT conditions in Lebesgue space $L^2(\mu_Y)$. The proof details are provided in
\ifreport
Appendix \ref{app5_1}.
\fi
}
\end{IEEEproof}
\ifreport
The policy spaces $\Pi$, $\Pi_{\text{SR}}$, $\Pi_{\text{SD}}$, and the obtained optimal policy $\pi_{\text{opt}}$ are depicted in Fig. \ref{policy_spaces}.
\fi


\section{When Is It Better to Wait than to Update?}\label{sec:whenoptimal}
When ${f_{\max}}=\infty$, the constraint \eqref{eq_SD_equavilent1} is always satisfied. In this case, a logical policy is the \emph{zero-wait} policy: the source node submits a fresh update once the prior update is delivered, i.e., $\pi_{\text{zero-wait}}=(0,0,\ldots)$. According to the example in the introduction, this zero-wait policy is not always optimal. 
In this section, we will study when it is optimal to submit updates with the minimum average waiting time and when it is not.
\ifreport
\begin{figure}
\ifjournal
\centering \includegraphics[width=0.35\textwidth]{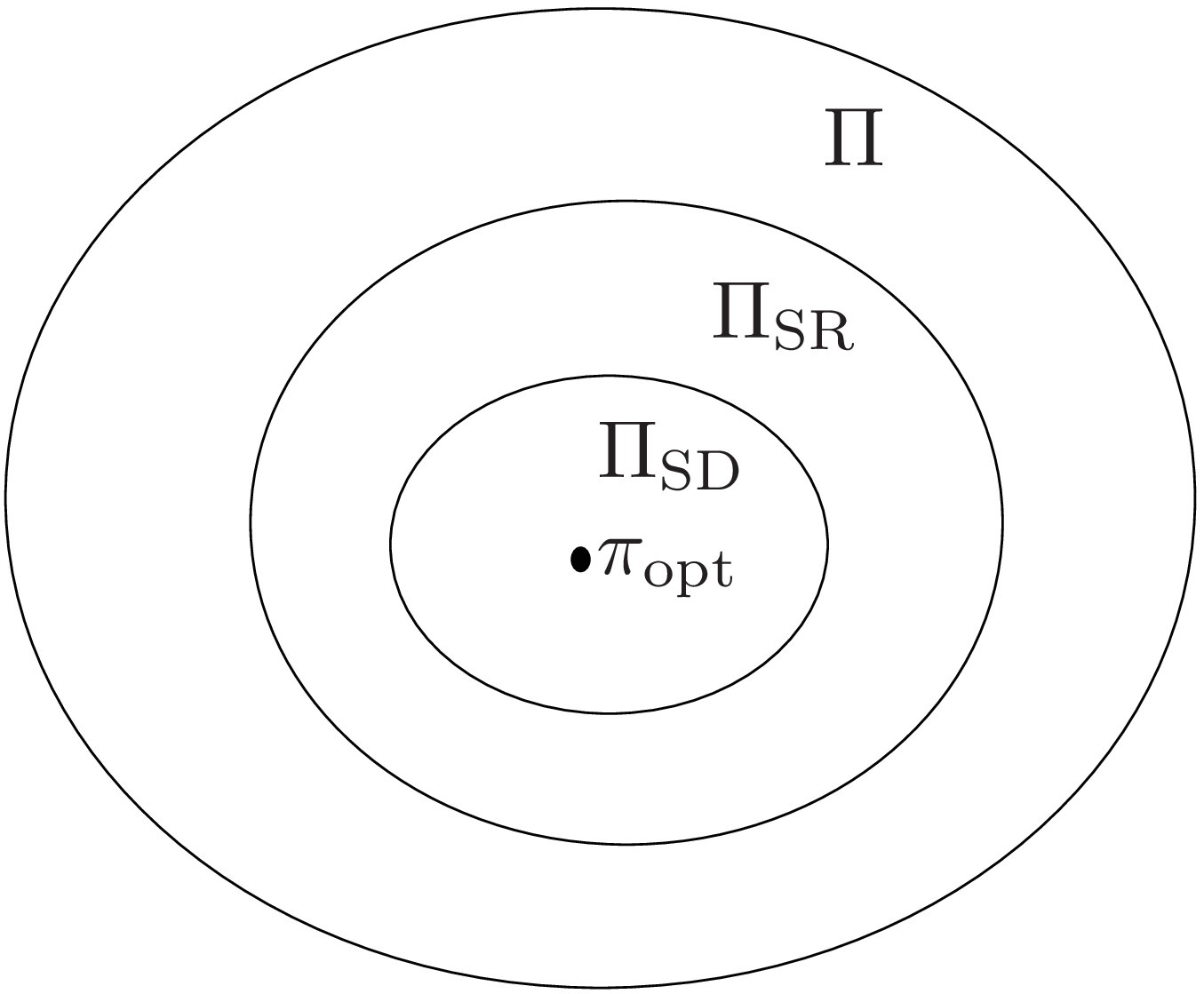} 
\else
\centering \includegraphics[width=0.25\textwidth]{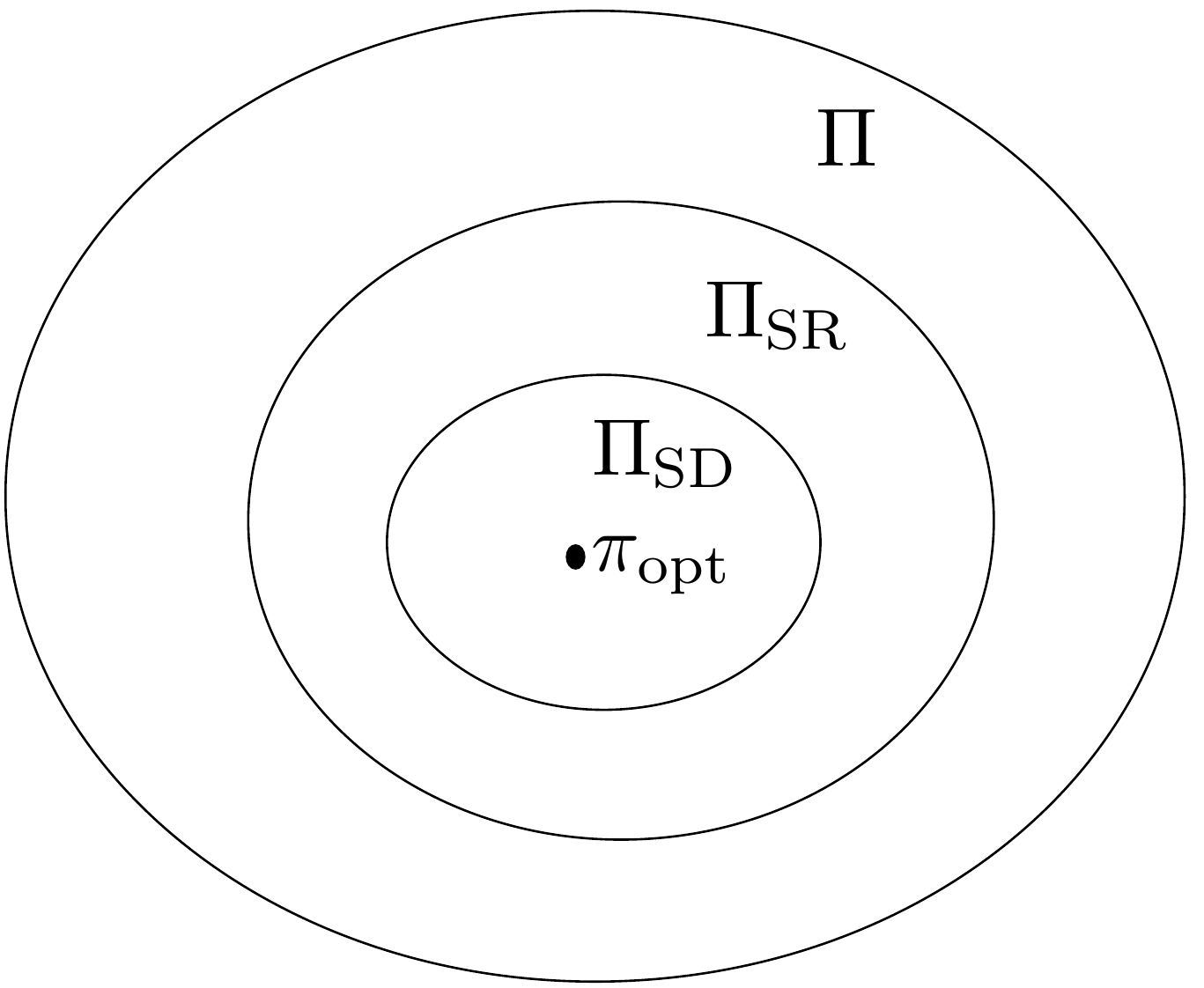} 
\fi
\caption{Illustration of the set of causally feasible  policies $\Pi$, the set of stationary randomized policies $\Pi_{\text{SR}}$, the set of stationary deterministic policies $\Pi_{\text{SD}}$, and the obtained optimal policy $\pi_{\text{opt}}$.}
\label{policy_spaces} \vspace{-0.5cm}
\end{figure}
\else
\fi
\subsection{A Special Case of $g(\age)=\age$ with i.i.d. Transmission Times}
Consider the case that $g(\age)=\age$ and the $Y_i$'s are \emph{i.i.d.} with $0<\mathbb{E}[Y]<\infty$. In this case, Problem \eqref{eq_SD} has a simpler solution than that provided by Algorithm \ref{alg1}. Interestingly, this solution explicitly characterizes whether the optimal control $z(\cdot)$ can have the minimum average waiting time such that $\mathbb{E}[Y+ z(Y)]= {1}/{f_{\max}}$.

As shown in Fig. \ref{fig:age1}, $Q_n = \frac{1}{2}\left[(Y_n+Z_n+Y_{n+1})^2-Y_n^2\right]$ is the area of a trapezoid. This corresponds to
\begin{align}
q(y,z,y') = \frac{1}{2}\left[(y+z+y')^2-y^2\right]. \nonumber
\end{align}
Because the $Y_i$'s are \emph{i.i.d.}, $Y$ and $Y'$ in Problem \eqref{eq_SD} are also \emph{i.i.d.} Using this, we can obtain
\begin{align}
&\mathbb{E}[q(Y,z(Y),Y')] \nonumber\\
=&\mathbb{E}\left[\frac{1}{2}(Y+z(Y)+Y')^2 -\frac{1}{2}{Y'}^2\right]\label{eq_step}\\
=&\frac{1}{2}\mathbb{E}\left[(Y+z(Y))^2\right] + \mathbb{E}\left[Y+z(Y)\right]\mathbb{E}\left[Y'\right],\nonumber
\end{align}
where in \eqref{eq_step} we have used that $\mathbb{E}[Y^2]=\mathbb{E}[Y'^2]$.
Hence, Problem \eqref{eq_SD} can be reformulated as
\begin{align}\label{eq_SD_average}
\min_{\substack{z\in L^2(\mu_Y)}}&~ \frac{\mathbb{E}[(Y+z(Y))^2]}{2\mathbb{E}[Y+z(Y)]} + \mathbb{E}[Y_{}] \\
~\text{s.t.}~~~&~ \mathbb{E}[Y+z(Y)]\geq \frac{1}{f_{\max}}\label{eq_SD_average_con}\\
&~0\leq z(y)\leq M,~\forall~y\geq0.\nonumber
\end{align}
The following lemma tells us that Problem \eqref{eq_SD_average} is a functional convex optimization problem. 
\ifreport
\begin{algorithm}
\caption{Bisection method for solving Problem \eqref{eq_SD_average}} \label{alg2}
\begin{algorithmic}[]
\STATE \textbf{given} $l=0$, sufficiently large $u$, tolerance $\epsilon$.
\REPEAT
\STATE $\beta:= (l+u)/2$.
\STATE $o :=\mathbb{E}\left[(Y+z(Y))\right]-\max\left(\frac{1}{f_{\max}},\frac{\mathbb{E}\left[(Y+z(Y))^2\right]}{2\beta}\right)$, where $z(\cdot)$ is given by \eqref{z_solution}.
\STATE \textbf{if} $o\geq 0$, $u:=\beta$; \textbf{else}, $l:=\beta$.
\UNTIL $u-l\leq \epsilon$.
\STATE Compute $z(\cdot)$ by \eqref{z_solution}.
\STATE \textbf{return} $z(\cdot)$.
\end{algorithmic}
\end{algorithm}
\fi

\begin{lemma}\label{thm_convex}
The functional $h_1: L^2(\mu_Y) \rightarrow \mathbb{R}$ defined by $$h_1(z) = \frac{\mathbb{E}[(Y+z(Y))^2]}{\mathbb{E}[Y+z(Y)]}$$ is convex on the domain
$$\text{\textbf{dom} } h_1 = \left\{z\in L^2(\mu_Y):  z(y)\in[0,M], ~\forall y\geq0 \right\}.$$
\end{lemma}
\begin{IEEEproof}
\ifreport
See Appendix \ref{app5}.
\fi
\end{IEEEproof}
Using the KKT theorem for infinite dimensional space and the calculus of variations, we can obtain

\begin{theorem}\label{beta}
If $\mathbb{E}[Y]>0$, the optimal solution to Problem \eqref{eq_SD_average} is
\begin{equation}\label{z_solution}
z(y)=(\beta-y)_0^M,
\end{equation}
where $(x)_0^M \triangleq \min\{\max\{x,0\},M\}$ and $\beta>0$ satisfies
\begin{align}\label{Ebeta}
\mathbb{E}\left[Y+z(Y)\right]&=\max\bigg(\frac{1}{f_{\max}},\frac{\mathbb{E}[(Y+z(Y))^2]}{2\beta}\bigg).
\end{align}
\end{theorem}

\begin{IEEEproof}
\ifreport
See Appendix \ref{app6}.
\fi
\end{IEEEproof}

Equation \eqref{z_solution} has the form of a \emph{water-filling} solution, where the water-level $\beta$ is given by the root of equation \eqref{Ebeta}. One can observe that \eqref{z1_solution} reduces to \eqref{z_solution} if $g(\age)=\age$, the $Y_i$'s are \emph{i.i.d.}, and $\nu$ is replaced by $\beta+\mathbb{E}[Y]$. The root $\beta$ of equation \eqref{Ebeta} can be simply solved by the bisection search method in Algorithm \ref{alg2}. We note that Algorithm \ref{alg2} has lower complexity than Algorithm \ref{alg1} in the special case of $g(\age)=\age$ and \emph{i.i.d.} transmission process, while Algorithm \ref{alg1} can obtain the optimal policy in more general scenarios. 

Theorem \ref{beta} provides a closed-form criterion on whether the optimal $z(\cdot)$ satisfies $\mathbb{E}[Y+z(Y)]= {1}/{f_{\max}}$.
Specifically, \eqref{z_solution} and \eqref{Ebeta} tell us that if ${1}/{f_{\max}}\geq \frac{\mathbb{E}[(Y+z(Y))^2]}{2\beta}$, then the optimal control $z(\cdot)$ satisfies
\begin{align}\label{Ebeta1}
\mathbb{E}[Y+z(Y)]= \frac{1}{f_{\max}}\geq \frac{\mathbb{E}[(Y+z(Y))^2]}{2\beta},
\end{align}
such that the optimal policy achieves the minimum possible average waiting time;
otherwise, if ${1}/{f_{\max}}< \frac{\mathbb{E}[(Y+z(Y))^2]}{2\beta}$, the optimal control $z(\cdot)$ satisfies
\begin{align}\label{Ebeta2}
\mathbb{E}[Y+z(Y)]=\frac{\mathbb{E}[(Y+z(Y))^2]}{2\beta}>\frac{1}{f_{\max}}, 
\end{align}
such that the optimal policy can not achieve the minimum possible average waiting time.
{\orange In \cite{2015ISITYates}, the author solved a slightly different version of Problem \eqref{eq_SD_average}: an equality constraint on the updating frequency was considered in \cite{2015ISITYates}, while an inequality constraint is adopted in Problem \eqref{eq_SD_average}. It was observed in \cite{2015ISITYates} that the optimal time-average age is not necessarily decreasing in the update frequency. The solution to Problem \eqref{eq_SD_average} in Theorem \ref{beta} further allows us to obtain the optimal update frequency.} 


Furthermore, if $\mathbb{E}[Y]\geq  {1}/{f_{\max}}$ (e.g., $f_{\max}=\infty$), the constraint \eqref{eq_SD_average_con} is always satisfied and can be removed. 
In this case, the optimality of the zero-wait policy is characterized by a sufficient and necessary condition in the following theorem.
\begin{theorem}\label{iffcondition}
If $\mathbb{E}[Y]>0$ and $f_{\max}=\infty$, then the zero-wait policy is optimal for  Problem \eqref{eq_SD_average}  if, and only if,
\begin{align}\label{eq_iffcondition}
{\mathbb{E}[Y^2]}{} \leq 2 y_{\inf} \mathbb{E}[Y],
\end{align}
where $y_{\inf}=\inf \{ y\in[0,\infty): \Pr[Y\leq y]>0\}$. 

\end{theorem}
\begin{IEEEproof}
\ifreport
See Appendix \ref{app8}.
\fi
\end{IEEEproof}
Informally speaking, $y_{\inf}$ is the smallest possible value of the random transmission time $Y$. 
From Theorem \ref{iffcondition}, it immediately follows that:
\begin{coro}\label{coro1}
If $\mathbb{E}[Y]>0$ and the zero-wait policy is feasible, then the following assertions are true:
\begin{itemize}
\item[(a).] If the transmission times are positive and constant (i.e., $Y = const>0$), the zero-wait policy is optimal for  Problem \eqref{eq_SD_average}. 

\item[(b).] If the transmission times satisfy $y_{\inf}=0$, the zero-wait policy is \emph{not} optimal for  Problem \eqref{eq_SD_average}. 
\end{itemize}
\end{coro}
{\bf Remark:} As one can readily see from Corollary \ref{coro1}(b), the zero-wait policy is not optimal for many commonly used distributions in communication and queueing theory, such as exponential distribution, geometric distribution, Erlang distribution, hyperexponential distribution, etc.


\ifjournal
\begin{figure}
\centering \includegraphics[width=0.6\textwidth]{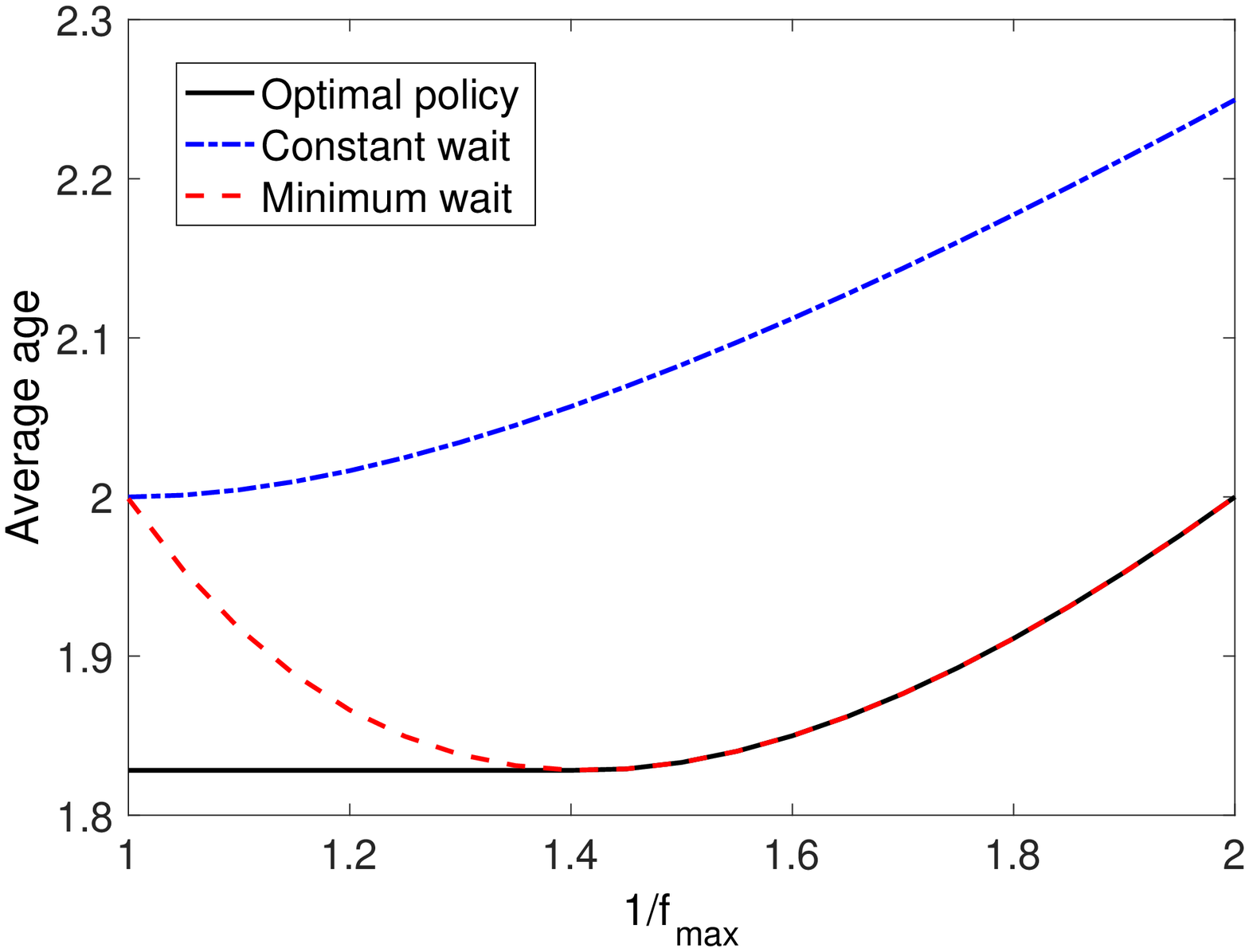}
 \caption{Average age vs. ${1}/{f_{\max}}$ with $i.i.d.$ discrete transmission times.}
\label{fig1} \vspace{-0.cm}
\end{figure}
\begin{figure}
\centering \includegraphics[width=0.6\textwidth]{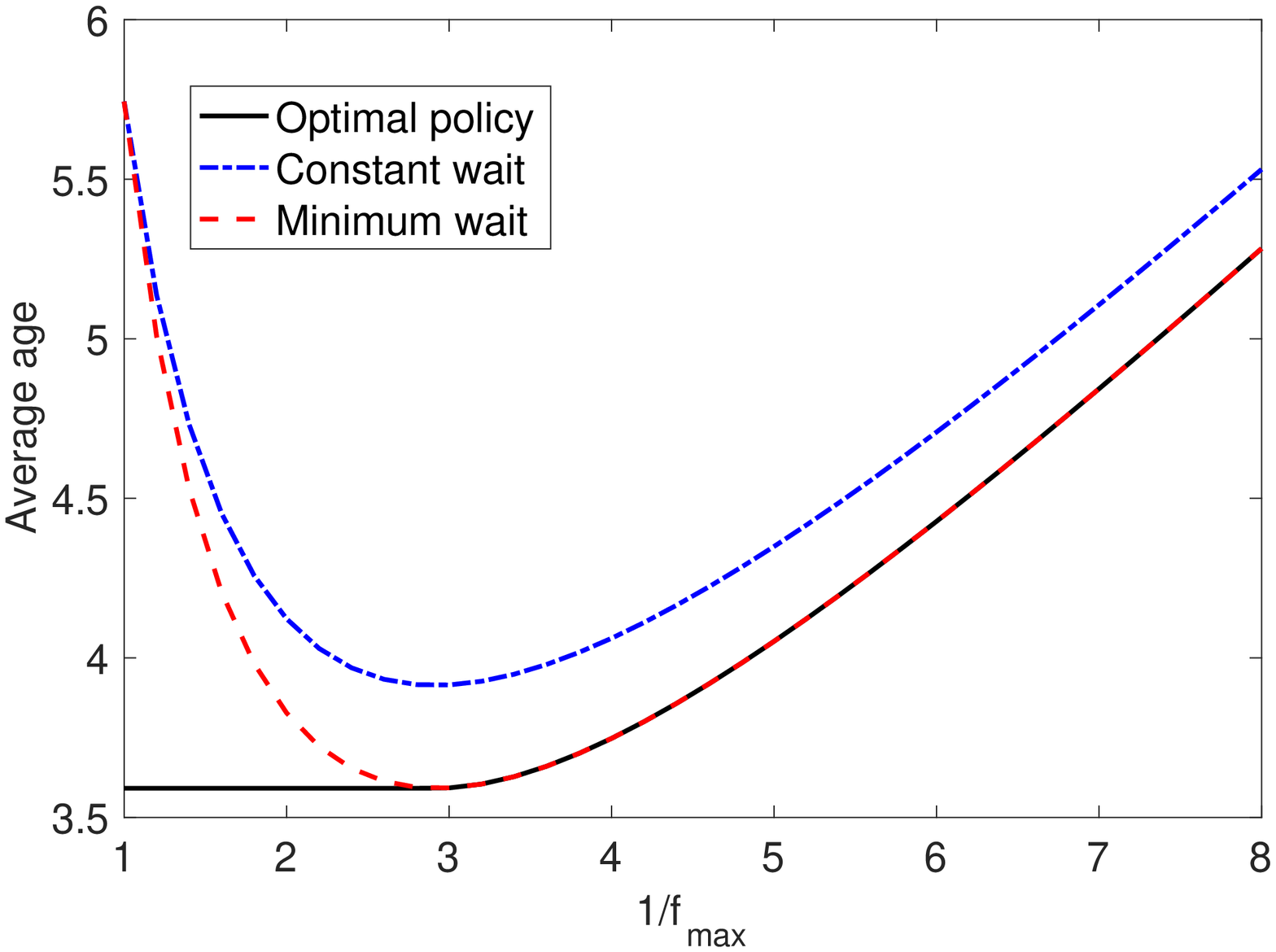}
\caption{Average age vs. ${1}/{f_{\max}}$ with $i.i.d.$ log-normal distributed transmission times.}
\label{fig2} \vspace{-0.cm}
\end{figure}
\fi

\subsubsection{Numerical Results}
We use ``optimal policy" to refer to the policy provided in Theorem \ref{lem_SD_solution} (or its special case in Theorem \ref{beta}), and compare it with three reference policies:
\begin{itemize}
\item ``Constant wait'': Each update is followed by a constant waiting time $Z= {1}/{f_{\max}}-\mathbb{E}[Y]$ before submitting the next update.

\item ``Minimum wait'': The update waiting time is determined by  $Z=z(Y)$, where  $z(\cdot)$ is given by \eqref{z_solution} and $\beta$ in \eqref{Ebeta} is chosen to satisfy $\mathbb{E}[z(Y)]= {1}/{f_{\max}}-\mathbb{E}[Y]$.\footnote{This policy was called ``$\beta$-minimum'' in \cite{2015ISITYates}.}

\end{itemize}
When $\mathbb{E}[Y]={1}/{f_{\max}}$, both the constant wait and minimum wait policies reduce to the zero-wait policy.

Two transmission time models are considered: The first  is a discrete Markov chain with a probability mass function $\Pr[Y_i = 0]=\Pr[Y_i=2]=0.5$ and a transition matrix
\begin{align}
P = \left[\begin{array}{c c} p &1-p\\1-p&p  \end{array}\right].\nonumber
\end{align}
Hence, the $Y_i$'s are \emph{i.i.d.} when $p = 0.5$, and the correlation coefficient between $Y_i$ and $Y_{i+1}$ is
$\rho_{\text{}} = 2p-1$. The second  is a log-normal distributed Markov chain, where $Y_i = e^{\sigma X_i}/ \mathbb{E}[e^{\sigma X_i}]$ and $(X_0,X_1,\ldots)$ is a Gaussian Markov process satisfying the first-order autoregressive (AR) equation
\begin{align}
X_{i+1} = \eta X_i + \sqrt{1-\eta^2}W_i,\nonumber
\end{align}
where $\sigma>0$ is the scale parameter of log-normal distribution, $\eta\in[-1,1]$ is the parameter of the AR model, and the $W_i$'s are \emph{i.i.d.} Gaussian random variables with zero mean and unit variance. The log-normal distributed Markov chain is normalized such that $\mathbb{E}[Y_i]=1$.  According to the properties of log-normal distribution, the correlation coefficient between $Y_i$ and $Y_{i+1}$ is
$\rho = (e^\eta-1)/(e-1)$. Then, the $Y_i$'s are \emph{i.i.d.} when $\eta = 0$. {\blue The value of $M$ is set to be $10$.}
\ifreport
\ifjournal
\else
\begin{figure}
\centering \includegraphics[width=0.3\textwidth]{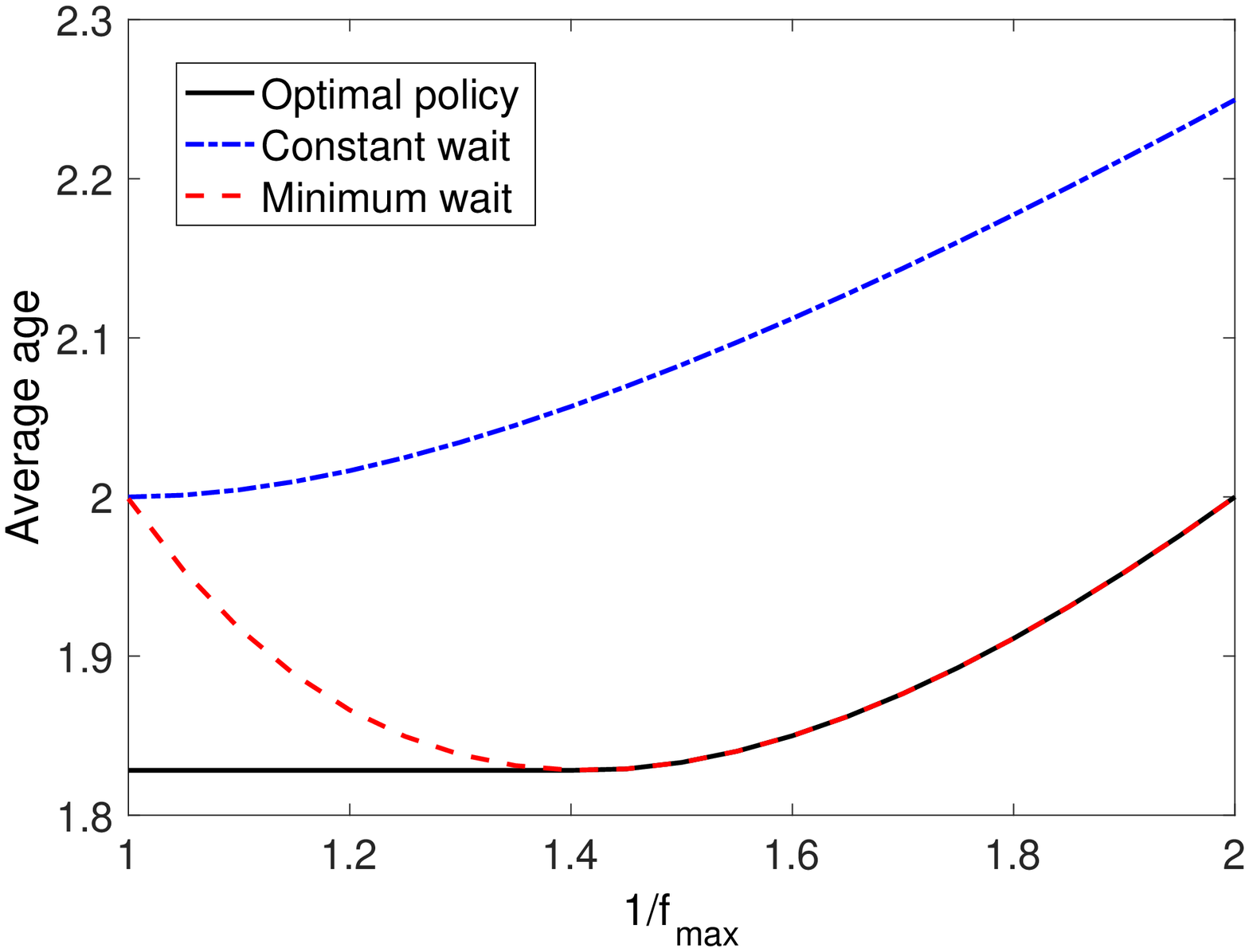} \caption{Average age vs. ${1}/{f_{\max}}$ with $i.i.d.$ discrete transmission times.}
\label{fig1} \vspace{-0.cm}
\end{figure}
\begin{figure}
\centering \includegraphics[width=0.3\textwidth]{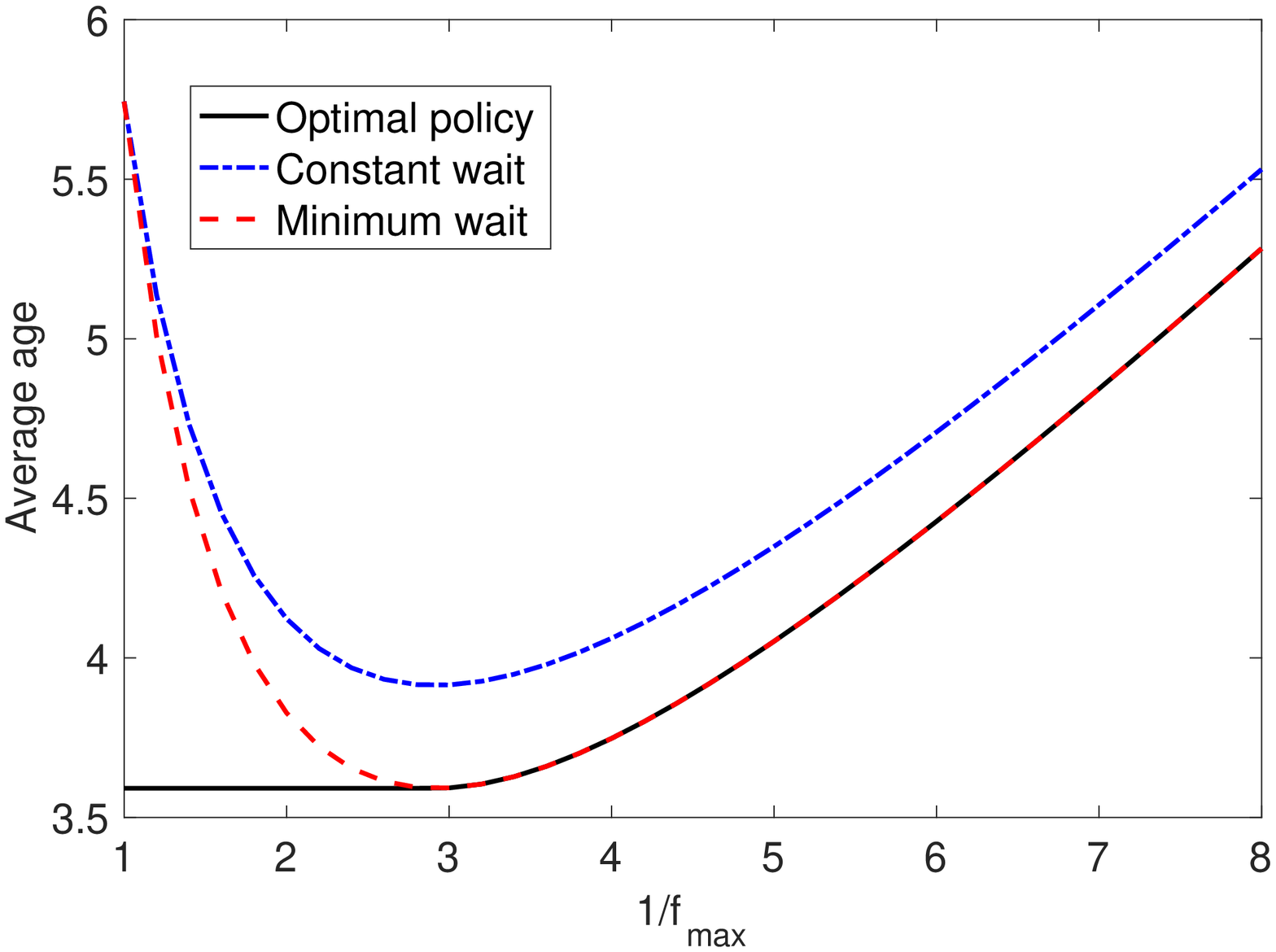} \caption{Average age vs. ${1}/{f_{\max}}$ with $i.i.d.$ log-normal distributed transmission times.}
\label{fig2} \vspace{-0.5cm}
\end{figure}
\fi
\fi

Figures \ref{fig1} and \ref{fig2} illustrate the average age vs. ${f_{\max}}$ for \emph{i.i.d.} discrete and log-normal distributed transmission times, respectively, where $\sigma=1.5$ In both figures, one can observe that the constant wait policy always incurs a larger average age than the optimal policy. 
In addition, as expected from \eqref{Ebeta1} and \eqref{Ebeta2}, as ${1}/{f_{\max}}$ exceeds a certain threshold, the optimal policy meets the constraint \eqref{eq_SD_average_con} with equality. For smaller values of ${1}/{f_{\max}}$, the constraint \eqref{eq_SD_average_con} is not active in the optimal solution. Consequently, the minimum wait policy deviates from the optimal policy for small values of ${1}/{f_{\max}}$, which is in accordance with Corollary \ref{coro1}(b).

%

\ifjournal
\begin{figure}
\centering \includegraphics[width=0.6\textwidth]{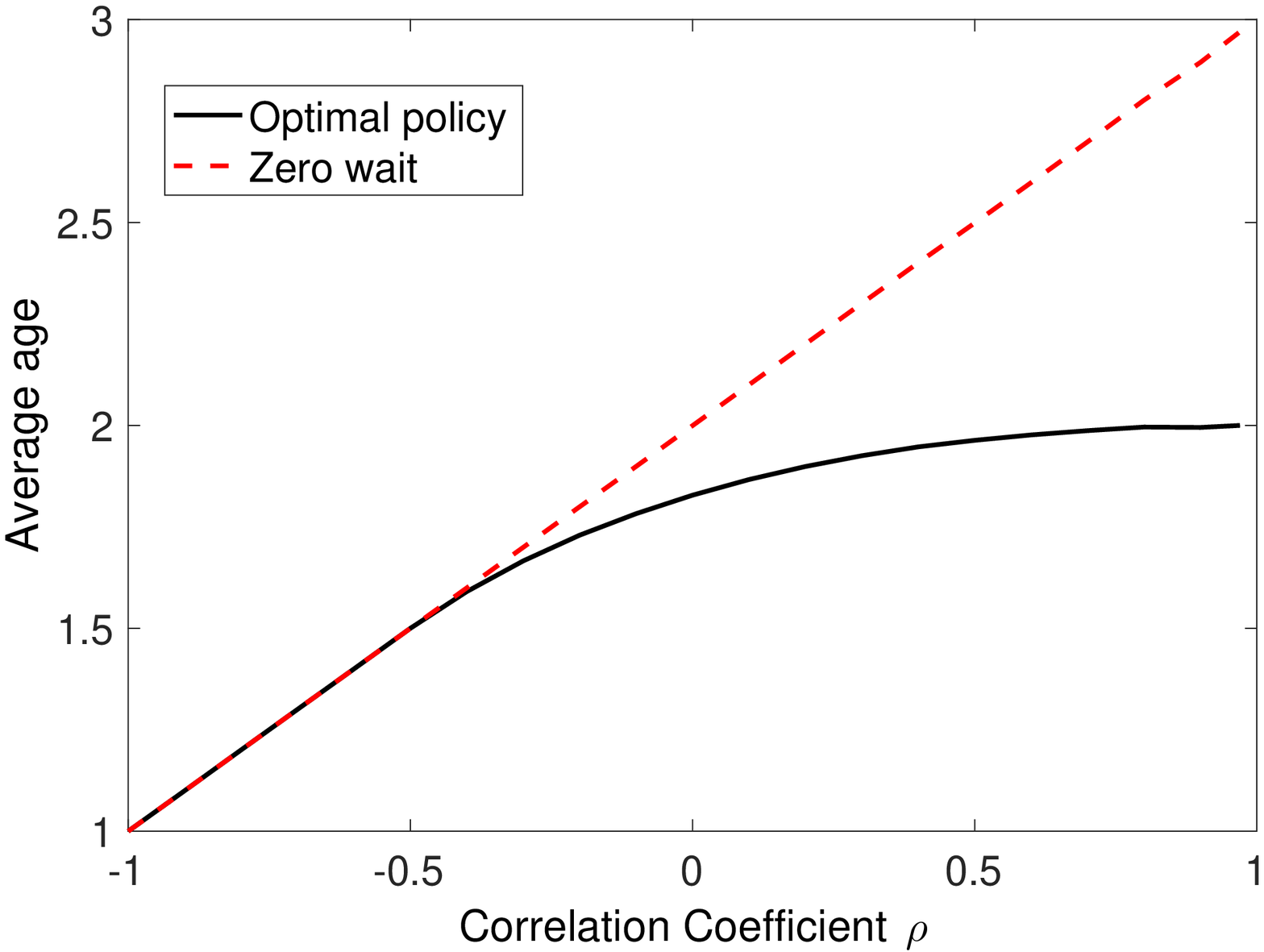} 
\caption{Average age vs. the correlation coefficient $\rho$ between $Y_i$ and $Y_{i+1}$ for discrete transmission times, where $\mathbb{E}[Y]\geq{1}/{f_{\max}}$.}
\label{fig5} \vspace{-0.3cm}
\end{figure}
\begin{figure}
\centering \includegraphics[width=0.6\textwidth]{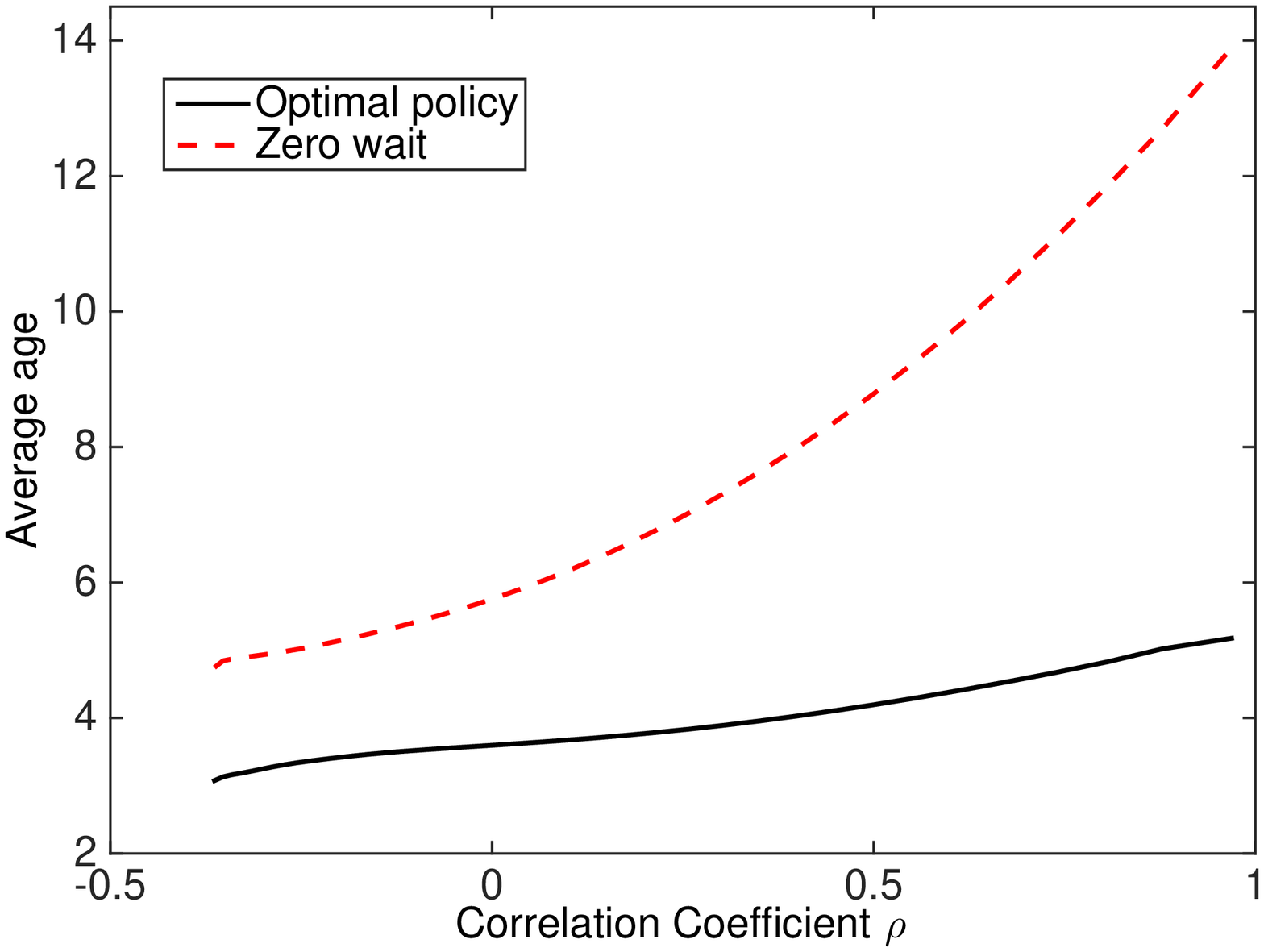} 
\caption{Average age vs. the correlation coefficient $\rho$ between $Y_i$ and $Y_{i+1}$ for log-normal distributed transmission times, where $\mathbb{E}[Y]\geq{1}/{f_{\max}}$.}
\label{fig6} \vspace{-0.5cm}
\end{figure}
\begin{figure}
\centering \includegraphics[width=0.6\textwidth]{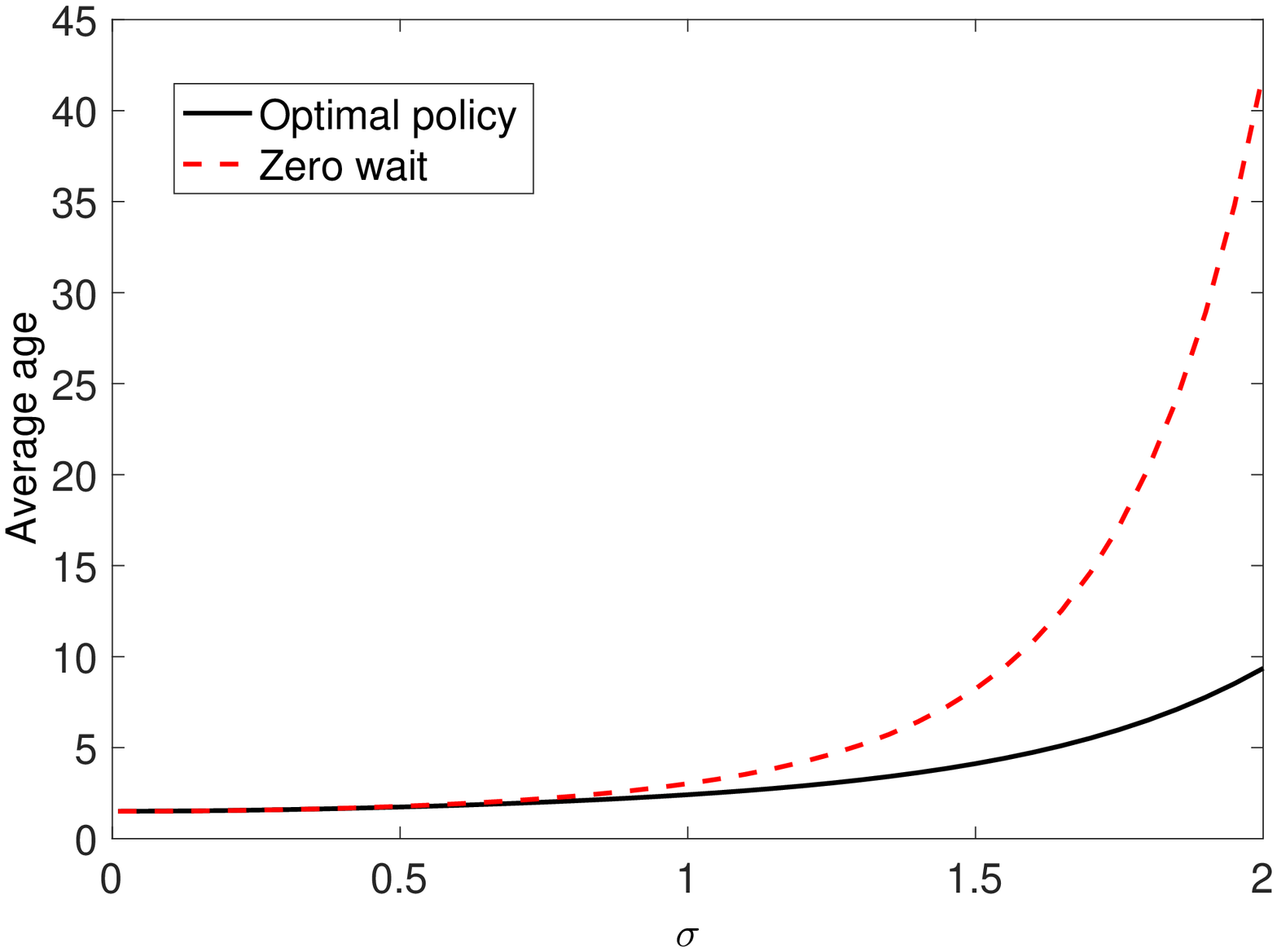} \caption{Average age vs. the distribution parameter $\sigma$ of log-normal distributed transmission times, where $\mathbb{E}[Y]\geq{1}/{f_{\max}}$ and $\rho = (e^{0.5}-1)/(e-1)$.}
\label{fig4} \vspace{-0.cm}
\end{figure}
\fi

\ifreport
\ifjournal
\else
\begin{figure}
\centering \includegraphics[width=0.3\textwidth]{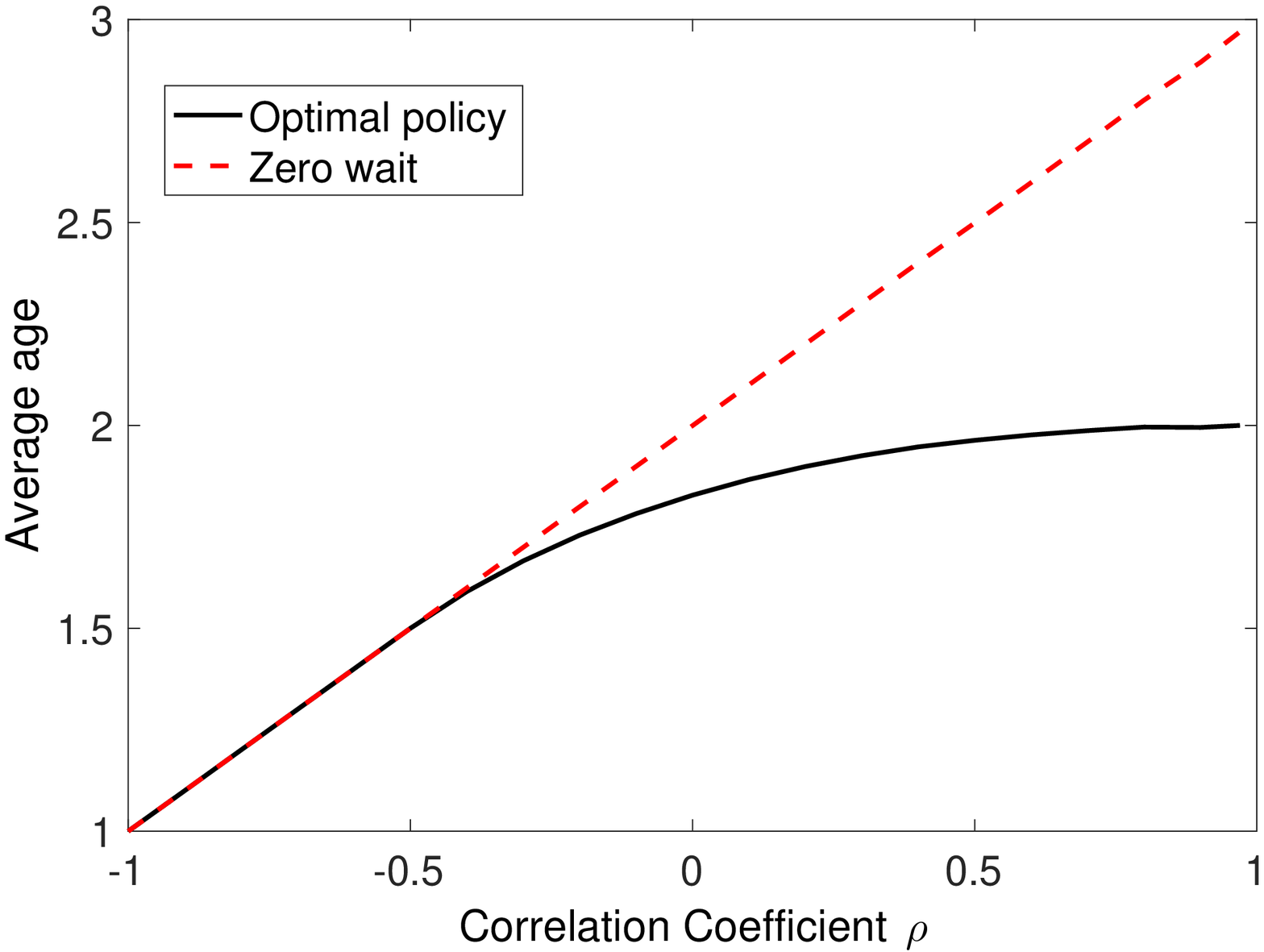} 
\caption{Average age vs. the correlation coefficient $\rho$ between $Y_i$ and $Y_{i+1}$ for discrete transmission times, where $\mathbb{E}[Y]\geq{1}/{f_{\max}}$.}
\label{fig5} \vspace{-0.cm}
\end{figure}
\begin{figure}
\vspace{0.cm}
\centering 
\centering \includegraphics[width=0.3\textwidth]{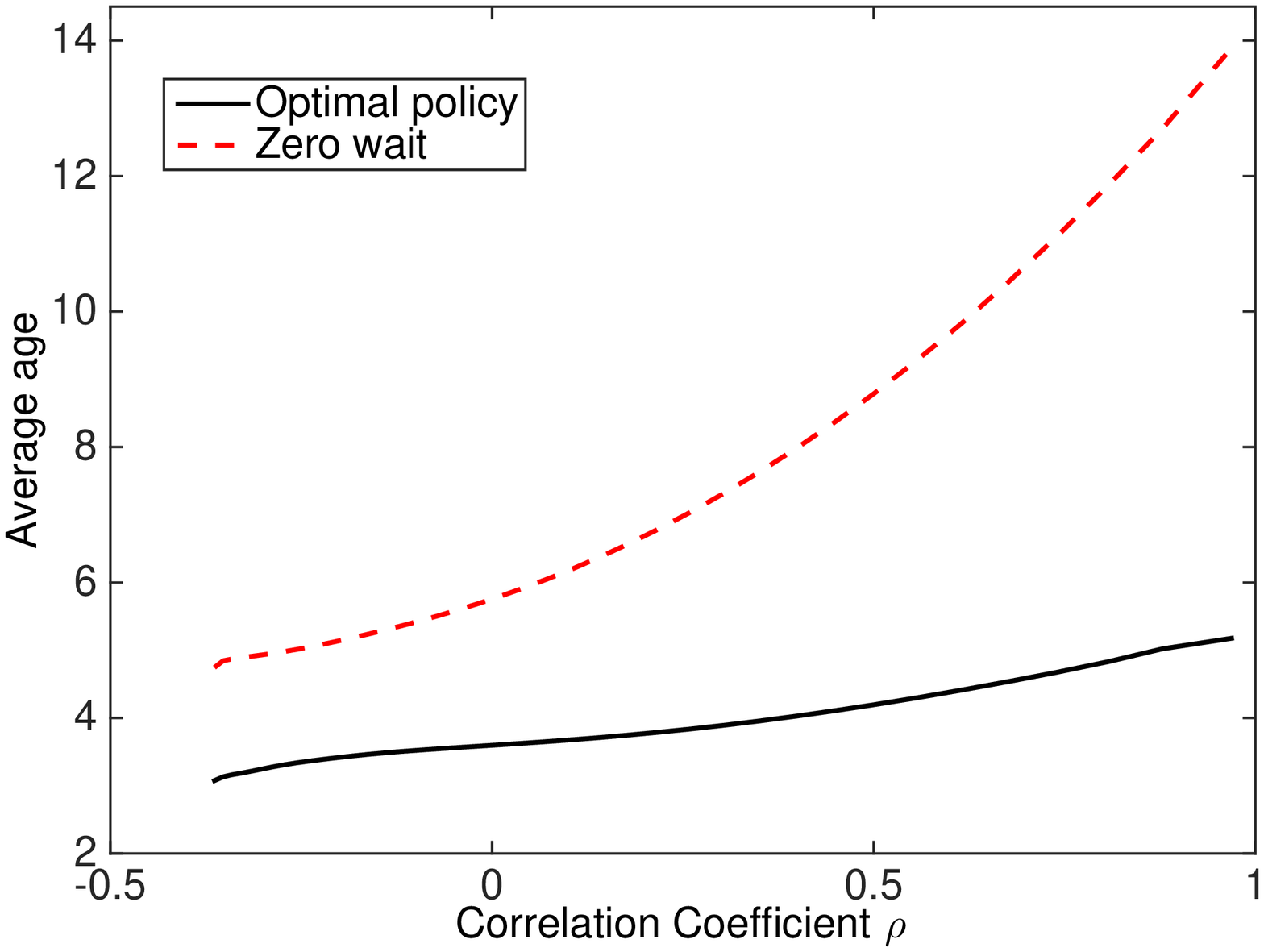}
\caption{Average age vs. the correlation coefficient $\rho$ between $Y_i$ and $Y_{i+1}$ for log-normal distributed transmission times, where $\mathbb{E}[Y]\geq{1}/{f_{\max}}$.}
\label{fig6} 
\end{figure}
\fi
\fi

\subsection{General Age Penalties and Correlated  Transmission Times}
For general age penalties and correlated transmission time processes, it is essentially difficult to find closed-form characterization on whether the optimal control $z(\cdot)$ can have the minimum average waiting time such that $\mathbb{E}[Y+z(Y)]= {1}/{f_{\max}}$.
Therefore, we focus on the case of $\mathbb{E}[Y]\geq {1}/{f_{\max}}$ (this is equivalent to removing the update frequency constraint \eqref{eq_SD_equavilent1}) and study when the zero-wait policy minimizes the average age penalty. Sufficient conditions for the optimality of the zero-wait policy are provided as follows:
\begin{lemma}\label{just_in_time}
Suppose that $\mathbb{E}[Y]\geq {1}/{f_{\max}}$, $g(\cdot)$ is measurable, non-negative, and non-decreasing. The zero-wait policy is optimal for Problem \eqref{eq_SD} if one of the following conditions is satisfied:
\begin{itemize}
\item[1).] The correlation coefficient between $Y_i$ and $Y_{i+1}$ is $-1$;
\item[2).] The $Y_i$'s are equal to a constant value;
\item[3).] $g(\cdot)$ is a constant function.
\end{itemize}
\end{lemma}
\begin{IEEEproof}
\ifreport
See Appendix \ref{app7}.
\fi
\end{IEEEproof}

\subsubsection{Numerical Results}
We now provide some Numerical Results for general age penalties and/or correlated transmission time processes.
Figures \ref{fig5} and \ref{fig6} depict the average age vs. the correlation coefficient $\rho$ between $Y_i$ and $Y_{i+1}$ for discrete and log-normal distributed transmission times, respectively. In Fig. \ref{fig5}, the range of $\rho$ is $[-1, 1)$. We observe that the zero-wait policy is optimal when $\rho\in[-1,-0.5]$, and the performance gap between the optimal policy and the zero-wait policy grows with $\rho$ when $\rho\geq-0.5$. This is in accordance with the example in the introduction: As $\rho$ grows, the occurrence of two consecutive zero transmission times (i.e., $(Y_i,Y_{i+1})=(0,0)$) increases. Therefore, more and more updates are wasted in the zero-wait policy, leading to a larger gap from the optimum. In Fig. \ref{fig6}, the range of $\rho$ is $[(e^{-1}-1)/(e-1), 1)$ and the sub-optimality gap of the zero-wait policy also increases with $\rho$.  The point $\rho=1$ is not plotted in these figures because the corresponding Markov chains are not ergodic.

Figure \ref{fig4} considers the average age vs. the parameter $\sigma$ of log-normal distributed transmission times, where $\rho = (e^{0.5}-1)/(e-1)$. We observe that the zero-wait policy is optimal for small $\sigma$ and is not optimal for large $\sigma$. When $\sigma=0$, the transmission times are constant, i.e., $Y_i=1$ for all $i$, and hence by Lemma \ref{just_in_time}, the zero-wait policy is optimal. For large $\sigma$, the time-average age of the zero-wait policy is significantly larger than the optimum. This implies that the sub-optimality gap of the zero-wait policy can be quite large for heavy-tail transmission time distributions. 
 
\ifreport
\ifjournal
\else
\begin{figure}
\centering \includegraphics[width=0.3\textwidth]{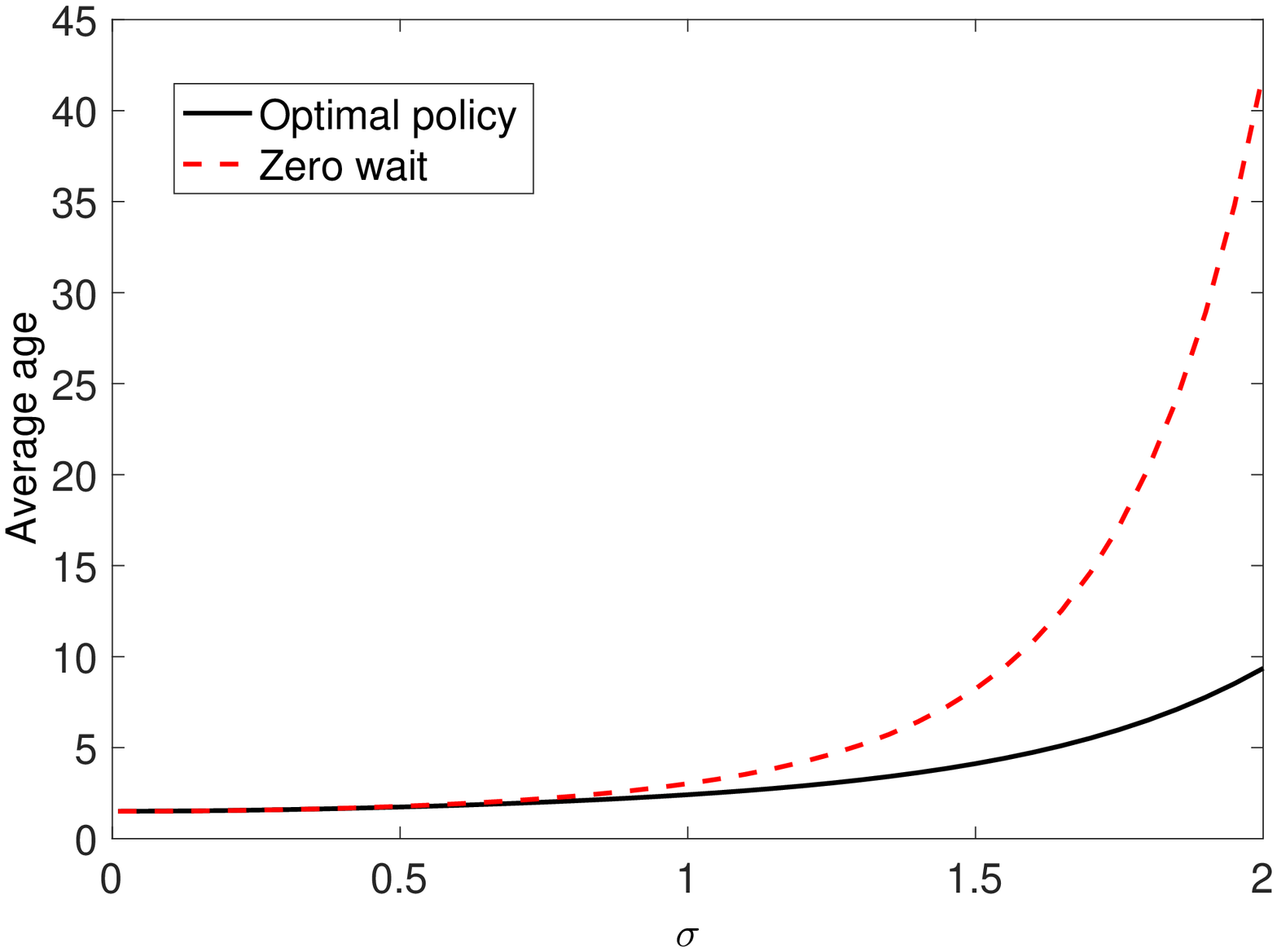}
\caption{Average age vs. the distribution parameter $\sigma$ of $i.i.d.$ log-normal distributed transmission times, where $\mathbb{E}[Y]\geq{1}/{f_{\max}}$.}
\label{fig4} \vspace{-0.cm}
\end{figure}
\begin{figure}
\centering \includegraphics[width=0.3\textwidth]{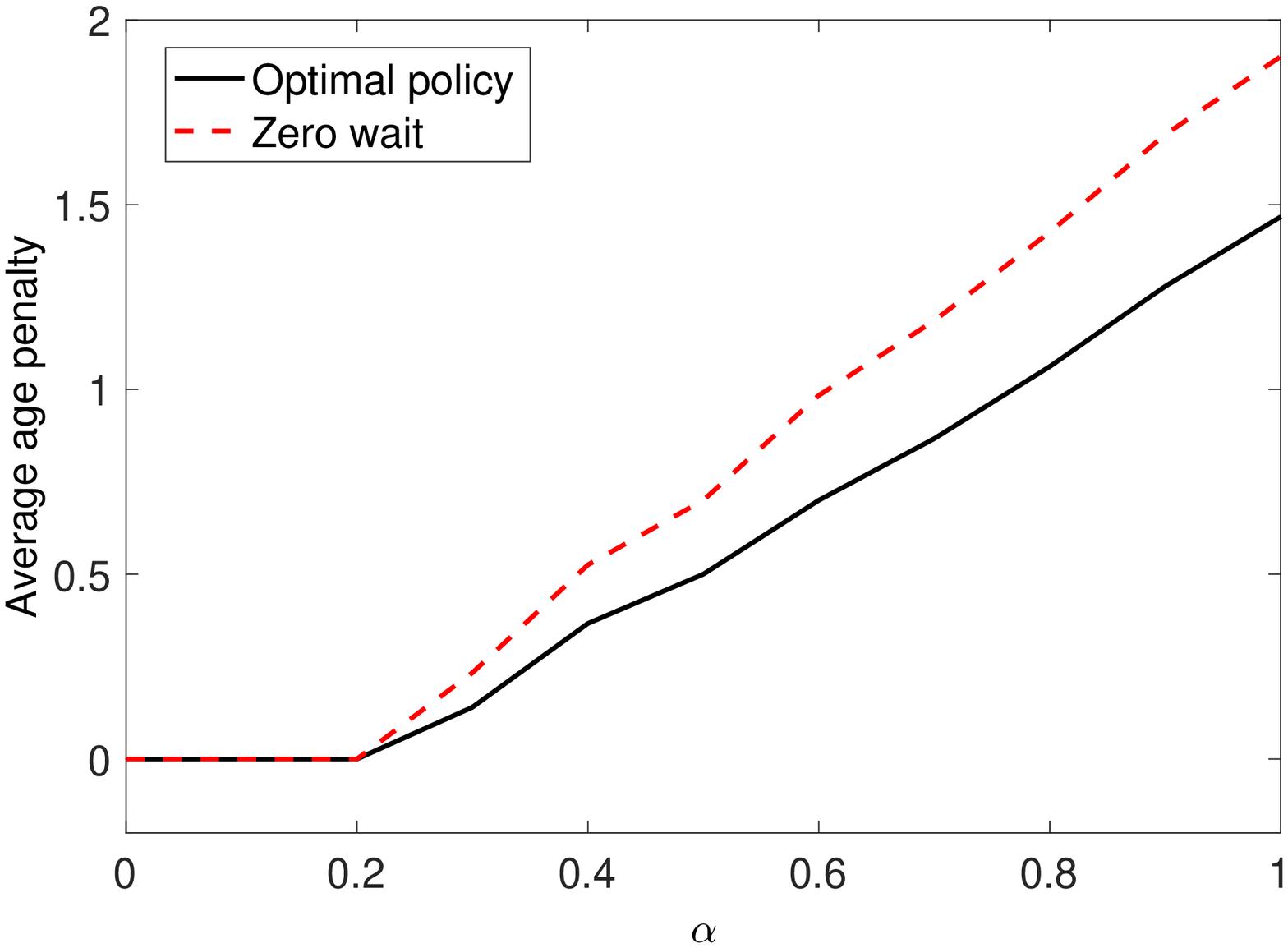} \caption{Average age penalty vs. the parameter $\alpha$ of stair-step penalty functions with discrete transmission times, where $\mathbb{E}[Y]\geq{1}/{f_{\max}}$, $g(\age)=\lfloor\alpha\age\rfloor$, and $\rho = 0.4$.}
\label{fig11} \vspace{-0.cm}
\end{figure}
\fi
\fi

\ifreport
\ifjournal
\begin{figure}
\centering \includegraphics[width=0.6\textwidth]{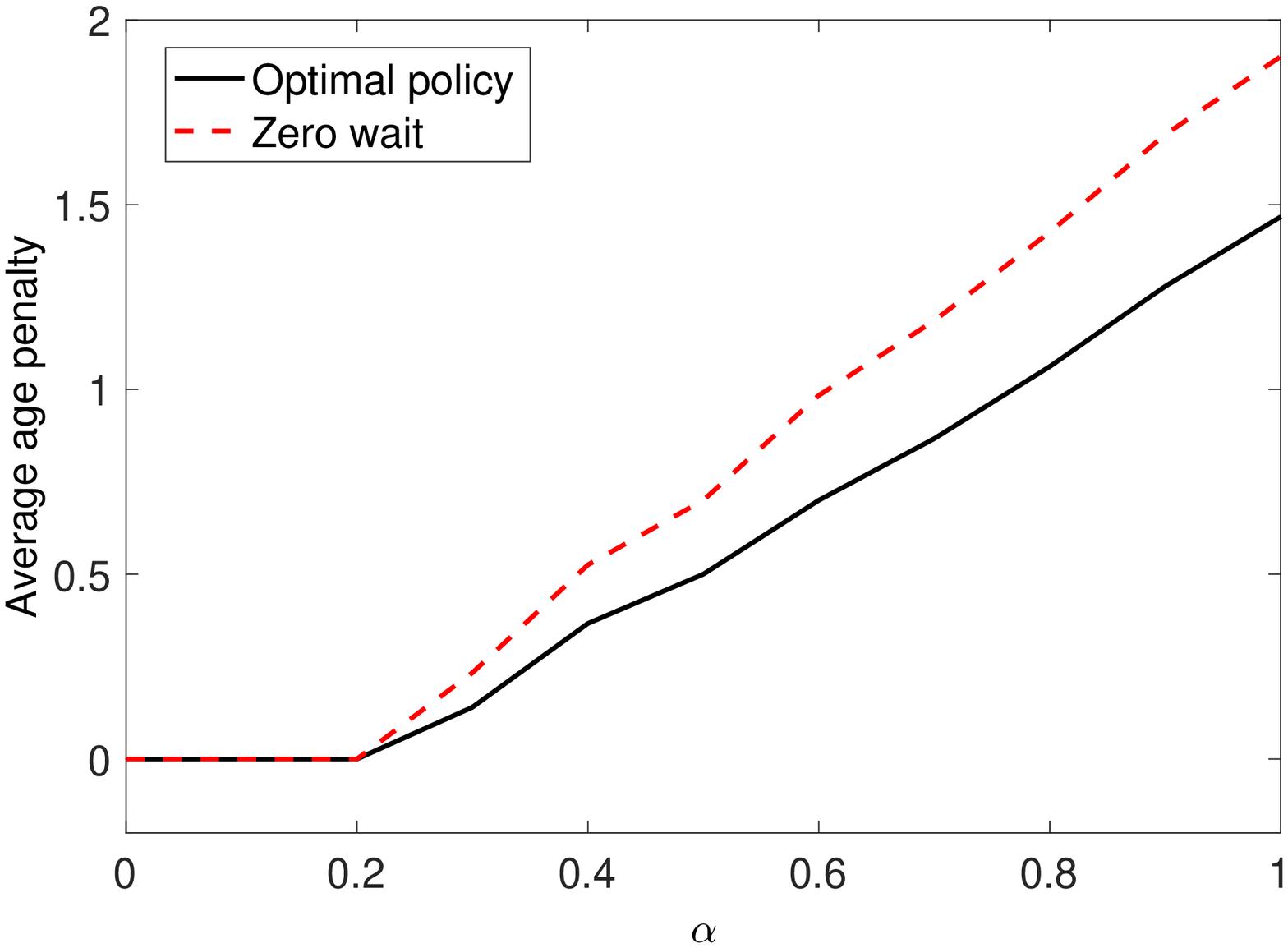} \caption{Average age penalty vs. the parameter $\alpha$ of stair-step penalty functions with discrete transmission times, where $\mathbb{E}[Y]\geq{1}/{f_{\max}}$, $g(\age)=\lfloor\alpha\age\rfloor$, and $\rho = 0.4$.}
\label{fig11} \vspace{-0.cm}
\end{figure}

\begin{figure}
\centering \includegraphics[width=0.6\textwidth]{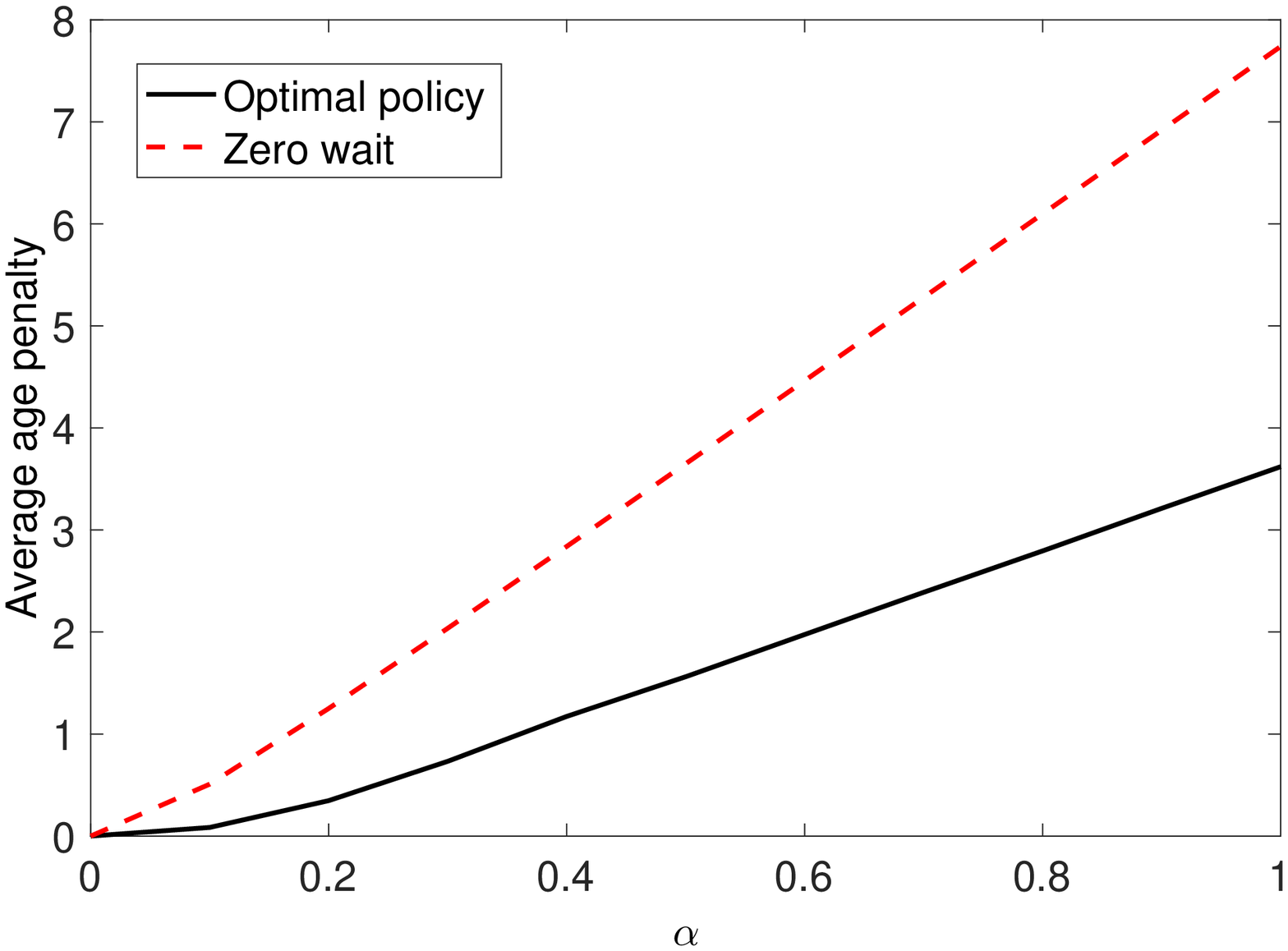} \caption{Average age penalty vs. the parameter $\alpha$ of stair-step penalty functions with log-normal distributed transmission times, where $\mathbb{E}[Y]\geq{1}/{f_{\max}}$, $g(\age)=\lfloor\alpha\age\rfloor$, and $\rho = (e^{0.5}-1)/(e-1)$.}
\label{fig12} \vspace{-0.cm}
\end{figure}

Figures \ref{fig11}-\ref{fig10} show the average age penalty vs. the parameter $\alpha$ of three types of age penalty functions, where the stair-step function $g(\age)=\lfloor\alpha\age\rfloor$ is considered in Fig. \ref{fig11} and \ref{fig12}, the power function $g(\age)=\age^\alpha$ is considered in Fig. \ref{fig7} and \ref{fig8}, and the exponential function $g(\age)=e^{\alpha\age}-1$ is considered in Fig. \ref{fig9} and \ref{fig10}. The correlation coefficient is $\rho = 0.4$ for discrete transmission times, and is $\rho = (e^{0.5}-1)/(e-1)$ for log-normal distributed transmission times. We find that the zero-wait policy is optimal if $\alpha =0$, in which case $g(\age)$ is a constant function. When $\alpha$ is large, the age penalty function grows quickly with respect to the age. In this case, the sub-optimality gap of the zero-wait policy can be quite large.
\else

Figures \ref{fig11}-\ref{fig10} show the average age penalty vs. the parameter $\alpha$ of three types of age penalty functions, where the stair-step function $g(\age)=\lfloor\alpha\age\rfloor$ is considered in Fig. \ref{fig11} and \ref{fig12}, the power function $g(\age)=\age^\alpha$ is considered in Fig. \ref{fig7} and \ref{fig8}, and the exponential function $g(\age)=e^{\alpha\age}-1$ is considered in Fig. \ref{fig9} and \ref{fig10}. The correlation coefficient is $\rho = 0.4$ for discrete transmission times, and is $\rho = (e^{0.5}-1)/(e-1)$ for log-normal distributed transmission times. We find that the zero-wait policy is optimal if $\alpha =0$, in which case $g(\age)$ is a constant function. When $\alpha>0$, the zero-wait policy may not be optimal.
\fi
\fi

These numerical results suggest that the conditions in Lemma \ref{just_in_time} are sufficient but not necessary. 

\ifreport
\ifjournal
\else
\begin{figure}
\centering \includegraphics[width=0.3\textwidth]{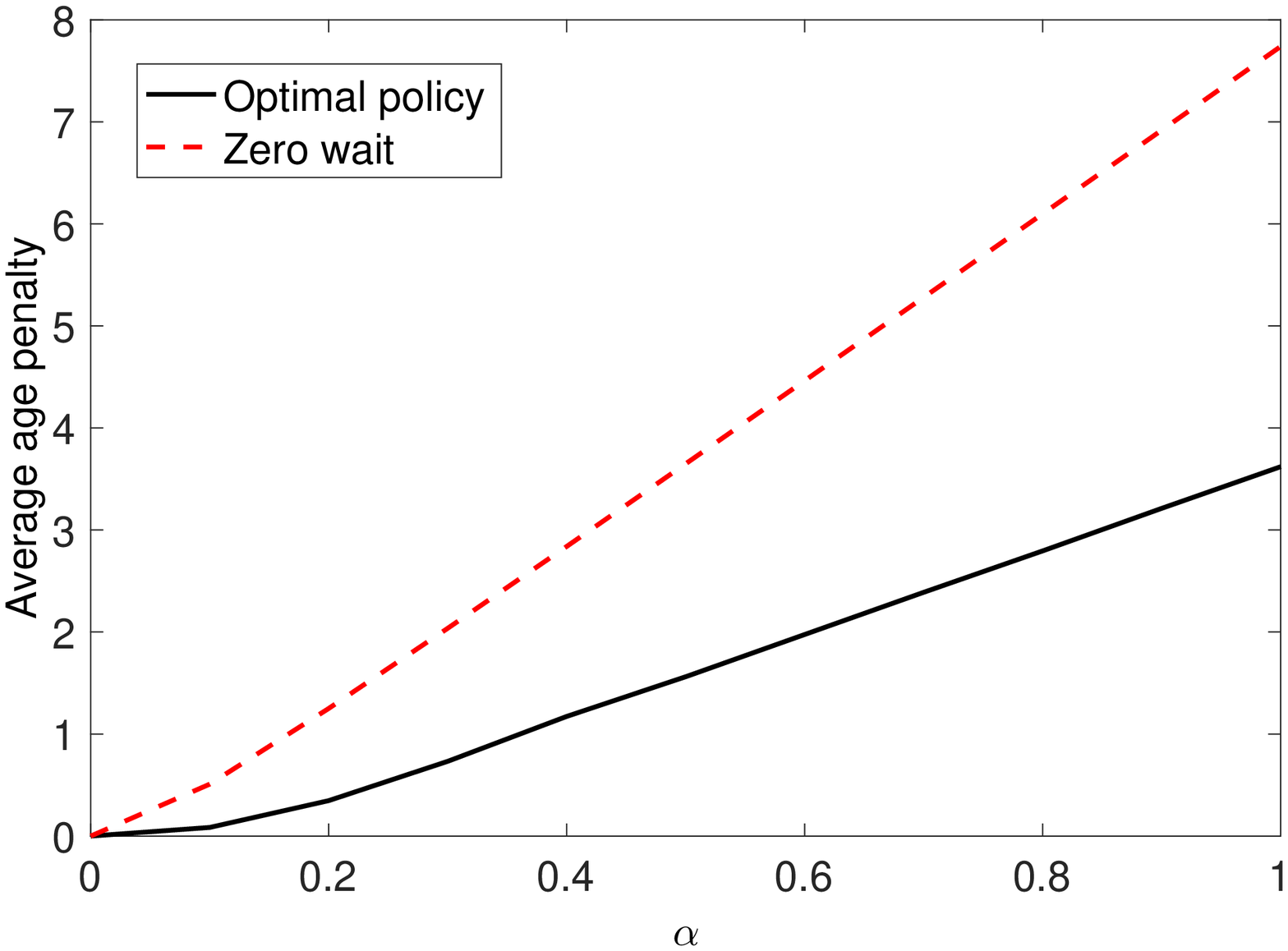} \caption{Average age penalty vs. the parameter $\alpha$ of stair-step penalty functions with log-normal distributed transmission times, where $\mathbb{E}[Y]\geq{1}/{f_{\max}}$, $g(\age)=\lfloor\alpha\age\rfloor$, and $\rho = (e^{0.5}-1)/(e-1)$.}
\label{fig12} \vspace{-0.cm}
\end{figure}
\begin{figure}
\centering \includegraphics[width=0.3\textwidth]{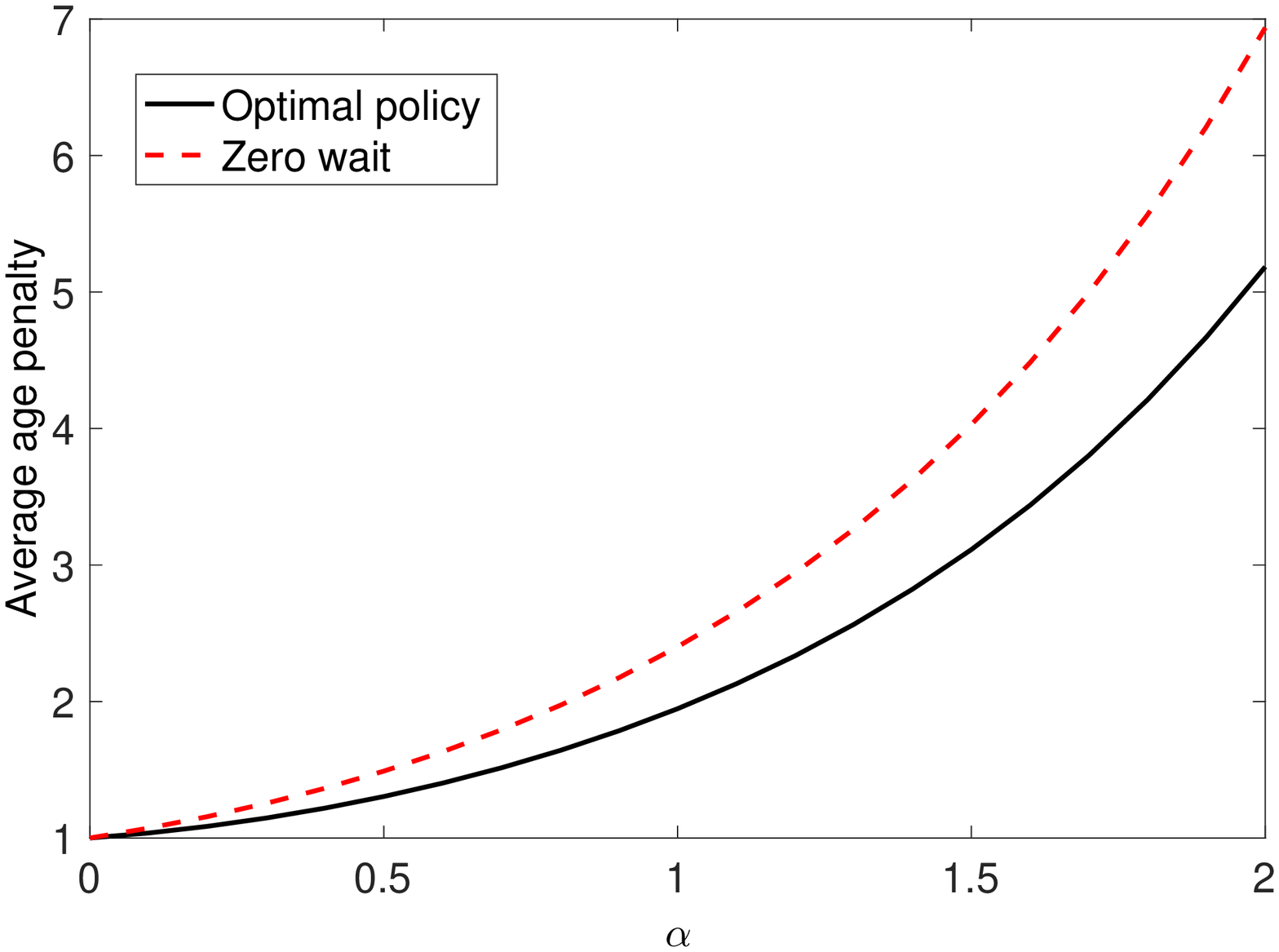} \caption{Average age penalty vs. the parameter $\alpha$ of  power penalty functions with discrete transmission times, where $\mathbb{E}[Y]\geq{1}/{f_{\max}}$, $g(\age)=\age^{\alpha}$, and $\rho = 0.4$.}
\label{fig7} \vspace{-0.cm}
\end{figure}
\fi
\fi

\ifreport
\ifjournal
\begin{figure}
\centering \includegraphics[width=0.6\textwidth]{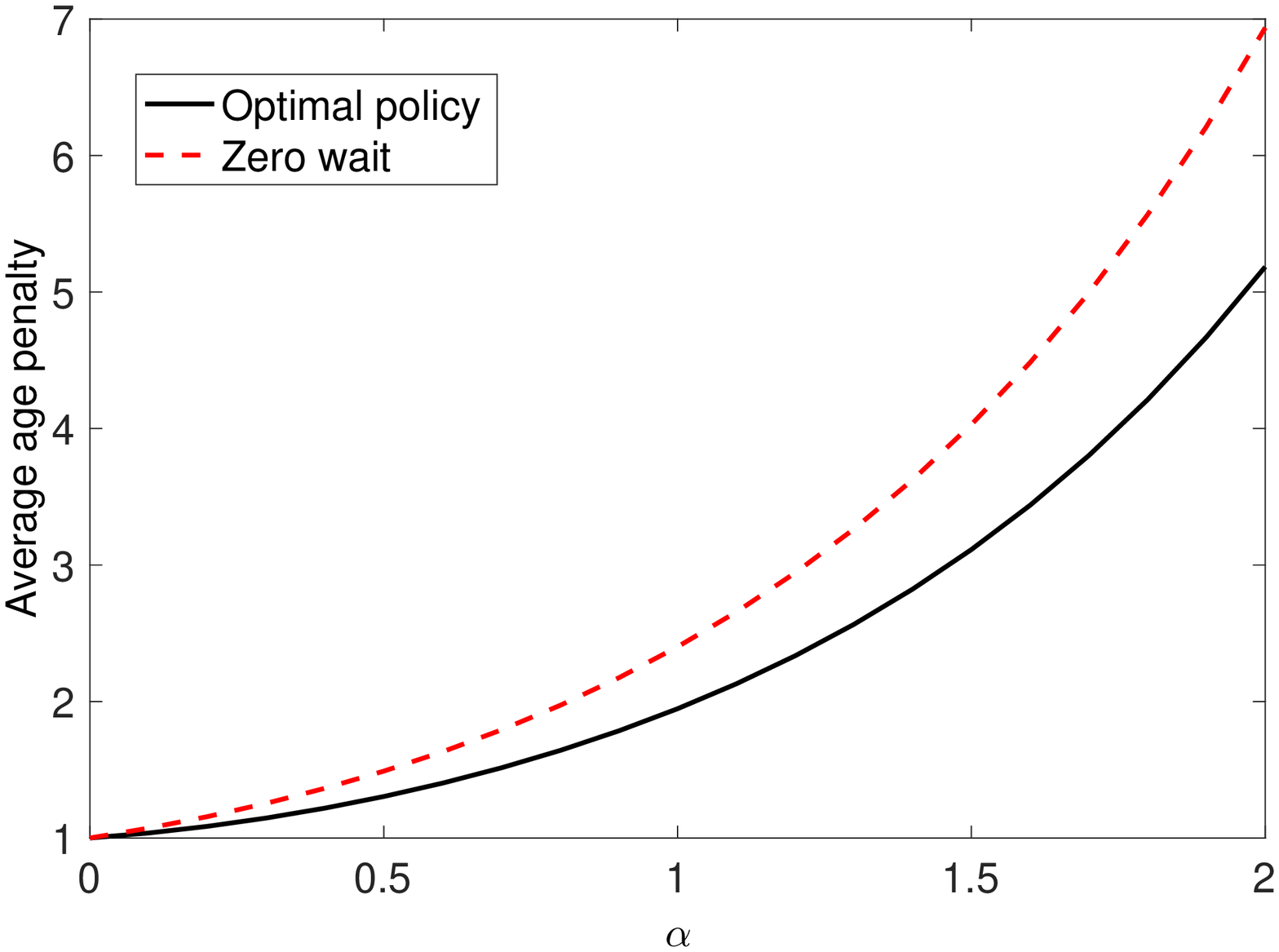} \caption{Average age penalty vs. the parameter $\alpha$ of  power penalty functions with discrete transmission times, where $\mathbb{E}[Y]\geq{1}/{f_{\max}}$, $g(\age)=\age^{\alpha}$, and $\rho = 0.4$.}
\label{fig7} \vspace{-0.cm}
\end{figure}
\begin{figure}
\centering \includegraphics[width=0.6\textwidth]{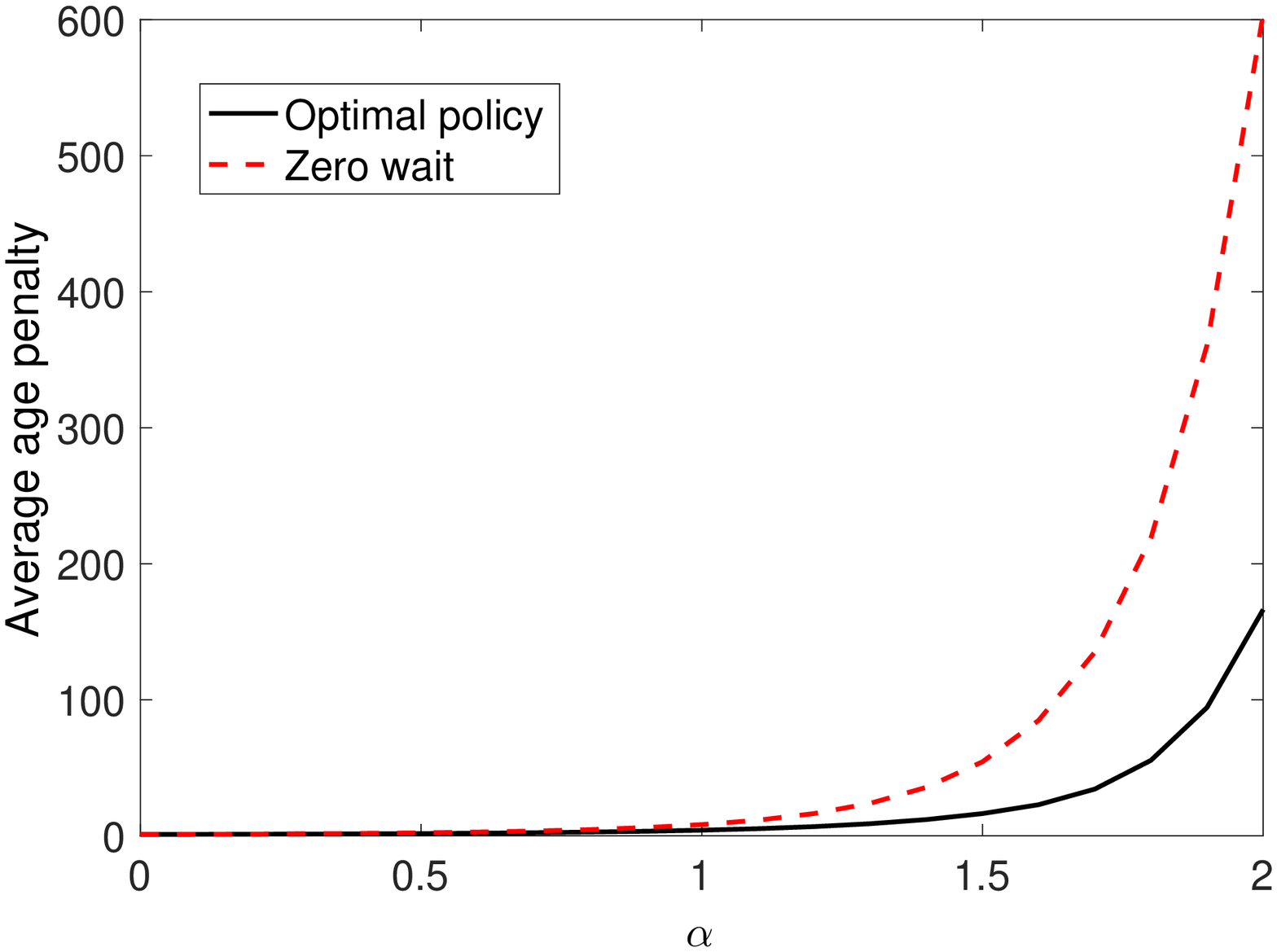} \caption{Average age penalty vs. the parameter $\alpha$ of  power penalty functions with log-normal distributed transmission times, where $\mathbb{E}[Y]\geq{1}/{f_{\max}}$, $g(\age)=\age^{\alpha}$, and $\rho = (e^{0.5}-1)/(e-1)$.}
\label{fig8} \vspace{0.1cm}
\end{figure}
\else
\fi

\fi


\ifjournal
\begin{figure}
\centering \includegraphics[width=0.6\textwidth]{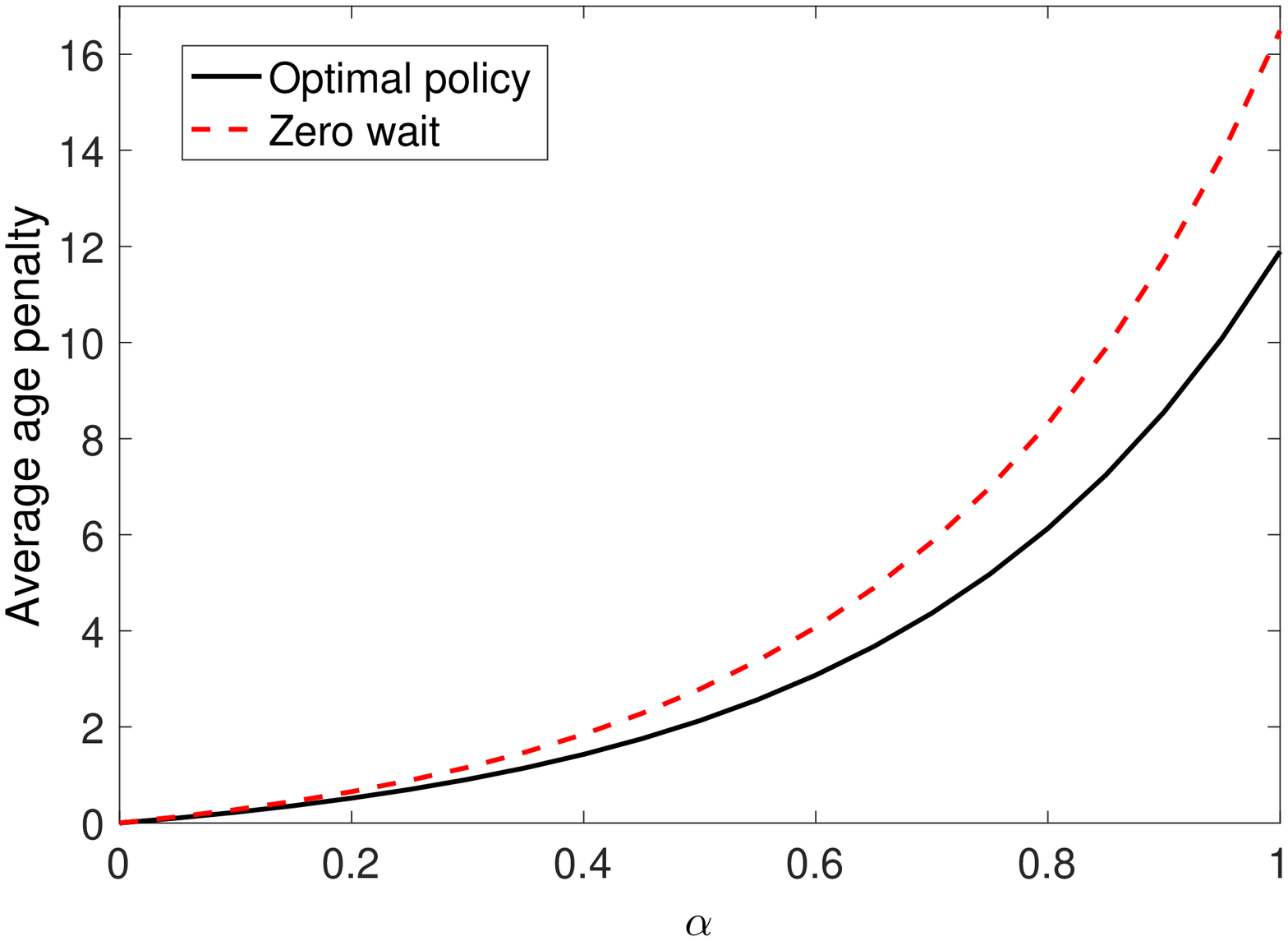} \caption{Average age penalty vs. the parameter $\alpha$ of exponential penalty functions with discrete transmission times, where $T_{\min}\leq\mathbb{E}[Y]$, $g(\age)=e^{\alpha\age}-1$, and $\rho = 0.4$.}
\label{fig9} \vspace{-0.cm}
\end{figure}

\begin{figure}
\centering \includegraphics[width=0.6\textwidth]{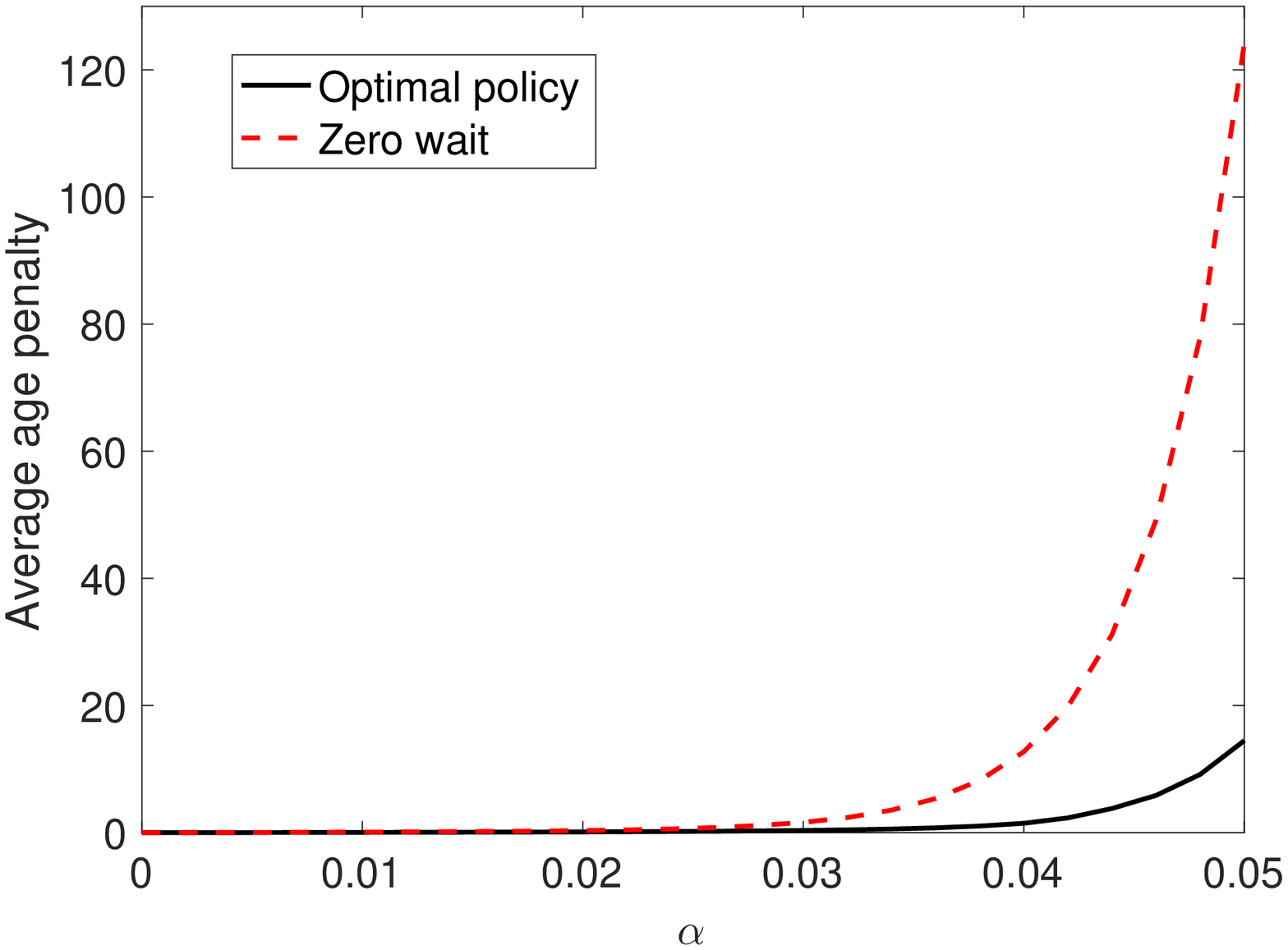} \caption{Average age penalty vs. the parameter $\alpha$ of exponential penalty functions with log-normal distributed transmission times, where $T_{\min}\leq\mathbb{E}[Y]$, $g(\age)=e^{\alpha\age}-1$, and $\rho = (e^{0.5}-1)/(e-1)$.}
\label{fig10} \vspace{-0.cm}
\end{figure}
\fi
\section{Conclusion}\label{conclusion}
We studied the optimal control of information updates sent from a source node to a remote destination via a communication server. We generalized prior study on the age-of-information by considering general age penalty functions and non-\emph{i.i.d.} transmission time processes. 
We developed efficient algorithms to find the optimal update policy for minimizing the average age penalty among all causal update policies. We showed that, surprisingly, the optimal policy is to wait for a certain amount of time before submitting a new update in many scenarios. Sufficient and necessary conditions were established to characterize  the optimality of the zero-wait policy. In particular, the zero-wait policy is far from the optimum if (i) the age penalty function grows quickly with respect to the age, (ii) the packet transmission times over the channel are positively correlated over time, or (iii) the packet transmission times are highly random (e.g., following a heavy-tail distribution). In our future work, we will further investigate how to improve the freshness of real-time signals transmitted over a channel. Some interesting initial result has been obtained in  \cite{SunISIT2017}.


\appendices
\section{Proof of Theorem \ref{lem_SRoptimal}}\label{app3}
\subsection{An Upper Bound of $\overline{g}_{\text{opt}}$}
By restricting $\Pi$ in Problem \eqref{eq_DPExpected} to $\Pi_{\text{SR}}$, we obtain the following problem:
\begin{align}\label{eq_DPExpected_upper}
\overline{g}_{\text{SR}}=&\min_{\pi\in\Pi_{\text{SR}}}~ \limsup_{n\rightarrow \infty}\frac{\mathbb{E}\left[\sum_{i=0}^{n-1} q(Y_i,Z_i,Y_{i+1})\right]}{\mathbb{E}[\sum_{i=0}^{n-1} (Y_{i}+Z_{i})]} \\
&~\text{s.t.}~~ \liminf_{n\rightarrow \infty} \frac{1}{n} \mathbb{E}\left[\sum_{i=0}^{n-1} (Y_{i}+Z_{i})\right]\geq \frac{1}{f_{\max}},\nonumber
\end{align}
where $\overline{g}_{\text{SR}}$ is the optimum objective value of Problem \eqref{eq_DPExpected_upper}. Since $\Pi_{\text{SR}} \subseteq \Pi$, we can obtain
\begin{align}\label{eq_time_average_exp0}
\overline{g}_{\text{SR}}\geq\overline{g}_{\text{opt}}.
\end{align}

\ifreport
\ifjournal
\else
\begin{figure}
\centering \includegraphics[width=0.3\textwidth]{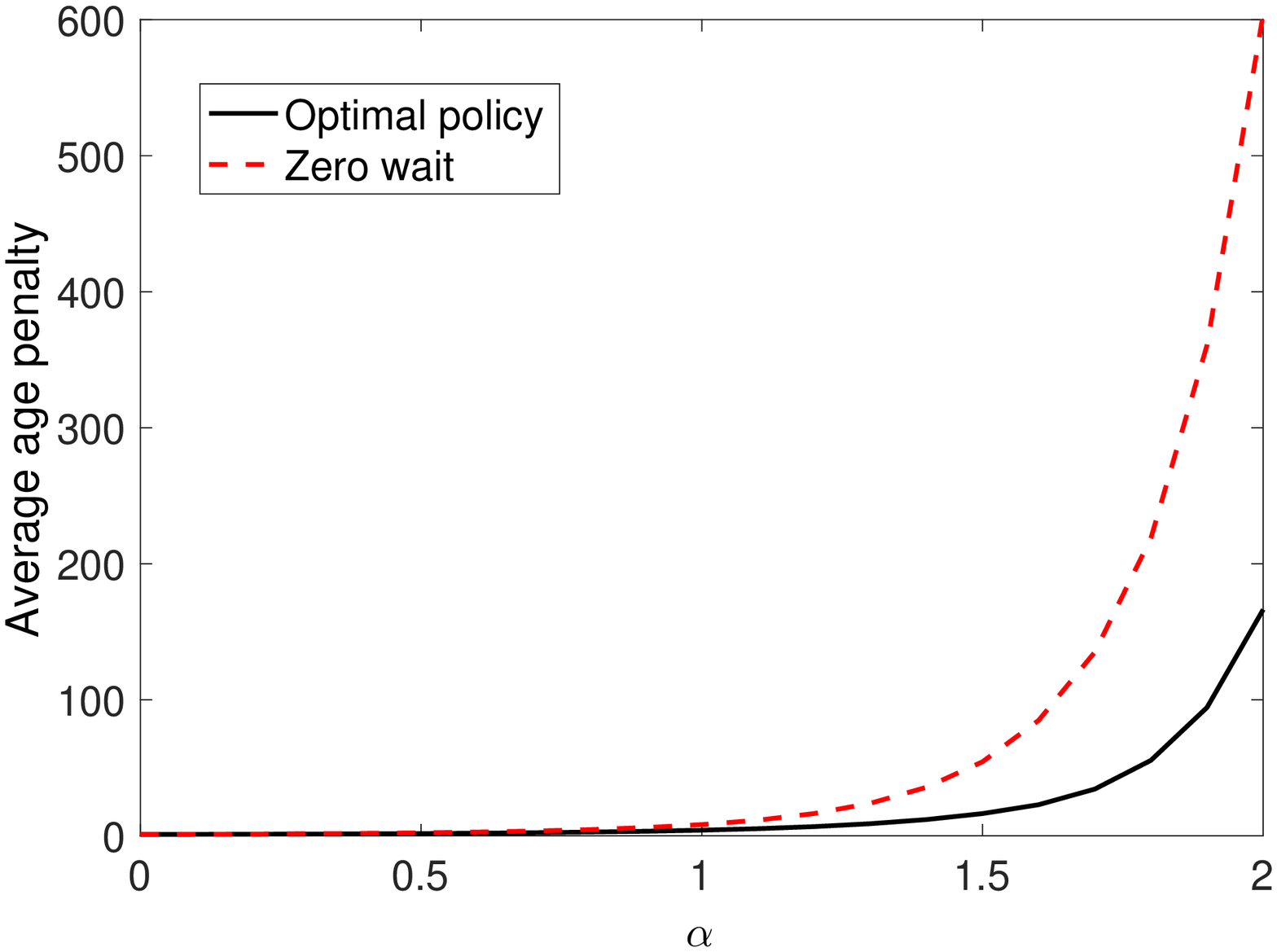} \caption{Average age penalty vs. the parameter $\alpha$ of  power penalty functions with log-normal distributed service times, where $\frac{1}{f_{\max}}\leq\mathbb{E}[Y]$, $g(\age)=\age^{\alpha}$, and $\rho = (e^{0.5}-1)/(e-1)$.}
\label{fig8} \vspace{0.1cm}
\end{figure}
\begin{figure}
\centering \includegraphics[width=0.3\textwidth]{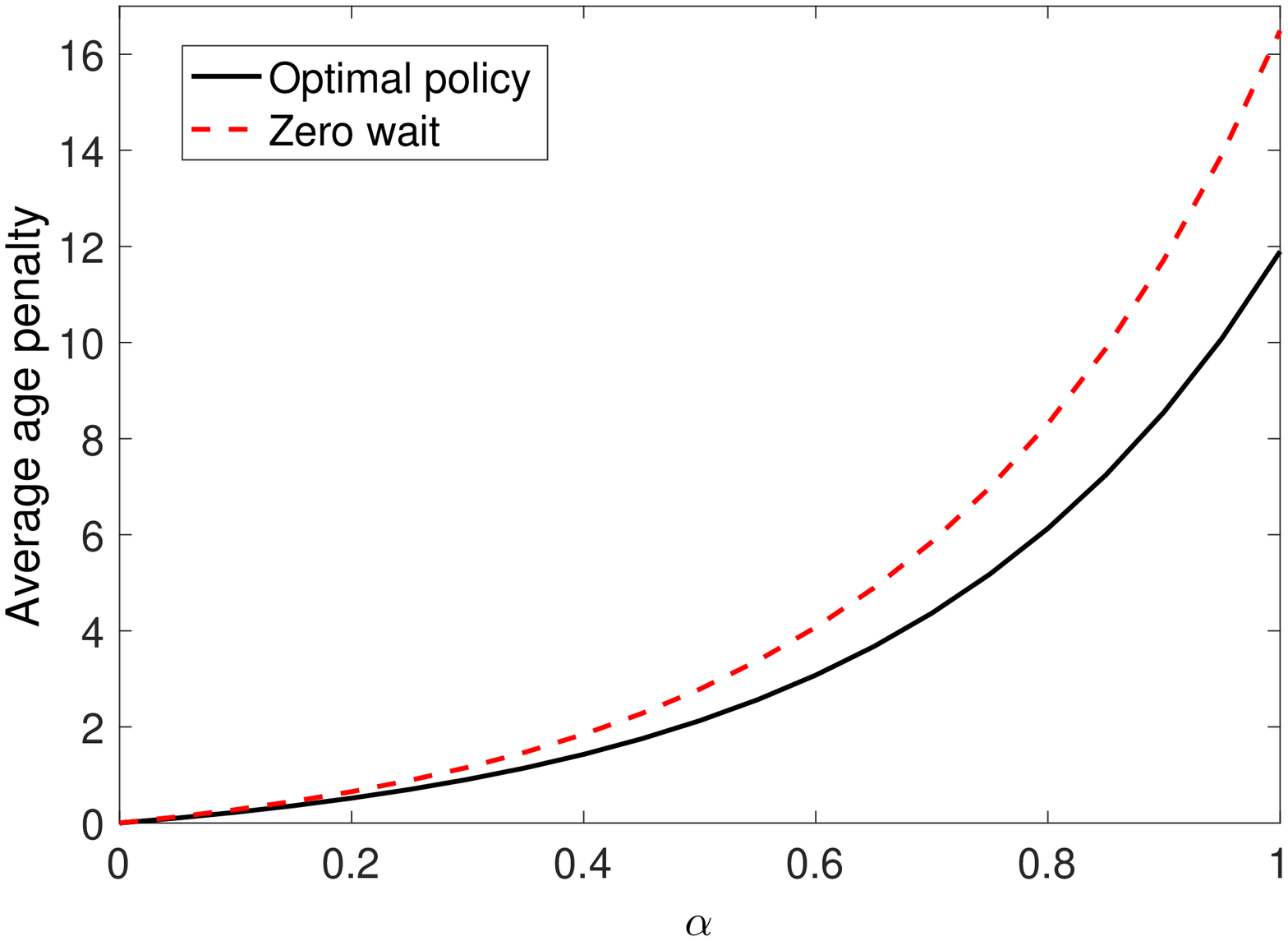} \caption{Average age penalty vs. the parameter $\alpha$ of exponential penalty functions with discrete service times, where $\frac{1}{f_{\max}}\leq\mathbb{E}[Y]$, $g(\age)=e^{\alpha\age}-1$, and $\rho = 0.4$.}
\label{fig9} \vspace{-0.cm}
\end{figure}

\fi
\fi

It is easy to show that the $(Y_i,Z_i,Y_{i+1})$'s are stationary and ergodic for all stationary randomized policies. This, together with the condition that $g(\cdot)$ is measurable, tells us that $q(Y_i,Z_i,Y_{i+1})$ is stationary and ergodic \cite[Theorems 7.1.1 and 7.1.3]{Durrettbook10}. For any stationary randomized policy $\pi=(Z_0,Z_1,\ldots)\in \Pi_{\text{SR}}$, we obtain
\begin{align}\label{eq_convergence1}
&\frac{1}{n}\mathbb{E}\left[\sum_{i=0}^{n-1} q(Y_i,Z_i,Y_{i+1})\right]=\mathbb{E}[q(Y_{0},Z_{0},Y_{1})],\\
&\frac{1}{n}\mathbb{E}\left[\sum_{i=0}^{n-1} (Y_{i}+Z_{i})\right]=\mathbb{E}[Y_0+Z_0].\label{eq_convergence2}
\end{align}
Hence, Problem \eqref{eq_DPExpected_upper} can be reformulated as Problem \eqref{eq_SR}.

\subsection{The Upper Bound of $\overline{g}_{\text{opt}}$ is Tight, i.e., $\overline{g}_{\text{SR}}=\overline{g}_{\text{opt}}$}
We will show $\overline{g}_{\text{SR}}=\overline{g}_{\text{opt}}$ in 4 steps. The following definitions are needed:
Since $\overline{g}_{\text{opt}}$ is finite, for each causally feasible policy $\pi=(Z_0,Z_1,\ldots)\in\Pi$ we can define ${a}_{n,\pi}$ and ${b}_{n,\pi}$ as in \eqref{eq_time_average_exp1} and \eqref{eq_time_average_exp2}, respectively.

Further, define $\Gamma_{\text{SR}}$ as the set of limit points of sequences $\big(({a}_{n,\pi_{}},{b}_{n,\pi}),n=1,2,\ldots\big)$ associated with all stationary randomized policies $\pi\in\Pi_{\text{SR}}$. 
Because the  reward $q(Y_i,Z_i,Y_{i+1})$ and  interval $Y_{i}+Z_i$ are stationary and ergodic for all stationary randomized policies $\pi\in\Pi_{\text{SR}}$, the sequence $({a}_{n,\pi},{b}_{n,\pi})$ has a unique limit point in the form of
\begin{align}\label{eq_form}
\left(\mathbb{E}[q(Y,Z,Y')]-\overline{g}_{\text{opt}}\mathbb{E}[Y+Z], \mathbb{E}[Y+Z]\right).
\end{align}
Hence, $\Gamma_{\text{SR}}$ is the set of all points $(\mathbb{E}[q(Y,Z,Y')]- $ $\overline{g}_{\text{opt}}\mathbb{E}[Y+Z], \mathbb{E}[Y+Z])$, where each point is associated with a conditional probability measure $p(y,A)=\Pr[Z\in A|Y=y]$, and the measure of $(Y,Y')$ is the same as that of $(Y_0,Y_1)$. Note that $\mathbb{E}[Y]=\mathbb{E}[Y']$.

\ifreport
\ifjournal
\else
\begin{figure}
\centering \includegraphics[width=0.3\textwidth]{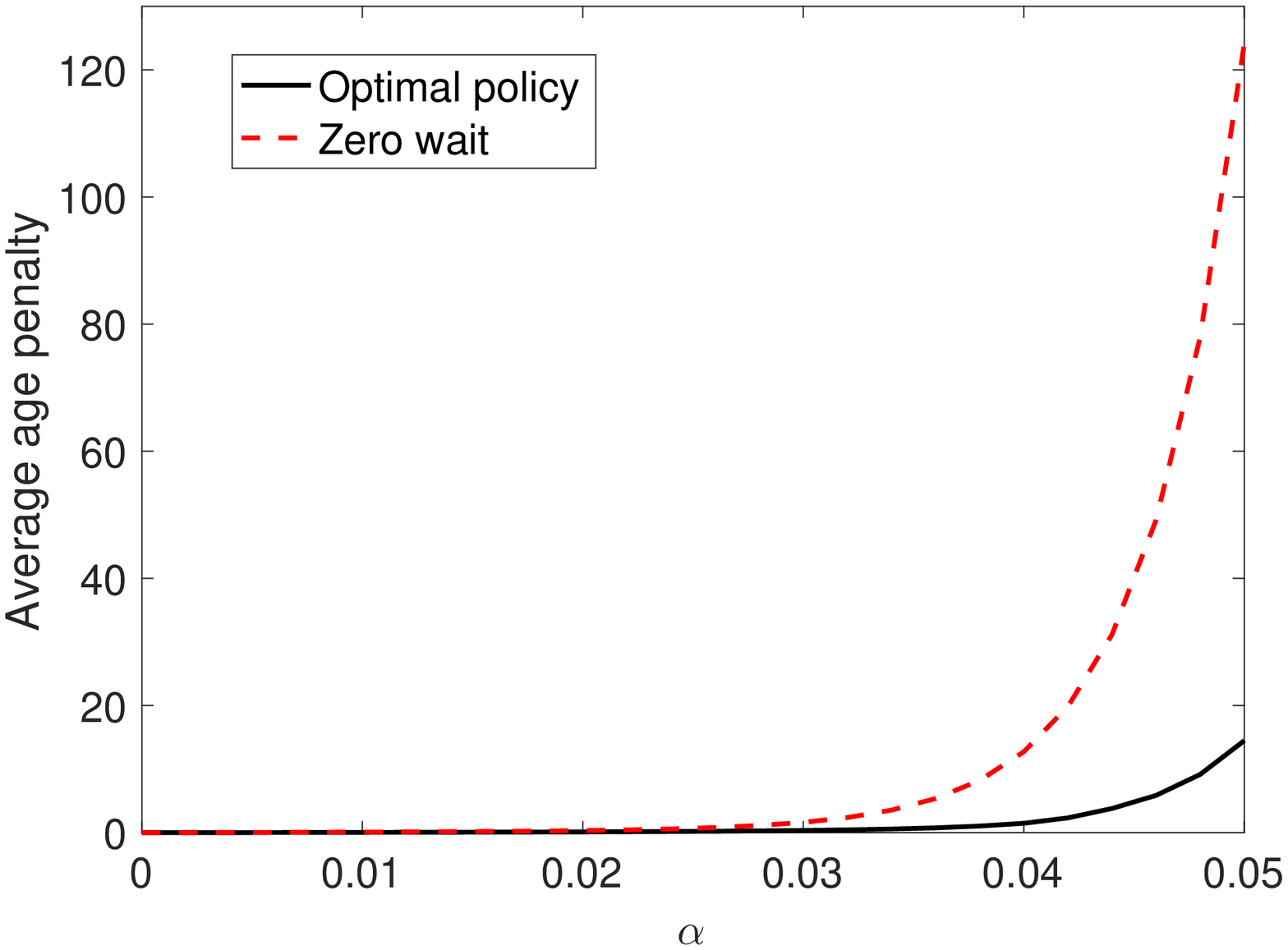} \caption{Average age penalty vs. the parameter $\alpha$ of exponential penalty functions with log-normal distributed service times, where $\frac{1}{f_{\max}}\leq\mathbb{E}[Y]$, $g(\age)=e^{\alpha\age}-1$, and $\rho = (e^{0.5}-1)/(e-1)$.}
\label{fig10} \vspace{-0.cm}
\end{figure}
\fi
\fi

\emph{Step 1: We will show that $\Gamma_{\text{SR}}$ is a convex and compact set}.

It is easy to show that $\Gamma_{\text{SR}}$ is convex by considering a stationary randomized policy that is a mixture of two stationary randomized policies.

For compactness, let $((d_j,e_j),j=1,2,\cdots)$ be any sequence of points in $\Gamma_{\text{SR}}$, we need to show that there is a convergent subsequence $(d_{j_k},e_{j_k})$ whose limit is also in $\Gamma_{\text{SR}}$.
Since $(d_j,e_j)\in\Gamma_{\text{SR}}$, there must exist $(Y,Z_{(j)},Y')$ with conditional probability $p_{j}(y,A)=\Pr[Z_{(j)}\in A|Y=y]$, such that $d_j=\mathbb{E}[q(Y,Z_{(j)},Y')]-\overline{g}_{\text{opt}}\mathbb{E}[Y+Z_{(j)}]$, $e_j=\mathbb{E}[Y+Z_{(j)}]$.
Let $\mu_j$ be the joint probability measure of $(Y,Z_{(j)},Y')$, then $(d_j,e_j)$ is uniquely determined by $\mu_j$.
For any $L$ satisfying $L\geq M$, we can obtain
\begin{align}
&\mu_j(Y\leq L, Z_{(j)}\leq L,Y'\leq L)\nonumber\\
=&\Pr(Y\leq L,Y'\leq L)\nonumber\\
\geq& \Pr(Y+Y'\leq L)\nonumber\\
\geq& 1- \frac{\mathbb{E}[Y+Y']}{L},~\forall~j,\nonumber
\end{align}
where the equality is due to the fact that $Z_{(j)}\leq M\leq L$ and the last inequality is due to Markov's inequality.
Therefore, for any $\epsilon$, there is an $L$ such that $$\liminf_{j\rightarrow\infty} \mu_j(|Y|\leq L, |Z_{(j)}|\leq L,|Y'|\leq L)\geq 1-\epsilon.$$ Hence, the sequence of measures $\mu_j$ is tight. By Helly's selection theorem \cite[Theorem 3.9.2]{Durrettbook10}, there is a subsequence of measures $\mu_{j_k}$ that converges weakly to a limit measure $\mu_\infty$.

Let $(Y,Z_{(\infty)},Y')$ and $p_\infty(y,A)=\Pr[Z_\infty\in A|Y=y]$ denote the random vector and conditional probability corresponding to the limit measure $\mu_\infty$, respectively. We can define $d_\infty=\mathbb{E}[q(Y,Z_{(\infty)},Y')]-\overline{g}_{\text{opt}}\mathbb{E}[Y+Z_{(\infty)}]$, $e_\infty=\mathbb{E}[Y+Z_{(\infty)}]$.
Since the function $q(y,z,y')$ is in the form of an integral, it is continuous and thus measurable. Using the continuous mapping theorem \cite[Theorem 3.2.4]{Durrettbook10}, we can obtain that $q(Y,Z_{(j_k)},Y')$ converges weakly to $q(Y,Z_{(\infty)},Y')$. Then, using the condition \eqref{eq_bound}, together with the dominated convergence theorem (Theorem 1.6.7 of \cite{Durrettbook10}) and Theorem 3.2.2 of \cite{Durrettbook10}, we can obtain
$\lim_{k\rightarrow\infty}(d_{j_k},e_{j_k})=(d_\infty,e_\infty)$. 
Hence, $((d_j,e_j),j=1,2,\cdots)$ has a convergent subsequence.
Further, we can generate a stationary randomized policy $\pi_{\infty,\text{SR}}$ by using the conditional probability $p_\infty(y,A)$ corresponding to $\mu_\infty$. Then, $(d_\infty,e_\infty)$ is the limit point generated by the stationary randomized policy $\pi_{\infty,\text{SR}}$, which implies
$(d_\infty,e_\infty)\in\Gamma_{\text{SR}}$. In summary, any sequence $(d_j,e_j)$ in $\Gamma_{\text{SR}}$ has a convergent subsequence $(d_{j_k},e_{j_k})$ whose limit $(d_\infty,e_\infty)$ is also in $\Gamma_{\text{SR}}$. Therefore, $\Gamma_{\text{SR}}$ is a compact set.

\emph{Step 2:} {\emph{We will show that there exists an optimal policy $\pi_{\text{opt}}\in\Pi$ of Problem \eqref{eq_DPExpected} such that the sequence $({a}_{n,\pi_\text{opt}},{b}_{n,\pi_\text{opt}})$ associated with policy $\pi_{\text{opt}}$ has at least one limit point in $\Gamma_{\text{SR}}$}.

Since the sequence $(Y_0,Y_1,\ldots)$ is a Markov chain, the observation $Y_{i+1}$ depends only on the immediately preceding state $Y_i$ and not on the history state and control $Y_0,\ldots, Y_{i-1}$, $Z_0,\ldots, Z_{i-1}$. Therefore, $Y_i$ is the \emph{sufficient statistic} \cite[p. 252]{Bertsekas2005bookDPVol1} for solving Problem \eqref{eq_DPExpected}. This tells us that there exists an optimal policy $\pi_{\text{opt}}=(Z_0,Z_1,\ldots)\in\Pi$ of Problem \eqref{eq_DPExpected} in which the control action $Z_i$ is determined based on only $Y_i$, but not the history state and control $Y_0,\ldots, Y_{i-1}$, $Z_0,\ldots, Z_{i-1}$ \cite{Bertsekas2005bookDPVol1}. We will show that the sequence $({a}_{n,\pi_\text{opt}},{b}_{n,\pi_\text{opt}})$ associated with this policy $\pi_{\text{opt}}$ has at least one limit point in $\Gamma_{\text{SR}}$.}

It is known that $Z_{i}$ takes values in the standard Borel space $(\mathbb{R},\mathcal{{R}})$, where $\mathcal{{R}}$ is the Borel $\sigma$-field. According to \cite[Thoerem 5.1.9]{Durrettbook10}, for each ${i}$ there exists a conditional probability measure $p'_i(y,A)$ such that $p'_i(y,A)=\Pr(Z_{i}\in A|Y_i=y)$ for almost all $y$. That is, the control action $Z_{i}$ is determined based on $Y_i$ and the  conditional probability measure $p'_i(y,A)=\Pr(Z_{i}\in A|Y_i=y)$.
One can use this conditional probability $p'_i(y,A)$ to generate a stationary randomized policy $\pi'_{i,\text{SR}}\in \Pi_{\text{SR}}$. Then, the one-stage expectation $(\mathbb{E}[q(Y_i,Z_i,Y_{i+1})] - \overline{g}_{\text{opt}}\mathbb{E}[Y_{i}+Z_{i}], \mathbb{E}[Y_{i}+Z_{i}])$ is exactly the limit point generated by the stationary randomized policy $\pi'_{i,\text{SR}}$. Thus, $(\mathbb{E}[q(Y_i,Z_i,Y_{i+1})] - \overline{g}_{\text{opt}}\mathbb{E}[Y_{i}+Z_{i}], \mathbb{E}[Y_{i}+Z_{i}])\in \Gamma_{\text{SR}}$ for all $i=0,1,2,\ldots$
Using \eqref{eq_time_average_exp1}, \eqref{eq_time_average_exp2}, and the fact that $\Gamma_{\text{SR}}$ is convex, we can obtain $({a}_{n,\pi_\text{opt}},{b}_{n,\pi_\text{opt}})\in \!\Gamma_{\text{SR}}$ for all $n=1,2,3\ldots$ In other words, the sequence $({a}_{n,\pi_\text{opt}},{b}_{n,\pi_\text{opt}})$ is within $\Gamma_{\text{SR}}$.
Since $\Gamma_{\text{SR}}$ is a compact set, the sequence $({a}_{n,\pi_\text{opt}},{b}_{n,\pi_\text{opt}})$ must have a convergent subsequence, whose limit is in $\Gamma_{\text{SR}}$.


\emph{Step 3: Let $(a^*,b^*)\in\Gamma_{\text{SR}}$ be one limit point of the sequence $(a_{n,\pi_\text{opt}},b_{n,\pi_\text{opt}})$ associated with policy $\pi_{\text{opt}}$. We will show that $a^*\leq0$ and $b^*\geq \frac{1}{f_{\max}}$.}


Policy $\pi_{\text{opt}}$ is feasible for Problem \eqref{eq_DPExpected} and meanwhile achieves the optimum objective value $\overline{g}_{\text{opt}}$.
Hence,
\begin{align}\label{eq_feasible_cond1}
&\limsup_{n\rightarrow\infty}\frac{c_{n,\pi_\text{opt}}}{b_{n,\pi_\text{opt}}} = \overline{g}_{\text{opt}},\\
&\liminf_{n\rightarrow\infty}b_{n,\pi_\text{opt}}\geq \frac{1}{f_{\max}},\label{eq_feasible_cond2}
\end{align}
where
\begin{align}
c_{n,\pi_\text{opt}}\triangleq\frac{1}{n}\mathbb{E}\bigg[\sum_{i=0}^{n-1} q(Y_i,Z_i,Y_{i+1})\bigg].\nonumber
\end{align}
By \eqref{eq_time_average_exp2}, $b_{n,\pi_\text{opt}}$ is upper bounded by
\begin{align}
b_{n,\pi_\text{opt}}\leq M+\mathbb{E}[Y]<\infty.\nonumber
\end{align}
Hence, by \eqref{eq_time_average_exp1}, we have
\begin{align}
a_{n,\pi_\text{opt}} &= c_{n,\pi_\text{opt}} - \overline{g}_{\text{opt}} b_{n,\pi_\text{opt}}\nonumber\\
&\leq \max\{c_{n,\pi_\text{opt}} - \overline{g}_{\text{opt}} b_{n,\pi_\text{opt}},0\}\nonumber\\
&= \max\{\frac{c_{n,\pi_\text{opt}}}{b_{n,\pi_\text{opt}}} - \overline{g}_{\text{opt}} ,0\} b_{n,\pi_\text{opt}}\nonumber\\
&\leq \max\{\frac{c_{n,\pi_\text{opt}}}{b_{n,\pi_\text{opt}}} - \overline{g}_{\text{opt}} ,0\} (M+\mathbb{E}[Y]).\nonumber
\end{align}
Taking the $\limsup$ in this inequality and using \eqref{eq_feasible_cond1}, yields
\begin{align}
\limsup_{n\rightarrow\infty} a_{n,\pi_\text{opt}}\leq 0.
\end{align}
Because $(a^*,b^*)$ is one limit point of $(a_{n,\pi_\text{opt}},b_{n,\pi_\text{opt}})$, we have
\begin{align}\label{eq_feasible_cond3}
a^*\leq \limsup_{n\rightarrow\infty} a_{n,\pi_\text{opt}}, b^*\geq \liminf_{n\rightarrow\infty}b_{n,\pi_\text{opt}}.
\end{align}
By \eqref{eq_feasible_cond2}-\eqref{eq_feasible_cond3}, we have  $a^*\leq0$ and $b^*\geq \frac{1}{f_{\max}}$.


\emph{Step 4: We will show that there exists a stationary randomized policy that is optimal for Problems \eqref{eq_DPExpected} and \eqref{eq_SR}, and thus $\overline{g}_{\text{SR}}=\overline{g}_{\text{opt}}$.}
By the definition of ${\Gamma}_{\text{SR}}$, $(a^*,b^*)\in{\Gamma}_{\text{SR}}$ must be the limit point generated by a stationary randomized policy $\pi^*\in \Pi_{\text{SR}}$. Let $(Y,Z^*,Y')$ be a random vector with the stationary distribution of policy $\pi^*$. 
Then, \eqref{eq_form} implies
\begin{align}
(a^*,b^*)=\left(\mathbb{E}[q(Y,Z_{}^*,Y')]-\overline{g}_{\text{opt}}\mathbb{E}[Y+Z^*], \mathbb{E}[Y+Z^*]\right).\nonumber
\end{align}
Using $a^*\leq0$ and $b^*\geq \frac{1}{f_{\max}}$, we can obtain
\begin{align}\label{eq_optimality}
&\mathbb{E}[q(Y,Z_{}^*,Y')]-\mathbb{E}[Y+Z^*]\overline{g}_{\text{opt}}\leq0,\\
&\mathbb{E}[Y+Z^*]\geq \frac{1}{f_{\max}}.\label{eq_feasible_condition5}
\end{align}
By \eqref{eq_optimality} and $\mathbb{E}[Y+Z^*]>0$, we have
\begin{align}
&\frac{\mathbb{E}[q(Y,Z_{}^*,Y')]}{\mathbb{E}[Y+Z^*]}  \leq \overline{g}_{\text{opt}}.\nonumber
\end{align}
Further, the inequality \eqref{eq_feasible_condition5} suggests that the stationary randomized policy $\pi^*$ is feasible for Problem \eqref{eq_SR}. Hence,
\begin{align}
\frac{\mathbb{E}[q(Y,Z_{}^*,Y')]}{\mathbb{E}[Y+Z^*]}\geq \overline{g}_{\text{SR}}.\nonumber
\end{align}
Therefore, $\overline{g}_{\text{SR}}\leq\overline{g}_{\text{opt}}$. This and \eqref{eq_time_average_exp0} suggest that
\begin{align}
\frac{\mathbb{E}[q(Y,Z_{}^*,Y')]}{\mathbb{E}[Y+Z^*]}=\overline{g}_{\text{SR}}=\overline{g}_{\text{opt}}. \nonumber
\end{align}
This completes the proof.

\ifreport
\section{Proof of Theorem \ref{lem_SD}}\label{app4}
Consider an arbitrarily chosen stationary randomized policy $\pi_1\in \Pi_{\text{SR}}$ that is feasible for Problem \eqref{eq_SR}. We will show that there exists a feasible stationary deterministic policy that is no worse than policy $\pi_1$.

For any $y$, we can use the conditional probability $p(y,A)$ associated with policy $\pi_1$ to compute the conditional expectation $\mathbb{E}[Z|Y=y]$ by
\begin{align}
\mathbb{E}[Z|Y=y] = \int_0^M z p(y,d z).\nonumber
\end{align}
Since the conditional expectation $\mathbb{E}[Z|Y]$ is unique w.p.1 \cite[Section 5.1]{Durrettbook10}, there is a deterministic function $z(\cdot)$ such that $z(y)=\mathbb{E}[Z|Y=y]$ w.p.1.
Consider the set $\Lambda\subset\Pi_{\text{SR}}$ of all stationary randomized policies that satisfy $\mathbb{E}[Z|Y=y] = z(y)$ w.p.1. Then, the stationary randomized policy $\pi_1$ is in $\Lambda$. It is also easy to show that the stationary deterministic policy $(Z_i=z(Y_i),i=1,2,\ldots)$ is also in $\Lambda$.

Using the iterated expectation, for any policy in $\Lambda$
\begin{align}\label{eq_constraint4}
\mathbb{E}[Y+Z] = \mathbb{E}\big[Y+ \mathbb{E}[Z|Y] \big]=\mathbb{E}\big[Y+ z(Y) \big].
\end{align}
Because $\pi_1\in \Lambda$ is feasible for Problem \eqref{eq_SR}, any policy in $\Lambda$ is feasible for Problem \eqref{eq_SR}.

Since $q(y,z,y')$ is the integral of a non-decreasing function $g$, it is easy to show that the function
$q(y,\cdot,y')$ is convex.
For any policy $\pi\in\Lambda$, Jensen's inequality tells us that
\begin{align}
&~~~\mathbb{E}[q(Y,Z,Y')|Y,Y'] \nonumber\\
& \geq q(Y,\mathbb{E}[Z|Y,Y'],Y') \nonumber\\
& = q(Y,\mathbb{E}[Z|Y],Y') \label{eq_cond_exp}\\
& = q(Y,z(Y),Y'), ~(\text{w.p.1}),\nonumber
\end{align}
where \eqref{eq_cond_exp} is due to the fact that $Z$ is determined based on $Y$, but not $Y'$.
Taking the expectation over $(Y,Y')$, yields
\begin{align}
\mathbb{E}[q(Y,z(Y),Y')]\leq \mathbb{E}[q(Y,Z,Y')] \nonumber
\end{align}
for any policy $\pi\in\Lambda$, where equality holds if $Z=z(Y)$. This and \eqref{eq_constraint4} suggest that the stationary deterministic policy $(Z_i=z(Y_i),i=1,2,\ldots)$ achieves the smallest objective value for Problem \eqref{eq_SR} among all policies in $\Lambda$. In conclusion, for any feasible stationary randomized policy $\pi_1\in \Pi_{\text{SR}}$, we can find a feasible stationary deterministic policy that is no worse than policy $\pi_1$. This completes the proof.

\section{Proof of Lemma \ref{lem_quasiconvex}}\label{app5_0}
We need the following lemma:
\begin{lemma}\label{lem_expect_convex}
If $l: \mathbb{R} \rightarrow \mathbb{R}$ is a convex function, then the functional $w: L^2(\mu_Y) \rightarrow \mathbb{R}$ defined by
\begin{align}
w(z) = \int_0^\infty l(z(y)) d\mu_Y(y)
\end{align}
is also convex.
\end{lemma}
\begin{IEEEproof}
For any $\lambda\in[0,1]$ and $z_1,z_2\in L^2(\mu_Y)$, we have
\begin{align}
&w(\lambda z_1+(1-\lambda) z_2) \nonumber\\
=&\int_0^\infty l(\lambda z_1(y)+(1-\lambda) z_2(y) ) d\mu_Y(y)\nonumber\\
\leq &\int_0^\infty \left[\lambda l( z_1(y))+(1-\lambda) l(z_2(y))\right] d\mu_Y(y)\nonumber\\
=&\lambda w(z_1)+(1-\lambda) w(z_2).
\end{align}
By this, $w(z)$ is convex.
\end{IEEEproof}
We now prove Lemma \ref{lem_quasiconvex}.
Since $q(y,z,y')$ is the integral of a non-negative and non-decreasing function $g$, it is easy to show that the function
$q(y,\cdot,y')$ is non-negative and convex. Hence, the conditional expectation
$\mathbb{E}\left[q(y,\cdot,Y')|Y=y\right]$ is also convex.
We can obtain
\begin{align}
&\mathbb{E}\left[q(Y,z(Y),Y')\right]\nonumber\\
=&\int_0^\infty\mathbb{E}\left[q(y,z(y),Y')|Y=y\right]d\mu_Y(y).
\end{align}
According to Lemma \ref{lem_expect_convex}, $\mathbb{E}\left[q(Y,z(Y),Y')\right]$ is a convex functional of $z$ and
$\mathbb{E}[Y+z(Y)]$ is an affine functional of $z$. It is known that the ratio of a non-negative convex functional and positive affine functional is quasi-convex \cite[p. 103]{Boyd04}. Hence, $h(z)$ is quasi-convex, which completes the proof.

\section{Proof of Theorem \ref{lem_SD_solution}}\label{app5_1}
We use the Lagrangian duality approach to solve Problem \eqref{eq_SD_equavilent}.
The Lagrangian of Problem \eqref{eq_SD_equavilent} is
\begin{align}
&L(z,\zeta,\gamma,\tau)\nonumber\\ =& \int_0^\infty \mathbb{E}\left[q(y,z(y),Y')|Y=y\right] d\mu_Y(y) \nonumber\\
& -c\int_0^\infty [y+z(y)] d\mu_Y(y) \nonumber\\
 &+ \zeta\left[\frac{1}{f_{\max}}\!-\!\int_0^\infty [y+z(y)] d\mu_Y(y)\!\right] \nonumber\\
 &- \int_0^\infty \gamma(y)z(y) d\mu_Y(y)
 +\int_0^\infty \tau(y)(z(y)-M) d\mu_Y(y)\nonumber\\
 =& \int_0^\infty \bigg\{\mathbb{E}\left[q(y,z(y),Y')|Y=y\right] - (c +\zeta)[y+z(y)]\nonumber\\
 & - \gamma(y)z(y)
 +\tau(y)\Big[z(y)-M\Big]\bigg\} d\mu_Y(y) + \zeta \frac{1}{f_{\max}}.\label{eq_Lagrange}
\end{align}
Since Problem \eqref{eq_SD_equavilent} is feasible, all constraints are affine, the refined Slater's condition \cite[Sec. 5.2.3]{Boyd04} is satisfied. According to \cite[Proposition 3.3.2]{infinite_dimensional} and \cite[pp. 70-72]{convex_analysis}, the Karush-Kuhn-Tucker (KKT) theorem remains valid for the Lebesgue space $L^2(\mu_Y)$. Hence, if a vector $(z,\zeta,\gamma,\tau)$ satisfies the KKT conditions \eqref{eq1_KKT1}-\eqref{eq1_KKTend}, it is an optimal solution to \eqref{eq_SD_equavilent}. The KKT conditions are given by
\begin{align}
&z = \min_{x\in L^2(\mu_Y)} L(x,\zeta,\gamma,\tau),\label{eq1_KKT1}\\
&\zeta\geq0, \int_0^\infty [y+z(y)] d\mu_Y(y)\geq \frac{1}{f_{\max}},\label{eq1_KKT5}\\
& \gamma(y)\geq0, z(y)\geq0, \forall~y\geq0, \label{eq1_KKT6}\\
& \tau(y)\geq0, z(y)\leq M, \forall~y\geq0, \label{eq1_KKT7}\\
& \zeta\left[\frac{1}{f_{\max}}\!-\!\int_0^\infty [y+z(y)] d\mu_Y(y)\!\right] = 0,\label{eq1_KKT4}\\
&  \gamma(y)z(y)=0, \forall~y\geq0, \label{eq1_KKTend1}\\
& \tau(y)(z(y)-M) = 0, \forall~y\geq0. \label{eq1_KKTend}
\end{align}


We now solve the KKT conditions \eqref{eq1_KKT1}-\eqref{eq1_KKTend} by using the calculus of variations.
The one-sided G\^ateaux derivative (similar to sub-gradient in finite dimensional space) of a functional $h$ in the direction of $w\in L^2(\mu_Y)$ at $z\in L^2(\mu_Y)$ is defined as
\begin{align}\label{derivative}
\delta h(z;w)\triangleq& \lim_{\epsilon\rightarrow 0^+} \frac{h(z+\epsilon w)-h(z)}{\epsilon}.
\end{align}
If $h$ is a function on $\mathbb{R}$, then \eqref{derivative} becomes the common one-sided derivative.
Let $l(z,y,\zeta,\gamma,\tau)$ denote the integrand in \eqref{eq_Lagrange}, and $r(z,y)=\mathbb{E}\left[q(y,z(y),Y')|Y=y\right]$. According to Lemma \ref{lem_expect_convex}, the function $q(y,z,y')$ and functionals $r(z,y)$, $l(z,y,\zeta,\gamma,\tau)$, and $L(z,\zeta,\gamma,\tau)$ are all convex in $z$. Therefore, their one-sided G\^ateaux derivatives with respect to $z$ exist \cite[p. 709]{Bertsekas}.
Since $g(x)$ is right-continuous, for any given $(y,y')$, the one-sided derivative $\delta q(y,z;w,y')$ of function $q(y,z,y')$ with respect to $z$ is given by
\begin{align}
&\delta q(y,z;w,y')\nonumber\\
=& \lim_{\epsilon\rightarrow 0^+}\frac{1}{\epsilon}\big\{q(y,z+\epsilon w,y')\!-\! q(y,z,y')\big\}\nonumber\\
= &\left\{\begin{array}{l l}\lim\limits_{x\rightarrow z^+}g(y+x+y')w, & \text{if} ~w\geq0;\\
\lim\limits_{x\rightarrow z^-}g(y+x+y')w, & \text{if} ~w<0.\end{array}\right.\nonumber
\end{align}
Next, consider the one-sided Ga\^teaux derivative $\delta r(z;w,y)$ of functional $ r(z,y)$. Since the function $g: [0,\infty)\rightarrow [0,\infty)$ is non-decreasing, $z \rightarrow q(y,z,y')$ is convex and finite for all $z \in[0,M]$. Hence, the function
$\epsilon\rightarrow [q(y,z+\epsilon w,y')-q(y,z,y')]/\epsilon$ is non-decreasing and bounded from above on $(0,a]$ for some $a>0$ \cite[Proposition 1.1.2(i)]{infinite_dimensional}. 
By using the monotone convergence theorem \cite[Theorem 1.5.6]{Durrettbook10}, we can interchange the limit and integral operators in $\delta r(z;w,y)$ such that
\begin{align}
&\delta r(z;w,y) \nonumber\\
=&\lim_{\epsilon\rightarrow 0^+}\frac{1}{\epsilon}\mathbb{E}\left[ q(y,z(y)+\epsilon w(y),Y')- q(y,z(y),Y')|Y=y\right]\nonumber\\
=&\mathbb{E}\!\left[ \lim_{\epsilon\rightarrow 0^+}\frac{1}{\epsilon}\big\{q(y,z(y)+\epsilon w(y),Y')\!-\! q(y,z(y),Y')\big\}\bigg|Y=y\right]\nonumber\\
=& \mathbb{E}\left[\lim\limits_{x\rightarrow z(y)^+}g(y+x+Y')w(y) 1_{\{w(y)>0\}}\right.\nonumber\\
&\left.+\lim\limits_{x\rightarrow z(y)^-}g(y+x+Y')w(y) 1_{\{w(y)<0\}}\bigg|Y=y\right]\nonumber\\
=& \lim\limits_{x\rightarrow z(y)^+}\mathbb{E}\left[g(y+x+Y')w(y) 1_{\{w(y)>0\}}\bigg|Y=y\right]\nonumber\\
&+\lim\limits_{x\rightarrow z(y)^-}\mathbb{E}\left[g(y+x+Y')w(y) 1_{\{w(y)<0\}}\bigg|Y=y\right],\!\!\!\!\!\!\!\!\label{eq_inequal3}
\end{align}
where $1_{E}$ is the indicator function of event $E$.
By using the monotone convergence theorem again, we have
\begin{align}
&\delta L(z;w,\zeta,\gamma,\tau) \nonumber\\
=& \int_0^\infty \delta l(z;w,y,\zeta,\gamma,\tau) d\mu_Y(y)\nonumber\\
=& \int_0^\infty \delta r(z;w,y) d\mu_Y(y)   \nonumber\\
&+ \int_0^\infty \left[-(c +\zeta)- \gamma(y)  +\tau(y)\right]w(y) d\mu_Y(y).
\end{align}
According to \cite[p. 710]{Bertsekas}, $z$ is an optimal solution to \eqref{eq1_KKT1} if and only if
\begin{align}\label{eq_inequal0}
\delta L(z;w,\zeta,\gamma,\tau) \geq 0,~~\forall ~w\in L^2(\mu_Y).
\end{align}
Since $w(\cdot)$ is an arbitrary function in $L^2(\mu_Y)$, considering positive functions $w(y)>0$, we can obtain from \eqref{eq_inequal3}-\eqref{eq_inequal0} that for each $y\geq0$, $z(y)$ must satisfy
\begin{align}\label{eq_inequal1}
\lim\limits_{x\rightarrow z(y)^+}\mathbb{E}\left[ g(y\!+\!x\!+\!Y')|Y=y\right]\! - \!(c+\zeta)\!-\!\gamma(y)\!+\!\tau(y)\geq0.\!\!
\end{align}
Similarly, considering  negative functions $w(y)<0$, we can obtain that for each $y\geq0$, $z(y)$ must satisfy
\begin{align}
\lim_{x\rightarrow z(y)^-} \mathbb{E}\left[ g(y\!+\!x\!+\!Y')|Y=y\right]\! -\! (c+\zeta)\!-\!\gamma(y)\!+\!\tau(y)\leq0.\!\!\label{eq_inequal2}
\end{align}
Because $g(\cdot)$ is non-decreasing, we can obtain from \eqref{eq_inequal1} and \eqref{eq_inequal2} that for each $y\geq0$, $z(y)$ needs to satisfy
\begin{align}\label{eq_inequal3_1}
&\mathbb{E}\left[ g(y\!+\!x\!+\!Y')|Y=y\right]\! - \!(c+\zeta)\!-\!\gamma(y)\!+\!\tau(y)\geq0
\end{align}
for all $x>z(y)$, and
\begin{align}
&\mathbb{E}\left[ g(y\!+\!x\!+\!Y')|Y=y\right]\! -\! (c+\zeta)\!-\!\gamma(y)\!+\!\tau(y)\leq0\label{eq_inequal4}
\end{align}
for all $x<z(y)$.

We solve the optimal primal solution $z(\cdot)$ by considering the following three cases: 

\emph{Case 1:} $\gamma(y)=\tau(y)=0$. The solutions to \eqref{eq_inequal3_1} and \eqref{eq_inequal4} may not be unique, if function $g$ is not strictly increasing. In particular, there may exist an interval $[a(y),b(y)]$ such that each $z(y)\in [a(y),b(y)]$ satisfies \eqref{eq_inequal3_1} and \eqref{eq_inequal4} for some $y$. In this case, we choose the largest possible solution of $z(y)$ to make sure that the constraint \eqref{eq_constraint} is satisfied. The largest solution satisfying \eqref{eq_inequal3_1} and \eqref{eq_inequal4} is given by
\begin{align}
\!\!\!z(y)=\sup\{x\in[0,M]:\mathbb{E}\left[ g(y\!+\!x\!+\!Y')|Y=y\right]\leq c+\zeta\}, \!\!\!\!\!\!\nonumber\\
~\forall~y\geq0,\!\!\nonumber
\end{align}
which is exactly \eqref{z1_solution}.

\emph{Case 2:} $\gamma(y)>0$. By \eqref{eq1_KKTend1}, we have $z(y)=0$.

\emph{Case 3:} $\tau(y)>0$. By \eqref{eq1_KKTend}, we have $z(y)=M$.
 
In summary, the optimal primal solution $z(\cdot)$ is given by \eqref{z1_solution}. 

Next, we find the optimal dual variable $\zeta$. By \eqref{eq1_KKT5} and \eqref{eq1_KKT4}, the optimal $\zeta$ satisfies
\begin{align}
\zeta=0, \mathbb{E}\left[Y+z(Y)\right]\geq \frac{1}{f_{\max}}
\end{align}
or
\begin{align}\label{eq_alpha}
\zeta>0, \mathbb{E}\left[Y+z(Y)\right]= \frac{1}{f_{\max}},
\end{align}
where $\mathbb{E}\left[Y+z(Y)\right]$ is determined by the optimal primal solution \eqref{z1_solution}.
Since $\mathbb{E}\left[Y+z(Y)\right]$ is non-decreasing in $\zeta$, we can use bisection to search for the optimal $\zeta$. By this, an optimal solution to \eqref{eq_SD_equavilent} is obtained for any given $c$. Finally, according to Sections 4.2.5 and 11.4 of \cite{Boyd04}, the optimal $c$ is solved by an outer-layer bisection search.
Therefore, an optimal solution to Problem \eqref{eq_SD} is given by Algorithm \ref{alg1}. This completes the proof.

\section{Proof of Lemma \ref{thm_convex}}\label{app5}
Let us rewrite the functional $h_1$ as
\begin{align}
h_1(z) = \frac{\int_0^\infty [y+z(y)]^2 d\mu_Y(y)}{\int_0^\infty [y+z(y)] d\mu_Y(y)}.\nonumber
\end{align}
We need to prove that the functional
$h_1$ is convex when restricted to any line that intersects its domain.
For any $w\in L^2(\mu_Y)$, consider the function $u: \mathbb{R} \rightarrow \mathbb{R}$ defined as
\begin{align}
u(\epsilon) = \frac{\int_0^\infty [z(y)+\epsilon w(y)+y]^2 d\mu_Y(y)}{\int_0^\infty [z(y)+\epsilon w(y)+y] d\mu_Y(y))}\nonumber
\end{align}
with domain
$$\text{\textbf{dom} } u = \left\{\epsilon: z(y)+\epsilon w(y)\in[0,M], ~\forall y\geq0,\epsilon \in \mathbb{R} \right\}.$$
Since the function $\epsilon \rightarrow [z(y)+\epsilon w(y)+y]^2$ is convex, the function $x\rightarrow \{[z(y)+(\epsilon +x) w(y)+y]^2-[z(y)+\epsilon w(y)+y]^2\}/x$ is non-decreasing and bounded from above on $(0,a]$ for some $a>0$. By using monotone convergence theorem \cite[Theorem 1.5.6]{Durrettbook10}, we can interchange the limit and integral operators such that
\begin{align}
&\frac{d}{d\epsilon} \int_0^\infty [z(y)+\epsilon w(y)+y]^2 d\mu_Y(y) \nonumber\\
=&\int_0^\infty 2[z(y)+\epsilon w(y)+y]w(y) d\mu_Y(y). \nonumber
\end{align}
Similarly,
\begin{align}
\frac{d}{d\epsilon} \int_0^\infty [z(y)+\epsilon w(y)+y] d\mu_Y(y)=\int_0^\infty w(y) d\mu_Y(y). \nonumber
\end{align}
By this, we have
\begin{align}\label{eq_du}
&\frac{du}{d\epsilon} = \frac{\int_0^\infty 2[z(y)+\epsilon w(y)+y] w(y)d\mu_Y(y)}{\int_0^\infty [z(y)+\epsilon w(y)+y] d\mu_Y(y)} \nonumber\\
&~~~~-\frac{\int_0^\infty [z(y)\!+\!\epsilon w(y)+y]^2d\mu_Y(y)\int_0^\infty w(y)d\mu_Y(y)}{\left[\int_0^\infty [z(y)+\epsilon w(y)+y] d\mu_Y(y)\right]^2}.\!
\end{align}
After some additional manipulations, we can obtain
\begin{align}
&\!\!\frac{d^2u}{d^2\epsilon} = \frac{2\left[\int_0^\infty [y\!+\!z(y)] d\mu_Y(y) \int_0^\infty w(y) d\mu_Y(y)\right]^2}{\left[\int_0^\infty [z(y)+\epsilon w(y)+y] d\mu_Y(y)\right]^3} \nonumber\\
&\!\!\times\!\!\int_0^\infty\!\!\left[\frac{y\!+\!z(y)}{\int_0^\infty [y\!+\!z(y)] d\mu_Y(y)}-\frac{w(y)}{\int_0^\infty w(y) d\mu_Y(y)}\right]^2\!\!\!d\mu_Y(y).\nonumber
\end{align}
Since $z(y)+\epsilon w(y)\geq0$ for all $y$ on $\text{\textbf{dom} } u$, we have $\frac{d^2u}{d^2\epsilon}\geq0$. Hence, the function $u$ is convex for all $w\in L^2(\mu_Y)$. By this, the functional
$h_1$ is convex, which completes the proof.

\section{Proof of Theorem \ref{beta}}\label{app6}
The Lagrangian of Problem \eqref{eq_SD_average} is determined as
\begin{align}
&L_1(z,\zeta,\gamma,\tau)\nonumber\\ =& \frac{\int_0^\infty [y+z(y)]^2 d\mu_Y(y)}{2\int_0^\infty [y+z(y)] d\mu_Y(y)}
 \!+\! \zeta\left[\frac{1}{f_{\max}}\!-\!\int_0^\infty [y+z(y)] d\mu_Y(y)\!\right] \nonumber\\
 &- \int_0^\infty \gamma(y)z(y) d\mu_Y(y)
 +\int_0^\infty \tau(y)(z(y)-M) d\mu_Y(y),\nonumber
\end{align}
where $\zeta\in \mathbb{R}$, $\gamma,\tau\in L^2(\mu_Y)$ are dual variables. According to \cite[Proposition 3.3.2]{infinite_dimensional} and \cite[pp. 70-72]{convex_analysis}, the KKT theorem remains valid for the Lebesgue space $L^2(\mu_Y)$. Hence, if a vector $(z,\zeta,\gamma,\tau)$ satisfies the KKT conditions \eqref{eq_KKT1}-\eqref{eq_KKTend}, it is an optimal solution to \eqref{eq_SD_average}. The KKT conditions are given by:
\begin{align}
&z = \min_{x\in L^2(\mu_Y)} L_1(x,\zeta,\gamma,\tau),\label{eq_KKT1}\\
&\zeta\geq0, \int_0^\infty [y+z(y)] d\mu_Y(y)\geq \frac{1}{f_{\max}},\label{eq_KKT5}\\
& \gamma(y)\geq0, z(y)\geq0, \forall~y\geq0, \label{eq_KKT6}\\
& \tau(y)\geq0, z(y)\leq M, \forall~y\geq0, \label{eq_KKT7}\\
& \zeta\left[\frac{1}{f_{\max}}\!-\!\int_0^\infty [y+z(y)] d\mu_Y(y)\!\right] = 0,\label{eq_KKT4}\\
&  \gamma(y)z(y)= 0, \forall~y\geq0, \label{eq_KKTend1}\\
& \tau(y)(z(y)-M) = 0, \forall~y\geq0. \label{eq_KKTend}
\end{align}
We now solve the KKT conditions by using the calculus of variations. For any fixed $(\zeta,\gamma,\tau)$, the G\^ateaux derivative of the Lagrange $L_1$ in the direction of $w\in L^2(\mu_Y)$ at $z\in L^2(\mu_Y)$ is defined as
\begin{align}
\delta L_1(z;w,\zeta,\gamma,\tau)\triangleq& \lim_{\epsilon\rightarrow 0} \frac{L_1(z+\epsilon w,\zeta,\gamma,\tau)-L_1(z,\zeta,\gamma,\tau)}{\epsilon}.\nonumber
\end{align}
Similar to the derivations of \eqref{eq_du}, we can obtain
\begin{align}
&\delta L_1(z;w,\zeta,\gamma,\tau)\nonumber\\
=&\!\int_0^\infty\!\!\left[ \frac{y+z(y)}{\int_0^\infty [y+z(y)] d\mu_Y(y)}
-\frac{\int_0^\infty [y+z(y)]^2 d\mu_Y(y)}{2\left[\int_0^\infty [y+z(y)] d\mu_Y(y)\right]^2}\right.\nonumber\\
&-\zeta -\gamma(y) + \tau(y)\Bigg]w(y) d\mu_Y(y),~~\forall ~w\in L^2(\mu_Y). \nonumber
\end{align}
Then, $z(\cdot)$ is an optimal solution to \eqref{eq_KKT1} if and only if \cite[p. 710]{Bertsekas}
\begin{align}
\delta L_1(z;w,\zeta,\gamma,\tau) \geq 0,~~\forall ~w\in L^2(\mu_Y).\nonumber
\end{align}
By $\delta L_1(z;w,\zeta,\gamma,\tau)=-\delta L_1(z;-w,\zeta,\gamma,\tau)$, we deduce
\begin{align}
\delta L_1(z;w,\zeta,\gamma,\tau) = 0,~~\forall ~w\in L^2(\mu_Y).\nonumber
\end{align}
Since $w(\cdot)$ is arbitrary, we have
\begin{align}\label{eq_KKT2}
&\frac{y+z(y)}{\int_0^\infty [y+z(y)] d\mu_Y(y)}
-\frac{\int_0^\infty [y+z(y)]^2 d\mu_Y(y)}{2\left[\int_0^\infty [y+z(y)] d\mu_Y(y)\right]^2}\nonumber\\
&-\zeta -\gamma(y) + \tau(y)=0,~\forall~y\geq0.
\end{align}
For notational simplicity, let us define
\begin{align}\label{eq_KKT3}
\beta \triangleq \zeta \int_0^\infty [y\!+\!z(y)] d\mu_Y(y) \!+\! \frac{\int_0^\infty [y+z(y)]^2 d\mu_Y(y)}{2 \int_0^\infty [y\!+\!z(y)] d\mu_Y(y)}.
\end{align}
Since $\mathbb{E}[Y]>0$, we have $\beta>0$.
The optimal primal solution $z(\cdot)$ is obtained by considering the following three cases: 

\emph{Case 1:} If $\gamma(y)=\tau(y)=0$, then by \eqref{eq_KKT2} and \eqref{eq_KKT3}, we obtain $z(y)=\beta - y$. In this case, we require $\beta - y\in[0,M]$ by \eqref{eq_KKT6} and \eqref{eq_KKT7}.

\emph{Case 2:} If $\gamma(y)>0$, then by \eqref{eq_KKTend1}, $z(y)=0$.

\emph{Case 3:} If $\tau(y)>0$, then by \eqref{eq_KKTend}, $z(y)=M$. 

In summary, the optimal primal solution $z(\cdot)$ is given by \eqref{z_solution}. 

The optimal dual variable $\beta$ is obtained by considering two cases:

\emph{Case 1:} $\zeta>0$. Then, \eqref{eq_KKT4} and \eqref{eq_KKT3} imply that
\begin{align}\label{eq_case1}
\int_0^\infty [y\!+\!z(y)] d\mu_Y(y)\! =\!\frac{1}{f_{\max}},
\beta \geq \frac{\int_0^\infty [y+z(y)]^2 d\mu_Y(y)}{2 \int_0^\infty [y\!+\!z(y)] d\mu_Y(y)}.\!\!\!
\end{align}

\emph{Case 2:} $\zeta=0$. Then, \eqref{eq_KKT5} and \eqref{eq_KKT3} imply that
\begin{align}\label{eq_case2}
\int_0^\infty [y\!+\!z(y)] d\mu_Y(y)\!\geq\!\frac{1}{f_{\max}},
\beta = \frac{\int_0^\infty [y+z(y)]^2 d\mu_Y(y)}{2 \int_0^\infty [y\!+\!z(y)] d\mu_Y(y)}.\!\!\!
\end{align}
Combining \eqref{eq_case1} and \eqref{eq_case2}, yields
 \begin{align}
 \int_0^\infty\! [y\!+\!z(y)] d\mu_Y(y) \!=\! \max\left(\frac{1}{f_{\max}},\frac{\int_0^\infty\! [z(y)\!+\!y]^2 d\mu_Y(y)}{2\beta}\right)\!.\nonumber
 \end{align}
Then, \eqref{Ebeta} is obtained. This completes the proof.


\section{Proof of Theorem \ref{iffcondition}}\label{app8}
Since $\mathbb{E}[Y]\geq  \frac{1}{f_{\max}}$, the constraint \eqref{eq_SD_average_con} is always satisfied and can be removed.
If the zero-wait policy is optimal, we solve the KKT conditions of Problem \eqref{eq_SD_average} without the constraint \eqref{eq_SD_average_con}. By this, we can obtain that the optimal primal solution is given by \eqref{z_solution} almost everywhere, and the optimal dual solution $\beta$ must satisfy
\begin{align}\label{eq_iffcondition1}
\beta \leq y_{\inf},~~ \mathbb{E}[Y^2] = 2\beta \mathbb{E}[Y],
\end{align}
from which \eqref{eq_iffcondition} follows. 

Next, we prove the reverse direction. If \eqref{eq_iffcondition} holds, by $\mathbb{E}[Y]>0$, we can get $\mathbb{E}[Y^2]>0$ and $y_{\inf}>0$. By \eqref{eq_iffcondition} and choosing $\beta=\frac{\mathbb{E}[Y^2]}{2\mathbb{E}[Y]}>0$, we obtain \eqref{eq_iffcondition1}. Substituting \eqref{eq_iffcondition1} and $\mathbb{E}[Y]\geq  \frac{1}{f_{\max}}$ into Theorem \ref{beta}, yields that the zero-wait policy is optimal. This completes the proof.

\section{Proof of Lemma \ref{just_in_time}}\label{app7}
1). When the correlation coefficient between $Y_i$ and $Y_{i+1}$ is $-1$, $Y+Y'$ is equal a constant value with probability one. Choosing $z(y)=0$, $\zeta = 0$, $c=g(Y+Y')$, $\gamma(y)=\tau(y)=0$, one can show that the KKT conditions \eqref{eq1_KKT1}-\eqref{eq1_KKTend} are satisfied.

2). If the $Y_i$'s are equal to a constant value, $Y+Y'$ is equal a constant value. The remaining proof follows from part 1).

3). When $g(\cdot)$ is a constant function, all policies are optimal. This completes the proof.

\fi

\bibliographystyle{IEEEtran}
\bibliography{./v4/sueh}
\end{document}